\documentclass[aps,prd,reprint,superscriptaddress,showpacs]{revtex4-1}

\usepackage{mathtools} % CYC
\usepackage{hyperref} % CYC

\newcommand\lar{\leftarrow}
\newcommand\rar{\rightarrow}
\newcommand\lrar{\leftrightarrow}

\newcommand\qlar{{\lar q}}
\newcommand\qrar{{q\rar}}
\newcommand\qminus{{-q \lrar}}
\newcommand\qplus{{\lrar q+}}

\begin{document}

%\preprint{}

\title{Conservative 3+1 General Relativistic Variable Eddington Tensor \\ Radiation Transport Equations}


\author{Christian Y. Cardall}
\affiliation{Physics Division, Oak Ridge National Laboratory, Oak Ridge, TN 37831-6354, USA}
\affiliation{Department of Physics and Astronomy, University of Tennessee, Knoxville, TN 37996-1200, USA}
\author{Eirik Endeve}
\affiliation{Computer Science and Mathematics Division, Oak Ridge National Laboratory, Oak Ridge, TN 37831-6354, USA}
\author{Anthony Mezzacappa}
\affiliation{Physics Division, Oak Ridge National Laboratory, Oak Ridge, TN 37831-6354, USA}
\affiliation{Department of Physics and Astronomy, University of Tennessee, Knoxville, TN 37996-1200, USA}
\affiliation{Computer Science and Mathematics Division, Oak Ridge National Laboratory, Oak Ridge, TN 37831-6354, USA}
%\thanks{}


\date{\today}

\begin{abstract}
We present conservative 3+1 general relativistic variable Eddington tensor radiation transport equations, including greater elaboration of the momentum space divergence (that is, the energy derivative term) than in previous work.
These equations are intended for use in simulations involving numerical relativity, particularly in the absence of spherical symmetry.
The independent variables are the lab frame coordinate basis spacetime position coordinates and the particle energy measured in the comoving frame. 
With an eye towards astrophysical applications---such as core-collapse supernovae and compact object mergers---in which the fluid includes nuclei and/or nuclear matter at finite temperature, and in which the transported particles are neutrinos, we pay special attention to the consistency of four-momentum and lepton number exchange between neutrinos and the fluid, showing the term-by-term cancellations that must occur for this consistency to be achieved.    
\end{abstract}

\pacs{95.30.Jx, 05.20.Dd, 47.70.-n, 97.60.Bw}
%\keywords{}

\maketitle

\section{Introduction \label{sec:Introduction}}

Neutrino transport is a necessary ingredient of core-collapse supernova simulations \cite{Mezzacappa2005ASCERTAINING-TH,Kotake2006Explosion-mecha,Kotake2012Multimessengers,Kotake2012Core-Collapse-S,Janka2012Explosion-Mecha,Burrows2012Perspectives-on,Janka2012Core-collapse-s}. 
Determining the fate of the stellar material---for instance, does an explosion happen, and if so, how?---requires calculation of the four-momentum and lepton number exchange between the fluid (which includes nuclei and/or nuclear matter at finite temperature) and the neutrinos that stream from and through it. 
%For the purpose of studying the explosion mechanism, it is typical, and perhaps adequate \cite{Chakraborty2011No-Collective-N,Suwa2011Impacts-of-Coll,Dasgupta2012Role-of-collect,Sarikas2012Suppression-of-,Saviano2012Stability-analy,Sarikas2012Supernova-neutr,Pejcha2012Effect-of-colle}, to consider only massless neutrinos described by classical distribution functions (phase space densities) $f(t,\mathbf{x},\mathbf{p})$. (In contrast, calculation of the emerging neutrino signals---of intrinsic interest as an observational probe of the core-collapse supernova environment, and of the properties of the neutrinos themselves---requires treatment of the quantum effects induced by neutrino mass and flavor mixing \cite{Dasgupta2010Physics-and-Ast,Raffelt2011New-opportuniti,Dighe2011Signatures-of-s}.)
For the purpose of studying the explosion mechanism, we take the traditional approach and consider only massless neutrinos described by classical distribution functions (phase space densities) $f(t,\mathbf{x},\mathbf{p})$ 
\footnote{Calculation of the emerging neutrino signals---of intrinsic interest as an observational probe of the core-collapse supernova environment, and of the properties of the neutrinos themselves---definitely requires treatment of the quantum effects induced by neutrino mass and flavor mixing \cite{Dasgupta2010Physics-and-Ast,Raffelt2011New-opportuniti,Dighe2011Signatures-of-s}.
Recent explorations suggest that flavor mixing does not impact the explosion mechansim \cite{Chakraborty2011No-Collective-N,Suwa2011Impacts-of-Coll,Dasgupta2012Role-of-collect,Sarikas2012Suppression-of-,Saviano2012Stability-analy,Sarikas2012Supernova-neutr,Pejcha2012Effect-of-colle}, 
but consensus on the impacts of flavor mixing in supernovae has a fickle history,
and future more definitive simulations that include neutrino transport with quantum kinetics could surprise us with flavor mixing effects on the explosion mechanism as well.}.

Solution for the neutrino distributions $f(t,\mathbf{x},\mathbf{p})$ in their full dimensionality---1D time + 3D position space + 3D momentum space---is
%, and for all interactions and physical effects on all neutrino species, is 
beyond current computational capabilities.
Thus various approximations have been employed, in particular various permutations of reduction in dimensionality, and in many cases exclusion of effects that alter neutrino energies (energy-changing scattering interactions, and Doppler and gravitational shifts; on the importance of these, see for instance Refs.~\cite{Lentz2012On-the-Requirem,Lentz2012Interplay-of-Ne}).
We leave detailed discussions of these developments---especially in spherical and axial symmetry---to the above-cited reviews and overviews \cite{Mezzacappa2005ASCERTAINING-TH,Kotake2006Explosion-mecha,Kotake2012Multimessengers,Kotake2012Core-Collapse-S,Janka2012Explosion-Mecha,Burrows2012Perspectives-on,Janka2012Core-collapse-s}, noting also an additional recent report of explosions in axisymmetry across a range of progenitor masses with self-consistent neutrino transport \cite{Bruenn2013Axisymmetric-Ab}. 

For present purposes, we note that the focus of the field is turning towards simulations that are 3D in position space, and that in this context treatments of neutrino transport remain in early stages.
The simplest treatments---which are not neutrino transport per se---are `light bulb' approaches with externally imposed, parametrized neutrino heating and cooling functions (e.g. Refs.~\cite{Iwakami2009Effects-of-Rota,Hanke2012Is-Strong-SASI-,Dolence2013Dimensional-Dep}).
When it comes to more self-consistent neutrino transport, most approaches to date with 3D position space are `grey' (neutrino energy dependence integrated out), and/or `ray-by-ray' (solution along independent radial rays, neglecting lateral transport except perhaps in optically thick regions), and/or involve solution of only the lowest angular moments (see also the next paragraph). 
One approach going a step beyond light bulb schemes involves cooling by neutrino `leakage,' plus grey heating based on optical depths computed in a ray-by-ray fashion \cite{Ott2013General-Relativ}. 
Computing grey solutions for the zeroth moment are e.g. Ref.~ \cite{Wongwathanarat2013Three-dimension} (ray-by-ray and with a prescribed inner boundary luminosity at finite radius), and Ref.~\cite{Fryer2006SNSPH:-A-Parall} (in connection with smoothed-particle hydrodynamics). 
Grey solutions for both the zeroth and first moments are obtained in Ref.~\cite{Kuroda2012Fully-General-R}, in which the source terms describing neutrino-matter interactions---which normally induce significant computational costs due to the need for implicit solution---are simplified with a leakage-type approach.
Energy-dependent ray-by-ray simulations solving self-consistently for the zeroth moment in 3D have been reported as underway \cite{Bruenn20092D-and-3D-core-}.
The only energy-dependent transport results in 3D position space completed to date are ray-by-ray: one with the `isotropic diffusion source approximation' (IDSA), in which the neutrinos are separated into diffusive and free-streaming components, with a prescription for exchange between them \cite{Takiwaki2012Three-dimension};
and another solving self-consistently for the zeroth and first moments \cite{Hanke2013SASI-Activity-i}.
First tests of a full 3D+3D Boltzmann solver have been reported \cite{Sumiyoshi2012Neutrino-Transf,Kotake2012Core-Collapse-S}, 
but it will be some time before fully detailed and well-resolved simulations with such solvers are computationally tractable.
Other approaches that traditionally have not been used in core-collapse supernova studies, but that are being considered for future simulations with 3D position space, include a Monte Carlo scheme \cite{Abdikamalov2012A-New-Monte-Car} and a spherical harmonics expansion \cite{Radice2012A-New-Spherical}.

%The formidable challenge of solving for the neutrino distribution functions $f(t,\mathbf{x},\mathbf{p})$ can be ameliorated by solving only for their angular moments, with some form of closure 
%(e.g. flux-limited diffusion 
%\footnote{For additional detail on the scheme used in the simulations reported in Ref.~\cite{Bruenn20092D-and-3D-core-}, see Appendix~A of Ref.~\cite{Bruenn1985Stellar-core-co} and Sec.~4.7 of Ref.~\cite{Liebendorfer2004A-Finite-Differ}.},\footnote{See the Appendix of Ref.~\cite{Burrows2007Features-of-the}.},\cite{Swesty2009A-Numerical-Alg} or a variable Eddington tensor \cite{Rampp2002Radiation-hydro,Muller2010A-New-Multi-dim,Obergaulinger2011Magnetic-field-,Shibata2011Truncated-Momen,Endeve2012Conservative-Mu}) serving to truncate the scheme at low order.
Given the current state of the field vis-\`{a}-vis 3D position space simulations,
it seems likely that a viable choice for many practitioners will be an approach in which solutions to only angular moments of $f(t,\mathbf{x},\mathbf{p})$ are sought,
with some form of closure
serving to truncate the scheme at low order (see e.g. Ref. \cite{Smit2000Closure-in-flux}). 
Common examples include flux-limited diffusion \footnote{For additional detail on the scheme used in the simulations reported in Ref.~\cite{Bruenn20092D-and-3D-core-}, see Appendix~A of Ref.~\cite{Bruenn1985Stellar-core-co} and Sec.~4.7 of Ref.~\cite{Liebendorfer2004A-Finite-Differ}.},\footnote{See the Appendix of Ref.~\cite{Burrows2007Features-of-the}.},\cite{Swesty2009A-Numerical-Alg}, truncated at the zeroth moment $\mathcal{J}(t,\mathbf{x},\epsilon)$, with prescriptions for the first and second moments $\mathcal{H}^{\hat{\imath}}(t,\mathbf{x},\epsilon)$ and $\mathcal{K}^{\hat{\imath}\hat{\jmath}}(t,\mathbf{x},\epsilon)$;
and a variable Eddington tensor approach \cite{Rampp2002Radiation-hydro,Muller2010A-New-Multi-dim,Obergaulinger2011Magnetic-field-,Shibata2011Truncated-Momen,Kuroda2012Fully-General-R,OConnor2012The-Progenitor-,Endeve2012Conservative-Mu}, 
truncated at the first moment $\mathcal{H}^{\hat{\imath}}(t,\mathbf{x},\epsilon)$, with prescriptions for the second and third moments $\mathcal{K}^{\hat{\imath}\hat{\jmath}}(t,\mathbf{x},\epsilon)$ and $\mathcal{L}^{\hat{\imath}\hat{\jmath}\hat{k}}(t,\mathbf{x},\epsilon)$ (see Sec.~\ref{sec:VET}). 
%(Note that Ref.~\cite{Kuroda2012Fully-General-R} integrates out the neutrino energy as well as angle dependence, which entails several major simplifications: in addition to the further reduction of dimensionality, it eliminates neutrino-energy-changing interactions, velocity-dependent terms, and the third moment $\mathcal{L}^{\hat{\imath}\hat{\jmath}\hat{k}}$. 
%Energy dependence is retained in Refs.~\cite{Obergaulinger2011Magnetic-field-} and \cite{OConnor2012The-Progenitor-}, but neutrino energy-changing interactions are not included, and velocity-dependent terms are dropped in Ref.~\cite{OConnor2012The-Progenitor-} as well. 
%On the importance of the neutrino-energy-changing interactions and velocity-dependent terms, see Ref.~\cite{Lentz2012On-the-Requirem}.)

Relative to full Boltzmann simulations for $f(t,\mathbf{x},\mathbf{p})$, the reduction in momentum space dimensionality afforded by moments approaches yields important savings in the memory needed to run simulations, and the impact on the number of floating point operations (flops) required is even greater.
Memory needs grow quadratically (or even linearly, in a matrix-free approach) with the number $N_\mathbf{p} = N_\nu N_\epsilon N_\vartheta N_\varphi$ of neutrino momentum space cells arising from $N_\nu$ neutrino species, $N_\epsilon$ energy bins, and $N_\vartheta N_\varphi$ angle bins.
In contrast, the flop count---which is dominated by the inversion of dense blocks representing momentum space couplings---grows as a higher power, something like $N_\mathbf{p}^{2-3}$ (\footnote{A power of 3 would come from LU decomposition of dense matrices representing couplings among all neutrino species, energies, and angles; a lower power in principle might be obtained by methods that exploit the substructure of the dense matrices. Savings also can be achieved by restricting allowable couplings, for instance between species or energies.}; see also Refs.~\cite{Cardall2012Towards-the-Cor,Kotake2012Core-Collapse-S}). 
Therefore moments approaches can be expected to required a number of flops that is smaller by a factor of order $\left(N_\vartheta N_\varphi \right)^{2-3} \gg 1$.

Notably for the focus of this paper, conservative 3+1 general relativistic variable Eddington tensor radiation moment equations are presented by Shibata et al. \cite{Shibata2011Truncated-Momen}.
Their variables are functions of lab frame coordinate basis spacetime position coordinates $x^\mu$---that is, the spacetime coordinates that appear in the 3+1 metric---but the momentum space dependence is on the neutrino energy $\epsilon$ measured in the comoving frame.
The long-recognized freedom to choose different coordinate systems for spacetime and momentum space (e.g. Refs. \cite{Lindquist1966Relativistic-Tr,Mihalas1980Solution-of-the,Munier1986Radiation-trans,Munier1986Radiation-trans2,Riffert1986A-general-Euler,Mezzacappa1989Computer-simula,Schinder1988General-relativ}) allows particle/fluid interactions to be evaluated in the comoving frame in the context of Eulerian grid-based approaches to multidimensional spatial dependence.
One
difference between our presentation from that of Shibata et al. \cite{Shibata2011Truncated-Momen} is our starting point. 
%Like Endeve et al. \cite{Endeve2012Conservative-Mu}, we 
We
begin from conservative reformulations \cite{Cardall2003Conservative-fo} of the general relativistic Boltzmann equation \cite{Lindquist1966Relativistic-Tr,Ehlers1971General-Relativ,Israel1972The-Relativisti,Riffert1986A-general-Euler,Mezzacappa1989Computer-simula} rather than the moments formalism of Thorne \cite{Thorne1981Relativistic-ra}.
(The conservative reformulations of the general relativistic Boltzmann equation in Ref.~\cite{Cardall2003Conservative-fo}, and a special relativistic specialization \cite{Cardall2005Conservative-sp}, were inspired by previous conservative formulations in spherical symmetry, e.g. \cite{Bruenn1985Stellar-core-co,Mezzacappa1989Computer-simula,Liebendorfer2002Conservative-ge}.
While we hope to solve the conservative multidimensional general relativistic Boltzmann equation in future core-collapse supernova simulations, its significance to the present work is the straightforward path it provides for the derivation of more computationally feasible moments equations.) 
By angular integration of the four-momentum conservative reformulation of the Boltzmann equation, we obtain in Sec.~\ref{sec:Relativistic_VET} a general relativistic variable Eddington tensor formalism in which the relationship between the lab frame (denoted by unadorned indices) and the comoving frame (denoted by hatted indices) is expressed in terms of coordinate transformations ${L^\mu}_{\hat{\mu}}$ and comoving frame connection coefficients ${\Gamma^{\hat{\mu}}}_{\hat{\nu}\hat{\rho}}$. %(see also Ref. \cite{Endeve2012Conservative-Mu}).

In specializing to the 3+1 formulation of general relativity in Sec.~\ref{sec:31_Formulation},
we extend the treatment of Shibata et al. \cite{Shibata2011Truncated-Momen} with a full elaboration of the momentum space divergence (i.e. the energy derivative term, in this angle-integrated moments case).
Important aspects of our approach include (a) consistent use of what we call `Eulerian decompositions' and `Eulerian projections,' which are natural to the 3+1 approach; and relatedly, (b) a shift from conceptualizing the relationship between the lab and comoving frames from coordinate transformations ${L^\mu}_{\hat{\mu}}$ to the (covariant) relative three-velocity $v^\mu$ connecting the four-velocities $n^\mu$ and $u^\mu$ of Eulerian and Lagrangian observers.
Our approach in Sec.~\ref{sec:31_Formulation} is more geometric than that in Sec.~\ref{sec:Relativistic_VET} (in conception if not notation); 
indeed it allows us to obtain explicit results while almost completely avoiding encounters with connection coefficients.

We also add to the treatment by Shibata et al. \cite{Shibata2011Truncated-Momen} by showing, in both Sec.~\ref{sec:Relativistic_VET} and Sec.~\ref{sec:31_Formulation}, how the four-momentum exchange with the fluid expressed by a conservative variable Eddington tensor formalism is consistent with a conservative treatment of lepton number exchange. 
%(see also Ref. \cite{Endeve2012Conservative-Mu}).
A conservative treatment of four-momentum exchange with the fluid, properly discretized for consistency with conservative number exchange, is expected to facilitate simultaneous energy and lepton conservation in numerical simulations---an important check on the physical reliability of simulation outcomes \cite{Liebendorfer2004A-Finite-Differ}.
In this respect, there may be room for improvement over cases in which this consistency has not been considered (e.g. Refs.~\cite{Kuroda2012Fully-General-R,OConnor2012The-Progenitor-}), or in which consistency between conservative number exchange and {\em non-conservative} four-momentum exchange has been addressed (e.g. Ref.~\cite{Muller2010A-New-Multi-dim}).
The nature of the consistency of our conservative four-momentum transport equations (modulo gravitational sources) with a conservative number transport equation is made particularly explicit in Sec.~\ref{sec:31_Formulation}, in which we elucidate the term-by-term cancellations that must occur for this consistency to be achieved.
In Sec.~\ref{sec:Conclusion} we discuss the moment equations, bringing together the pieces worked out in Sec.~\ref{sec:31_Formulation} and presenting overview tables of the many variables appearing in the formalism.

\section{General relativatistic variable Eddington tensor formalism}
\label{sec:Relativistic_VET}

After exhibiting the Boltzmann equation and its conservative reformulations in terms of number and four-momentum exchange with the fluid,
we obtain from the latter a variable Eddington tensor formalism in which the relationship between the lab frame (denoted by unadorned indices) and the comoving frame (denoted by hatted indices) is expressed in terms of coordinate transformations ${L^\mu}_{\hat{\mu}}$ and comoving frame connection coefficients ${\Gamma^{\hat{\mu}}}_{\hat{\nu}\hat{\rho}}$, and show how it relates to the angle-integrated number-conservative reformulation.
%This ground is also covered in Ref.~\cite{Endeve2012Conservative-Mu}, but is considered here to establish notation and provide a self-contained presentation.

\subsection{The Boltzmann equation}
\label{sec:Boltzmann}

Classical neutrino distribution functions $f(t,\mathbf{x},\mathbf{p})$ are governed by the Boltzmann equation \cite{Lindquist1966Relativistic-Tr,Ehlers1971General-Relativ,Israel1972The-Relativisti,Riffert1986A-general-Euler,Mezzacappa1989Computer-simula}.
In its geometric form, it states that the change in $f$ along a phase space trajectory with affine parameter $\lambda$ is equal to the phase space density $C[f]$ of point-like collisions that add or remove particles from the trajectory: 
\begin{equation}
\frac{df}{d\lambda} = C[f].
\label{eq:BoltzmannGeometric}
\end{equation}
The phase space measure is defined in such a way that $f$ and $C[f]$ are both invariant scalars.
For practical computations it is necessary to introduce phase space coordinates: spacetime coordinates $x^\mu$, and momentum space coordinates $p^i$ (the timelike momentum component $p^0$ is fixed in terms of the spacelike components $p^i$ by the mass shell constraint).
In terms of these coordinates, Eq.~(\ref{eq:BoltzmannGeometric}) becomes
\begin{equation}
\frac{dx^\mu}{d\lambda} \frac{\partial f}{\partial x^\mu} 
+ \frac{dp^i}{d\lambda} \frac{\partial f}{\partial p^i}
= C[f].
\label{eq:BoltzmannCoordinates} 
\end{equation}
The geodesic equations describing the trajectory are
\begin{eqnarray}
\frac{dx^\mu}{d\lambda} &=& p^\mu, \label{eq:Geodesic_x} \\
\frac{dp^\mu}{d\lambda} &=& - \Gamma^\mu_{\nu\rho} p^\nu p^\rho,
  \label{eq:Geodesic_p}
\end{eqnarray}
so that Eq.~(\ref{eq:BoltzmannCoordinates}) becomes
\begin{equation}
p^\mu \frac{\partial f}{\partial x^\mu} 
- \Gamma^i_{\nu\mu} p^\nu p^\mu \frac{\partial f}{\partial p^i} 
= C[f],
\label{eq:BoltzmannLab}
\end{equation}
now an integro-partial differential equation (the integrals appearing on the right-hand side).

There is freedom in choosing the spacetime and momentum space coordinates.
Taking the unadorned indices to denote what we shall call a lab frame coordinate basis (also called a `natural' or `holonomic' basis), the connection coefficients $\Gamma^\mu_{\nu\rho}$ are given in terms of the spacetime metric $g_{\mu\nu}$ as
\begin{equation}
\Gamma^\mu_{\nu\rho} 
= \frac{1}{2} g^{\mu \sigma} \left(
\frac{\partial g_{\sigma \nu}}{\partial x^\rho}
+ \frac{\partial g_{\sigma \rho}}{\partial x^\nu}
- \frac{\partial g_{\nu \rho}}{\partial x^\sigma}
\right).
\end{equation}
However, it is most convenient to express the particle interactions entering $C[f]$ in terms of momentum components $p^{\hat{\imath}}$ reckoned with respect to an orthonormal reference frame comoving with the fluid (a `comoving frame').
We define a composite transformation
\begin{equation}
{L^\mu}_{\hat{\mu}}
= {e^\mu}_{\bar{\mu}} {\Lambda^{\bar{\mu}}}_{\hat{\mu}}
\label{eq:CompositeTransformation}
\end{equation}
consisting of a Lorentz boost ${\Lambda^{\bar{\mu}}}_{\hat{\mu}}$ from an orthonormal comoving frame (denoted by indices with a hat) to an orthonormal lab frame (denoted by indices with a bar), 
followed by a transformation to the lab frame coordinate basis with a local tetrad ${e^\mu}_{\bar{\mu}}$.
This tetrad is independent of the fluid velocity; it locally transforms the metric into the Lorentz form $\left(\eta_{\bar\mu \bar\nu} \right) = \mathrm{diag}[-1,1,1,1]$:
\begin{equation}
{e^\mu}_{\bar{\mu}} {e^\nu}_{\bar{\nu}} \, g_{\mu\nu} = \eta_{\bar\mu \bar\nu}.
\end{equation}
Of course, the boost ${\Lambda^{\bar{\mu}}}_{\hat{\mu}}$ preserves the Lorentz metric;
this implies that the composite transformation ${L^\mu}_{\hat{\mu}}$ is itself also a tetrad.
The inverse of Eq.~(\ref{eq:CompositeTransformation}) is
\begin{equation}
{L^{\hat{\mu}}}_{{\mu}}
= {\Lambda^{\hat{\mu}}}_{\bar{\mu}} {e^{\bar{\mu}}}_{{\mu}}, 
\label{eq:CompositeTransformationInverse}
\end{equation}
expressed in terms of the inverse tetrad ${e^{\bar{\mu}}}_{{\mu}}$ and inverse boost ${\Lambda^{\hat{\mu}}}_{\bar{\mu}}$.
In terms of lab frame coordinate basis spacetime components and comoving frame momentum components, the Boltzmann equation reads
\begin{equation}
{L^\mu}_{\hat{\mu}} p^{\hat{\mu}} \frac{\partial f}{\partial x^\mu} 
- \Gamma^{\hat{\imath}}_{\hat{\nu}\hat{\mu}} p^{\hat{\nu}} p^{\hat{\mu}} \frac{\partial f}{\partial p^{\hat{\imath}}} 
= C[f],
\label{eq:BoltzmannComoving}
\end{equation}
where the connection coefficients in the comoving frame are
\begin{equation}
\Gamma^{\hat{\mu}}_{\hat{\nu}\hat{\rho}}
=  {L^{\hat{\mu}}}_{{\mu}} {L^\nu}_{\hat{\nu}} {L^\rho}_{\hat{\rho}} \,\Gamma^\mu_{\nu\rho} 
+ {L^{\hat{\mu}}}_{{\mu}} {L^\rho}_{\hat{\rho}} 
\frac{\partial {L^\mu}_{\hat{\nu}}}{\partial x^\rho}.
\label{eq:ConnectionComoving}
\end{equation}
Finally, assuming particles of zero mass, it is convenient to express the comoving frame null momentum components in terms of energy, polar angle, and azimuthal angle, that is, in terms of momentum space spherical polar coordinates (denoted by indices with a tilde) $\left( p^{\tilde{\imath}} \right) = \left( \epsilon, \vartheta, \varphi \right)^T$: 
\begin{eqnarray}
\left( p^{\hat{\mu}} \right) &=& \epsilon \left( 1, \ell^{\hat{1}}, \ell^{\hat{2}}, \ell^{\hat{3}} \right)^T \nonumber \\
& = & \epsilon \left( 1, \cos\vartheta, \sin\vartheta \cos\varphi, \sin\vartheta \sin\varphi \right)^T,
\label{eq:ComovingMomentum}
\end{eqnarray}
which also defines the unit normal three-vector $\ell^{\hat\imath}$ tangent to the comoving-frame three-momentum $p^{\hat\imath}$.
In terms of these momentum space coordinates the Boltzmann equation now reads
\begin{equation}
{L^\mu}_{\hat{\mu}} p^{\hat{\mu}} \frac{\partial f}{\partial x^\mu} 
- \Gamma^{\hat{\imath}}_{\hat{\nu}\hat{\mu}} p^{\hat{\nu}} p^{\hat{\mu}} \frac{\partial p^{\tilde{\jmath}}} {\partial p^{\hat{\imath}}} \frac{\partial f}{\partial p^{\tilde{\jmath}}} 
= C[f],
\label{eq:BoltzmannSpherical}
\end{equation}
where 
\begin{equation}
\frac{\partial p^{\tilde{\jmath}}} {\partial p^{\hat{\imath}}} = \frac{1}{\epsilon}
\begin{pmatrix}
\epsilon \cos\vartheta && \epsilon \sin\vartheta \cos\varphi && \epsilon \sin\vartheta \sin\varphi \\
-\sin\vartheta && \cos\vartheta \cos\varphi && \cos\vartheta \sin\varphi  \\
0 && -\sin\varphi / \sin\vartheta && \cos\varphi / \sin\vartheta 
\end{pmatrix}
\label{eq:MomentumCoordinateTransformation}
\end{equation}
is the Jacobian relating momentum space spherical and Cartesian coordinates.

\subsection{Conservative reformulations of the Boltzmann equation}

Conservative reformulations of the Boltzmann equation are available \cite{Cardall2003Conservative-fo} that render plain its connection to number and four-momentum conservation (or balance, given the presence of source terms), and therefore may be helpful in attempts to maintain fidelity to global conservation laws in numerical simulations.

The number-conservative reformulation of Eq.~(\ref{eq:BoltzmannSpherical}) is
\begin{equation}
S_N + M_N = C[f],
\label{eq:NumberConservative}
\end{equation}
with spacetime divergence $S_N$ and momentum space divergence $M_N$ given by
\begin{eqnarray}
S_N &=& \frac{1}{\sqrt{-g}} \frac{\partial}{\partial x^\mu}
\left( \sqrt{-g} \,{L^\mu}_{\hat{\mu}} \, p^{\hat{\mu}} \, f\right), \\
M_N &=& \frac{1}{\epsilon \sin\vartheta} \frac{\partial}{\partial p^{\tilde{\jmath}}}
\left( - \epsilon \sin\vartheta\, \Gamma^{\hat{\imath}}_{\hat{\nu}\hat{\mu}} \, \frac{\partial p^{\tilde{\jmath}}} {\partial p^{\hat{\imath}}} \, p^{\hat{\nu}} p^{\hat{\mu}}  f \right).
\end{eqnarray}
When Eq.~(\ref{eq:NumberConservative}) is integrated over the invariant momentum space volume element (e.g. Ref.~\cite{Lindquist1966Relativistic-Tr})
\begin{eqnarray}
dP &=& \sqrt{-g} \, \varepsilon_{ijk} \,\frac{d_1 p^i \, d_2 p^j \, d_3 p^k}{\left(-p_0\right)} \\
&=& \epsilon \sin\vartheta \, d\epsilon \, d\vartheta \, d\varphi, 
\end{eqnarray}
the momentum space divergence term manifestly disappears, leaving the number balance equation
\begin{equation}
\frac{1}{\sqrt{-g}} \frac{\partial}{\partial x^\mu}
\left( \sqrt{-g} \,N^\mu \right) = \int C[f] \, \frac{dP}{(2\pi)^3},
\end{equation}
where 
\begin{equation}
N^\mu = \int p^\mu  f \, \frac{dP}{(2\pi)^3}
\label{eq:NumberVector}
\end{equation}
is the number flux vector (e.g. Ref.~\cite{Lindquist1966Relativistic-Tr}), expressed here in terms of the lab frame coordinate basis (note $p^\mu = {L^\mu}_{\hat{\mu}} \,p^{\hat{\mu}}$).
We use units in which $\hbar = c = 1$; relative to works in which instead $h = c =1$,  this leads to the factors of $(2\pi)^3$ in the preceding two equations.

Similarly, the four-momentum-conservative reformulation of Eq.~(\ref{eq:BoltzmannSpherical}) is
\begin{equation}
\left(S_T\right)^\rho + \left(M_T\right)^\rho = {L^\rho}_{\hat{\rho}}\,p^{\hat{\rho}} \,C[f],
\label{eq:MomentumConservative}
\end{equation}
with spacetime divergence $S_T$ and momentum space divergence $M_T$ given by
\begin{eqnarray}
\left(S_T\right)^\rho &=& \frac{1}{\sqrt{-g}} \frac{\partial}{\partial x^\mu}
\left( \sqrt{-g} \, {L^\rho}_{\hat{\rho}} {L^\mu}_{\hat{\mu}} \, p^{\hat{\rho}} p^{\hat{\mu}} \, f\right) \nonumber \\
& & + \, \Gamma^{\rho}_{\nu\mu} \,{L^\nu}_{\hat{\nu}} {L^\mu}_{\hat{\mu}} \, p^{\hat{\nu}} p^{\hat{\mu}} f, \\
\left(M_T\right)^\rho &=& \frac{1}{\epsilon \sin\vartheta} \frac{\partial}{\partial p^{\tilde{\jmath}}}
\left( - \epsilon \sin\vartheta\, \Gamma^{\hat{\imath}}_{\hat{\nu}\hat{\mu}} \, \frac{\partial p^{\tilde{\jmath}}} {\partial p^{\hat{\imath}}} \, {L^\rho}_{\hat{\rho}}\,p^{\hat{\rho}} p^{\hat{\nu}} p^{\hat{\mu}}  f \right). \nonumber \\
&&
\end{eqnarray}
When integrated over $dP$, Eq.~(\ref{eq:MomentumConservative}) yields the four-momentum balance equation
\begin{equation}
\frac{1}{\sqrt{-g}} \frac{\partial}{\partial x^\mu} \left( \sqrt{-g} \,T^{\rho \mu} \right) 
+ \, \Gamma^{\rho}_{\nu\mu}\, T^{\nu\mu}
= \int p^\rho \,C[f] \, \frac{dP}{(2\pi)^3},
\end{equation}
where 
\begin{equation}
T^{\mu \nu} = \int p^\mu p^\nu  f \, \frac{dP}{{(2\pi)^3}}
\label{eq:StressEnergyTensor}
\end{equation}
is the stress-energy tensor (e.g. Ref.~\cite{Lindquist1966Relativistic-Tr}), expressed here in terms of the lab frame coordinate basis.

\subsection{Variable Eddington tensor formalism}
\label{sec:VET}

Solving for $f$ in its full dimensionality being computationally overwhelming, the dimensionality of the problem can be reduced by considering only its lowest angular moments. 
A truncation of the hierarchy of moments must be performed by means of closure relations %(see e.g. the references in Sec.~\ref{sec:Introduction}).
(see e.g. Ref.~\cite{Smit2000Closure-in-flux}).

Just as it is most convenient to describe neutrino interactions with the fluid in terms of momentum components reckoned in the comoving frame, so also it seems sensible to define angular moments and prescribe closure relations in the comoving frame.
We define the lowest angular moments of $f\left(x^\mu, \epsilon, \Omega \right)$ as follows:
\begin{eqnarray}
\mathcal{J}\left(x^\mu, \epsilon \right) &=& \epsilon \int f\left(x^\mu, \epsilon, \Omega \right) \, d\Omega, \label{eq:AngularMoment_0}\\
\mathcal{H}^{\hat{\imath}}\left(x^\mu, \epsilon \right) &=& \epsilon \int \ell^{\hat{\imath}} \, f\left(x^\mu, \epsilon, \Omega \right) \, d\Omega, 
\label{eq:AngularMoment_1}\\
\mathcal{K}^{\hat{\imath}\hat{\jmath}}\left(x^\mu, \epsilon \right) &=& \epsilon \int \ell^{\hat{\imath}} \ell^{\hat{\jmath}} \, f\left(x^\mu, \epsilon, \Omega \right) \, d\Omega, \label{eq:AngularMoment_2} \\
\mathcal{L}^{\hat{\imath}\hat{\jmath}\hat{k}}\left(x^\mu, \epsilon \right) &=& \epsilon \int \ell^{\hat{\imath}} \ell^{\hat{\jmath}} \ell^{\hat{k}} \, f\left(x^\mu, \epsilon, \Omega \right) \, d\Omega, \label{eq:AngularMoment_3}
\end{eqnarray}
where $\ell^{\hat\imath}$ is the unit three-vector tangent to the three-momentum in the comoving frame, defined in connection with Eq.~(\ref{eq:ComovingMomentum}).
The integration over $d\Omega = \sin\vartheta \, d\vartheta \, d\varphi$ is performed over the unit sphere. 
Note that the energy dependence is retained; these monochromatic moments are functions of lab frame coordinate basis spacetime position components $x^\mu$ and the comoving frame energy $\epsilon$.
(This is the first instance of a convention we employ, of denoting monochromatic or energy-dependent quantities with script symbols.)
The flux-limited diffusion approximation entails truncation of the hierarchy at the zeroth moment $\mathcal{J}$, with prescriptions for the first and second moments $\mathcal{H}^{\hat{\imath}}$ and $\mathcal{K}^{\hat{\imath}\hat{\jmath}}$ in terms of $\mathcal{J}$.
In the variable Eddington tensor approach the hierarchy is truncated at $\mathcal{H}^{\hat{\imath}}$, with the next higher moments rewritten as
\begin{eqnarray}
\mathcal{K}^{\hat{\imath}\hat{\jmath}} &=& k^{\hat{\imath}\hat{\jmath}} \mathcal{J}, \label{eq:Closure_2} \\
\mathcal{L}^{\hat{\imath}\hat{\jmath}\hat{k}} &=& l^{\hat{\imath}\hat{\jmath}\hat{k}} \mathcal{J},
\label{eq:Closure_3}
\end{eqnarray}
that is,
\begin{eqnarray}
k^{\hat{\imath}\hat{\jmath}} &=& \frac{\int \ell^{\hat{\imath}} \ell^{\hat{\jmath}} \, f \, d\Omega}{\int f \, d\Omega}, 
\label{eq:EddingtonTensor_2}\\
l^{\hat{\imath}\hat{\jmath}\hat{k}} &=& \frac{\int \ell^{\hat{\imath}} \ell^{\hat{\jmath}} \ell^{\hat{k}} \, f \, d\Omega}{\int f \, d\Omega}.
\label{eq:EddingtonTensor_3}
\end{eqnarray}
A number of different approaches to computing the Eddington tensors $k^{\hat{\imath}\hat{\jmath}}$ and $l^{\hat{\imath}\hat{\jmath}\hat{k}}$ might be taken; 
%see for instance the references in Ref.~\cite{Endeve2012Conservative-Mu}.
several are reviewed in Ref.~\cite{Smit2000Closure-in-flux}.
Specific analytic choices used in some recent calculations are spelled out in Refs.~\cite{Obergaulinger2011Magnetic-field-,Shibata2011Truncated-Momen,Kuroda2012Fully-General-R,OConnor2012The-Progenitor-}.
An alternative method in e.g. Refs.~\cite{Rampp2002Radiation-hydro,Muller2010A-New-Multi-dim} involves Eddington factors numerically extracted from the solution of a simplified Boltzmann equation.
Full elaboration of closure schemes is beyond the scope of this paper, but we further discuss in Sec.~\ref{sec:LagrangianObservers} the general forms the Eddington tensors must take.

We choose the variable Eddington tensor approach, and note that we require four equations for the four unknowns $\mathcal{J}$, $\mathcal{H}^{\hat{\imath}}$ (in addition to whatever scheme is used to compute the Eddington tensors). 
Inspection of Eqs.~(\ref{eq:MomentumConservative})-(\ref{eq:StressEnergyTensor}) indicates that the four-momentum conservative formulation of the Boltzmann equation may serve as a suitable basis for the four equations we require. 
They are suggestive of conservative evolution of zeroth and first moments in the lab frame coordinate basis, which may prove helpful in maintaining numerical four-momentum conservation; 
yet the arguments of the spacetime and momentum space divergences can nevertheless be expressed in terms of the comoving-frame moments $\mathcal{J}$, $\mathcal{H}^{\hat{\imath}}$ we must take as our primitive unknowns.

We prepare to implement this strategy with some additional definitions.
We define a monochromatic stress energy, whose components are functions of lab frame coordinate basis spacetime position components $x^\mu$ and the comoving frame energy $\epsilon$:
\begin{equation}
\mathcal{T}^{{\hat{\mu}}{\hat{\nu}}}\left(x^\mu, \epsilon \right)
= \frac{1}{\epsilon} \int p^{\hat{\mu}} p^{\hat{\nu}}\,f\left(x^\mu,\epsilon,\Omega\right)\, d\Omega.
\label{eq:MonochromaticStressEnergyDefinition}
\end{equation}
Its components are related to the comoving frame moments by
\begin{equation}
\begin{pmatrix}
\mathcal{T}^{\hat{0}\hat{0}} && \mathcal{T}^{\hat{0}\hat{\jmath}} \\
\mathcal{T}^{\hat{\imath}\hat{0}} && \mathcal{T}^{\hat{\imath}\hat{\jmath}} \\
\end{pmatrix}
=
\begin{pmatrix}
\mathcal{J} && \mathcal{H}^{\hat{\jmath}} \\
\mathcal{H}^{\hat{\imath}} && \mathcal{K}^{{\hat{\imath}}{\hat{\jmath}}}
\end{pmatrix}
=
\begin{pmatrix}
\mathcal{J} && \mathcal{H}^{\hat{\jmath}} \\
\mathcal{H}^{\hat{\imath}} && k^{{\hat{\imath}}{\hat{\jmath}}} \mathcal{J}
\end{pmatrix}.
\label{eq:MonochromaticStressEnergy}
\end{equation}
Similarly we define
\begin{equation}
\mathcal{U}^{{\hat{\mu}}{\hat{\nu}}{\hat{\rho}}}\left(x^\mu, \epsilon \right)
= \frac{1}{\epsilon} \int p^{\hat{\mu}} p^{\hat{\nu}} p^{\hat{\rho}} \,f\left(x^\mu, \epsilon,\Omega \right)\, d\Omega,
\label{eq:U_Definition}
\end{equation}
whose components are given by
\begin{eqnarray}
\mathcal{U}^{{\hat{0}}{\hat{\mu}}{\hat{\nu}}} &=& \mathcal{U}^{{\hat{\mu}}{\hat{0}}{\hat{\nu}}} = \mathcal{U}^{{\hat{\mu}}{\hat{\nu}}{\hat{0}}} = \epsilon \, \mathcal{T}^{{\hat{\mu}}{\hat{\nu}}}, 
\label{eq:U_ComponentsTime}\\
\mathcal{U}^{{\hat{\imath}}{\hat{\jmath}}{\hat{k}}} &=& \epsilon \mathcal{L}^{\hat{\imath}\hat{\jmath}\hat{k}} = \epsilon \,l^{\hat{\imath}\hat{\jmath}\hat{k}} \mathcal{J}.
\label{eq:U_ComponentsSpace}
\end{eqnarray}
The bottom line is that all the components of both $\mathcal{T}^{{\hat{\mu}}{\hat{\nu}}}$ and $\mathcal{U}^{{\hat{\mu}}{\hat{\nu}}{\hat{\rho}}}$ are just our primitive unknowns $\mathcal{J}$ and $\mathcal{H}^{\hat{\imath}}$, modulo factors (taken to be known) of $\epsilon$, $k^{{\hat{\imath}}{\hat{\jmath}}}$, and $l^{\hat{\imath}\hat{\jmath}\hat{k}}$.

We obtain our equations for $\mathcal{J}$ and $\mathcal{H}^{\hat{\imath}}$ by integrating Eq.~(\ref{eq:MomentumConservative}) over $d\Omega$ and dividing by $\epsilon$:
\begin{equation}
\left({\mathsf{S}_T}\right)^\rho + \left({\mathsf{M}_T}\right)^\rho 
= {L^\rho}_{\hat{\rho}} \int \frac{p^{\hat{\rho}}}{\epsilon} \,C[f] \, d\Omega,
\label{eq:MomentumConservativeMoments}
\end{equation}
where the angle-integrated spacetime divergence $\left({\mathsf{S}_T}\right)^\rho$ and momentum space divergence $\left({\mathsf{M}_T}\right)^\rho$ (now denoted in a sans-serif font to distinguish them from the unintegrated $\left(S_T\right)^\rho$ and $\left(M_T\right)^\rho$) are given by
\begin{eqnarray}
\left({\mathsf{S}_T}\right)^\rho &=& \frac{1}{\sqrt{-g}} \frac{\partial}{\partial x^\mu}
\left( \sqrt{-g} \, {L^\rho}_{\hat{\rho}} {L^\mu}_{\hat{\mu}} \, \mathcal{T}^{{\hat{\rho}}{\hat{\mu}}} \right) \nonumber \\
& & + \, \Gamma^{\rho}_{\nu\mu} \,{L^\nu}_{\hat{\nu}} {L^\mu}_{\hat{\mu}} \, \mathcal{T}^{{\hat{\nu}}{\hat{\mu}}}, 
\label{eq:SpacetimeDivergenceMoments} \\
\left({\mathsf{M}_T}\right)^\rho &=& \frac{1}{\epsilon^2} \frac{\partial}{\partial \epsilon}
\left( - \epsilon \, {L^\rho}_{\hat{\rho}}\, 
\int \Gamma^{\hat{\imath}}_{\hat{\nu}\hat{\mu}} \, \frac{\partial p^{\tilde{1}}} {\partial p^{\hat{\imath}}} \, p^{\hat{\rho}} p^{\hat{\nu}} p^{\hat{\mu}} \, f \, d\Omega \right).\nonumber \\
& &
\end{eqnarray}
To further simplify $\left({\mathsf{M}_T}\right)^\rho$, note that 
\begin{equation}
\frac{\partial p^{\tilde{1}}} {\partial p^{\hat{\imath}}} = \frac{p_{\hat{\imath}} }{\epsilon}
\end{equation}
by virtue of Eq.~(\ref{eq:ComovingMomentum}) and the first row of Eq.~(\ref{eq:MomentumCoordinateTransformation}); and that %\cite{Endeve2012Conservative-Mu,Cardall2003Conservative-fo}
\cite{Cardall2003Conservative-fo}
\begin{equation}
\Gamma^{\hat{\imath}}_{\hat{\nu}\hat{\mu}}\, p_{\hat{\imath}}\,p^{\hat{\nu}} p^{\hat{\mu}}
= \epsilon \, \Gamma^{\hat{0}}_{\hat{\nu}\hat{\mu}} \,p^{\hat{\nu}} p^{\hat{\mu}}.
\end{equation}
This equation, which follows from $0 = d\left(p^{\hat{\mu}} p_{\hat{\mu}}\right)/d\lambda = 2\, p_{\hat{\mu}} \,dp^{\hat{\mu}}/d\lambda$ and Eq.~(\ref{eq:Geodesic_p}), is important because it makes only the third (rather than fourth) moment appear. %\cite{Endeve2012Conservative-Mu}.
With these relations we obtain
\begin{equation}
\left({\mathsf{M}_T}\right)^\rho = \frac{1}{\epsilon^2} \frac{\partial}{\partial \epsilon}
\left( - \epsilon^2 \, {L^\rho}_{\hat{\rho}}\, \Gamma^{\hat{0}}_{\hat{\nu}\hat{\mu}} \, \mathcal{U}^{{\hat{\rho}}{\hat{\nu}}{\hat{\mu}}} \right)
\label{eq:MomentumDivergenceMoments}
\end{equation}
for the angle-integrated momentum space divergence.

Equations~(\ref{eq:MomentumConservativeMoments}), (\ref{eq:SpacetimeDivergenceMoments}), and (\ref{eq:MomentumDivergenceMoments}) are the relations we seek.
They provide four equations ($\rho=0,1,2,3$) for our four primitive unknowns $\mathcal{J}$ and $\mathcal{H}^{\hat{\imath}}$, the comoving frame angular moments;
this is because the components of $\mathcal{T}^{{\hat{\mu}}{\hat{\nu}}}$ and $\mathcal{U}^{{\hat{\mu}}{\hat{\nu}}{\hat{\rho}}}$ are in fact $\mathcal{J}$ and $\mathcal{H}^{\hat{\imath}}$, modulo factors (taken to be known) of the comoving energy $\epsilon$ and Eddington tensors $k^{{\hat{\imath}}{\hat{\jmath}}}$ and $l^{\hat{\imath}\hat{\jmath}\hat{k}}$ (see Eqs.~(\ref{eq:MonochromaticStressEnergy}), (\ref{eq:U_ComponentsTime}), and (\ref{eq:U_ComponentsSpace})).
Moreover, Eqs.~(\ref{eq:MomentumConservativeMoments}), (\ref{eq:SpacetimeDivergenceMoments}), and (\ref{eq:MomentumDivergenceMoments}) are conservative in (that is, expressed in terms of divergences with respect to) our chosen phase space coordinates---the lab frame coordinate basis spacetime position coordinates $x^\mu$ and the comoving frame neutrino energy $\epsilon$.

\subsection{Four-momentum and lepton number exchange}
\label{sec:MomentumNumberExchange}

%Solution of Eqs.~(\ref{eq:MomentumConservativeMoments}), (\ref{eq:SpacetimeDivergenceMoments}), and (\ref{eq:MomentumDivergenceMoments}) provides source terms for the fluid energy and momentum equations.
The right-hand side of Eq.~(\ref{eq:MomentumConservativeMoments}) is intimately related to source terms for the fluid energy and momentum equations.
From the right-hand side of Eq.~(\ref{eq:MomentumConservativeMoments}), the monochromatic four-momentum source $q^\nu_a$ for neutrino species $a$ is
\begin{equation}
q^\nu_a = {L^\nu}_{\hat{\mu}} \int \frac{p^{\hat{\mu}}}{\epsilon} \,C_a[f] \, d\Omega.
\label{eq:MonochromaticMomentumSource}
\end{equation}
Integrating $\epsilon^2 / (2\pi)^3$ times Eqs.~(\ref{eq:MomentumConservativeMoments}), (\ref{eq:SpacetimeDivergenceMoments}), and (\ref{eq:MomentumDivergenceMoments})
over neutrino energy $\epsilon$ for a particular neutrino species $a$, we have
\begin{equation}
\nabla_\mu T_a^{\nu\mu} =  \int q^\nu_a \, \frac{\epsilon^2 \, d\epsilon}{(2\pi)^3} \equiv Q^\nu_a,
\label{eq:MomentumSource}
\end{equation}
where $\nabla_\mu$ denotes the covariant derivative.
The divergence of the total stress-energy---which includes the fluid and all species of neutrinos---must vanish:
\begin{equation}
\nabla_\mu \left( T^{\nu\mu}_{\mathrm{fluid}} + \sum_a T_a^{\nu\mu} \right) = 0.
\end{equation}
By virtue of Eq.~(\ref{eq:MomentumSource}) we have
\begin{equation}
\nabla_\mu T^{\nu\mu}_{\mathrm{fluid}} = -\sum_a Q^\nu_a,
\end{equation}
which expresses the four-momentum exchange between the fluid and the neutrinos.

To address the exchange of electron lepton number we must consider the number-conservative neutrino equation.
We begin by defining a monochromatic number flux
\begin{equation}
\mathcal{N}^{\hat{\mu}}\left(x^\mu, \epsilon \right) = \frac{1}{\epsilon} \int p^{\hat{\mu}} \,f\left(x^\mu, \epsilon, \Omega \right)\, d\Omega.
\label{eq:MonochromaticNumberFlux_0}
\end{equation} 
Inspection of Eqs.~(\ref{eq:ComovingMomentum}), (\ref{eq:MonochromaticStressEnergyDefinition}) and (\ref{eq:MonochromaticStressEnergy}) shows that it is related to the monochromatic stress energy and comoving moments by
\begin{equation}
\left(\mathcal{N}^{\hat{\mu}}\right) = \frac{1}{\epsilon} \left(\mathcal{T}^{{\hat{0}}{\hat{\mu}}}\right)
= \frac{1}{\epsilon} \left(\mathcal{T}^{{\hat{\mu}}{\hat{0}}}\right) 
= \frac{1}{\epsilon}\left( \mathcal{J}, \mathcal{H}^{\hat{\imath}} \right)^T.
\label{eq:MonochromaticNumberFlux}
\end{equation}
Following steps similar to those used to obtain Eqs.~(\ref{eq:MomentumConservativeMoments}), (\ref{eq:SpacetimeDivergenceMoments}), and (\ref{eq:MomentumDivergenceMoments}),
we integrate the number conservative Eq.~(\ref{eq:NumberConservative}) over $d\Omega$ and divide by $\epsilon$ to obtain
\begin{equation}
\mathsf{S}_N + \mathsf{M}_N 
= \frac{1}{\epsilon} \int C[f] \, d\Omega,
\label{eq:NumberConservativeMoments}
\end{equation}
where the spacetime divergence $\mathsf{S}_N$ and momentum space divergence $\mathsf{M}_N$ are given by
\begin{eqnarray}
\mathsf{S}_N &=& \frac{1}{\sqrt{-g}} \frac{\partial}{\partial x^\mu}
\left( \sqrt{-g} \, {L^\mu}_{\hat{\mu}} \, \mathcal{N}^{\hat{\mu}} \right), \\
\mathsf{M}_N &=& \frac{1}{\epsilon^2} \frac{\partial}{\partial \epsilon}
\left( - \epsilon^2 \, \Gamma^{\hat{0}}_{\hat{\nu}\hat{\mu}} \, \mathcal{T}^{{\hat{\nu}}{\hat{\mu}}} \right).
\label{eq:MomentumDivergenceMomentsNumber}
\end{eqnarray}
We define a monochromatic lepton number source $r_a$ for neutrino species $a$ from the right-hand side of Eq.~(\ref{eq:NumberConservativeMoments}):
\begin{equation}
r_a = \frac{1}{\epsilon}\int C_a[f] \, d\Omega.
\label{eq:MonochromaticNumberSource}
\end{equation}
Integrating $\epsilon^2 / (2\pi)^3$ times Eqs.~(\ref{eq:NumberConservativeMoments})-(\ref{eq:MomentumDivergenceMomentsNumber})
over neutrino energy $\epsilon$ for neutrino species $a$, we have
\begin{equation}
\nabla_\mu N_a^{\mu} =  \int r_a \, \frac{\epsilon^2 \, d\epsilon}{(2\pi)^3} 
\equiv R_a.
\label{eq:NumberSource}
\end{equation}
The divergence of the total electron lepton number vanishes:
\begin{equation}
\nabla_\mu \left( N^{\mu}_{e} + N_{\nu_e}^{\mu} - N_{\bar{\nu}_e}^{\mu} \right) = 0,
\end{equation}
where $N^\mu_e$ is the net electron number flux vector of the fluid.
By virtue of Eq.~(\ref{eq:NumberSource}) we have
\begin{equation}
\nabla_\mu N^{\mu}_{e} = - R_{\nu_e} + R_{\bar{\nu}_e},
\label{eq:NumberExchange}
\end{equation}
which expresses electron lepton number exchange between the fluid and the neutrinos.

Moreover, the lepton number source on the right-hand side of Eq.~(\ref{eq:NumberExchange}) is intimately related to the four-momentum sources.
Comparing Eqs.~(\ref{eq:MonochromaticNumberSource}) and (\ref{eq:MonochromaticMomentumSource}), we see that 
\begin{equation}
r_a = \frac{1}{\epsilon} {L^{\hat{0}}}_{\mu} q^\mu_a
\label{eq:NumberMomentumSources}
\end{equation}
(note that ${L^{\hat{0}}}_{\mu} {L^\mu}_{\hat{\nu}} = {\delta^{\hat{0}}}_{\hat{\nu}}$).
%Thus, given $q^\mu_a$ from a solution of Eqs.~(\ref{eq:MomentumConservativeMoments}), (\ref{eq:SpacetimeDivergenceMoments}), and (\ref{eq:MomentumDivergenceMoments}) for neutrino four-momentum, 
%Eq.~(\ref{eq:NumberMomentumSources}) provides the corresponding lepton number source $r_a$ entering Eq.~(\ref{eq:NumberSource}).

The numerical consistency of a scheme in which the lepton number exchange is expressed in terms of the four-momentum exchange via Eq.~(\ref{eq:NumberMomentumSources})---which consistency presumably has consequences for simultaneous conservation of four-momentum and lepton number \cite{Liebendorfer2004A-Finite-Differ}---depends on the discretizations chosen for Eqs.~(\ref{eq:MomentumConservativeMoments}), (\ref{eq:SpacetimeDivergenceMoments}), and (\ref{eq:MomentumDivergenceMoments}).
(This is true regardless of whether the fluid/neutrino couplings are handled in an operator split fashion, e.g. Refs.~\cite{Burrows2007Features-of-the,Rampp2002Radiation-hydro,Muller2010A-New-Multi-dim}; or simultaneously with the solution of the transport equations in a single implicit solve, e.g. Refs.~\cite{Bruenn20092D-and-3D-core-,Liebendorfer2004A-Finite-Differ}.)
To see this, note that the relation employed in Eq.~(\ref{eq:NumberMomentumSources}) between the right-hand sides of the number and four-momentum equations should apply to the left-hand sides as well.
In particular, the identity %(see also Ref.~\cite{Endeve2012Conservative-Mu})
\begin{equation}
\mathsf{S}_N + \mathsf{M}_N = \frac{1}{\epsilon} {L^{\hat{0}}}_{\rho} \left[ \left({\mathsf{S}_T}\right)^\rho + \left({\mathsf{M}_T}\right)^\rho \right] 
\label{eq:Identity}
\end{equation}
should hold for the discretized equations.
Examining the first term on the right-hand side, we find using Eqs.~(\ref{eq:SpacetimeDivergenceMoments}) and (\ref{eq:MonochromaticNumberFlux}) that
\begin{eqnarray}
\frac{1}{\epsilon} {L^{\hat{0}}}_{\rho} \left({\mathsf{S}_T}\right)^\rho
= \mathsf{S}_N 
+ \frac{1}{\epsilon} \left(\vphantom{\frac{\partial {L^{\hat{0}}}_{\rho}}{\partial x^\mu}}\right.&&\left.
{L^{\hat{0}}}_{\rho} {L^\nu}_{\hat{\nu}}  {L^\mu}_{\hat{\mu}} \Gamma^{\rho}_{\nu\mu} \right. \nonumber \\
&&\left. - {L^\rho}_{\hat{\nu}}  {L^\mu}_{\hat{\mu}} \frac{\partial {L^{\hat{0}}}_{\rho}}{\partial x^\mu}\right) \mathcal{T}^{{\hat{\nu}}{\hat{\mu}}}.
\end{eqnarray}
For the second term on the right-hand side of Eq.~(\ref{eq:Identity}), we find using Eqs.~(\ref{eq:MomentumDivergenceMoments}) and (\ref{eq:U_ComponentsTime}) that
\begin{equation}
\frac{1}{\epsilon} {L^{\hat{0}}}_{\rho} \left({\mathsf{M}_T}\right)^\rho
 = \mathsf{M}_N - \frac{1}{\epsilon} \, \Gamma^{\hat{0}}_{\hat{\nu}\hat{\mu}} \, \mathcal{T}^{{\hat{\nu}}{\hat{\mu}}}.
\end{equation}
Thus we see that in the sum of the above two equations, the `extra' terms on the right do indeed cancel by virtue of Eq.~(\ref{eq:ConnectionComoving}).
(Note that ${L^\rho}_{\hat{\nu}}\, \partial_\mu {L^{\hat{0}}}_{\rho} = - {L^{\hat{0}}}_{\rho}\, \partial_\mu {L^\rho}_{\hat{\nu}}$, thanks to $0 = \partial_\mu \left({\delta^{\hat{0}}}_{\hat{\nu}} \right) = \partial_\mu \left( {L^{\hat{0}}}_{\rho} {L^\rho}_{\hat{\nu}} \right)$.)   
Ideally, the analytic steps confirming Eq.~(\ref{eq:Identity}) can be followed in the discrete limit in order to find a discretization of Eqs.~(\ref{eq:MomentumConservativeMoments}), (\ref{eq:SpacetimeDivergenceMoments}), and (\ref{eq:MomentumDivergenceMoments}) that is consistent with respect to both four-momentum and lepton number exchange.

\section{Specialization to the 3+1 metric}
\label{sec:31_Formulation}

After briefly reviewing the 3+1 formulation of general relativity, we show how thinking in terms of three vectors---the four velocity $n^u$ of Eulerian observers, the four-velocity $u^\mu$ of Lagrangian observers, and the (covariant) relative three-velocity $v^\mu$ that connects them---facilitates a full elaboration of the 3+1 variable Eddington tensor moment equations, including the detailed relationship between these and the number exchange equation.

\subsection{Description and evolution of the geometry}
\label{sec:Geometry}

Numerical relativity often is built upon the 3+1 formulation of general relativity.
In this approach one considers a foliation of spacetime into spacelike slices, i.e. three-dimensional hypersurfaces $\Sigma_t$ labeled by coordinate time $t$ ($=x^0$ in our lab frame coordinate basis).
The summary below serves to establish notation and spells out only the results we need here. 
Pedagogical introductions include Refs.~\cite{Misner1973Gravitation,York1983The-initial-val,Gourgoulhon200731-Formalism-an}. 

Generic metric components in the 3+1 formulation are found from consideration of a `thin sandwich' of spacetime bounded by two spacelike slices $\Sigma_t$ and $\Sigma_{t+dt}$.
In particular we consider the spacetime interval $ds$ between two points: $x^\mu$ in $\Sigma_t$, and $x^\mu + dx^\mu$ in $\Sigma_{t+dt}$.
With proper time interval $d\tau$ orthogonal to $\Sigma_t$, and proper length interval $d\ell$ tangent to $\Sigma_t$, $ds$ is given by a Lorentzian version of the pythagorean theorem (note the signature -+++):
\begin{equation}
ds^2 = - d\tau^2 + d\ell^2.
\label{eq:SpacetimeInterval}
\end{equation}
Denote the orthogonal proper time at $x^\mu$ between $\Sigma_t$ and $\Sigma_{t+dt}$
\begin{equation}
d\tau = \alpha \, dt,
\label{eq:ProperTime}
\end{equation}
and call $\alpha$ the lapse function.
In considering the proper length $d\ell$ between $x^\mu$ in $\Sigma_t$ and the orthogonal projection of $x^\mu + dx^\mu$ in $\Sigma_{t+dt}$ onto $\Sigma_t$, we note that the curves $x^i = \mathrm{constant}$ traced out by the spatial coordinates need not be orthogonal to $\Sigma_t$.
That is, the spatial coordinates may be moving as seen by an observer at rest in $\Sigma_t$, such that they shift by a coordinate distance $\beta^i\,dt$ between $\Sigma_t$ and $\Sigma_{t+dt}$. 
Allowing for such a shift vector $\beta^i$ (which is tangent to $\Sigma_t$), and letting $\gamma_{ij}$ denote the three-metric within $\Sigma_t$, we have
\begin{equation}
d\ell^2 = \gamma_{ij} \left(dx^i + \beta^i\,dt \right) \left(dx^j + \beta^j\,dt \right).
\label{eq:ProperLength}
\end{equation}
Comparing Eqs.~(\ref{eq:SpacetimeInterval})-(\ref{eq:ProperLength}) with the line element $ds^2 = g_{\mu\nu}\, dx^\mu\,dx^\nu$, we read off the metric components
%The line element giving the proper spacetime interval between points $x^\mu$ in $\Sigma_t$ and $x^\mu + dx^\mu$ in $\Sigma_{t+dt}$ can be expressed
%\begin{equation}
%ds^2 = -\alpha^2\, dt^2 + \gamma_{ij} \left(dx^i + \beta^i\,dt \right) \left(dx^j + \beta^j\,dt \right).
%\end{equation}
%The first term, expressed in terms of the lapse function $\alpha$, gives the proper time $\alpha \, dt$ between $\Sigma_t$ and $\Sigma_{t+dt}$. 
%The second term gives the proper distance within $\Sigma_t$. 
%This proper distance is expressed in terms of the three-metric $\gamma_{ij}$, which governs the intrinsic three-geometry of $\Sigma_t$; and the shift vector $\beta^i$, which tells how the spatial coordinates $x^i$ `shift' between $\Sigma_t$ and $\Sigma_{t+dt}$ (i.e. coordinate curves $x^i=\mathrm{constant}$ need not be orthogonal to $\Sigma_t$).
%The three metric $\gamma_{ij}$ and its inverse $\gamma^{ij}$ lower and raise the indices of three-vectors within a spacelike slice, as in $\beta_i = \gamma_{ij} \beta^j$ and $\beta^i = \gamma^{ij} \beta_j$.
%The spacetime metric in our lab frame coordinate basis is
\begin{equation}
\left( g_{\mu\nu} \right) = 
\begin{pmatrix}
-\alpha^2 +\beta_k \beta^k && \beta_j \\
\beta_i && \gamma_{ij}
\label{eq:31_Metric}
\end{pmatrix}.
\end{equation}
The inverse metric is
\begin{equation}
\left( g^{\mu\nu} \right) = 
\begin{pmatrix}
-1/\alpha^2 && \beta_j / \alpha^2 \\
\beta_i / \alpha^2 && \gamma^{ij} - \beta^i \beta^j / \alpha^2
\label{eq:31_MetricInverse}
\end{pmatrix}.
\end{equation}
The three-metric $\gamma_{ij}$ and its inverse $\gamma^{ij}$ lower and raise the indices of three-vectors within (tangent to) the spacelike slice, as in $\beta_i = \gamma_{ij} \beta^j$ and $\beta^i = \gamma^{ij} \beta_j$.
Finally,
\begin{equation}
\sqrt{-g} = \alpha\sqrt{\gamma}
\label{eq:MetricDeterminant}
\end{equation}
expresses the determinant $g$ of the four-metric in terms of the lapse function and the determinant $\gamma$ of the three-metric.

Solution of the Einstein equations for the metric components---nonlinear partial differential equations, second order in space and time---constitutes knowledge of spacetime.
In the 3+1 approach, solution of the Einstein equations is transformed into a Cauchy problem: 
specify initial data (satisfying certain constraints from the Einstein equations) on an initial spacelike slice;
and with coordinate freedom fixed and spatial boundary conditions specified, evolve the geometry of the spacelike slices forward in time.
Of the ten degrees of freedom associated with the metric components (the number of independent elements in a $4\times 4$ symmetric matrix), four correspond to the freedom in general relativity to choose any spacetime coordinates whatsoever, leaving six physical degrees of freedom to be determined. 
The standard way to think about the coordinate freedom in the 3+1 approach is to regard the lapse function $\alpha$ and shift vector $\beta^i$ as freely specifiable functions in time and space, associated respectively with the choice of time coordinate (i.e. the foliation or spacetime slicing) and the choice of spatial coordinates (in particular the motion of these coordinates as seen by an observer at rest in a slice).
In order to facilitate practical solution by forward evolution in time, the second-order-in-time system is transformed to double the number of equations, first order in time, for double the number of dynamical variables.
In particular, the above phrase ``evolve the geometry of the spacelike slices forward in time'' corresponds to evolution of (a) the six independent components of the three-metric $\gamma_{ij}$ governing the geometry within a slice, and (b) the six independent components of the extrinsic curvature $K_{ij}$ (another symmetric tensor tangent to the spacelike slice) that describes the warp of the spacelike slices as embedded in spacetime.
Here we do not show these evolution equations; 
in what follows, we simply regard $\gamma_{ij}$ and $K_{ij}$ as given, for instance as having been obtained by numerical solution (often of even further transformed systems, as for instance in BSSN and related approaches); see e.g. Refs.~\cite{Gourgoulhon200731-Formalism-an,Alcubierre2008Introduction-to,Baumgarte2010Numerical-Relat}.

\subsection{Four-velocity of Eulerian observers and the spacetime divergence}
\label{sec:EulerianObservers}

In dealing with various forms of stress energy and the equations that govern them in the 3+1 context, two helpful tensors are the unit normal $n^\mu$ and the orthogonal projector $\gamma_{\mu\nu}$.
The unit normal $n^\mu$ to a spacelike slice at a given point can be regarded as the four-velocity of an Eulerian observer, i.e. one at rest in the lab frame. 
In the lab frame coordinate basis its components are
\begin{eqnarray}
\left( n^\mu \right) &=& (1/\alpha, -\beta^i / \alpha)^T, \label{eq:UnitNormalU} \\
\left( n_\mu \right) &=& (-\alpha,0,0,0). \label{eq:UnitNormalD} 
\end{eqnarray}
Note that indeed $n_\mu n^\mu = g_{\mu\nu} n^\mu n^\nu = -1$ as expected of a unit vector.
The orthogonal projector is
\begin{equation}
\gamma_{\mu\nu} = g_{\mu\nu} + n_\mu n_\nu. \label{eq:OrthogonalProjector}
\end{equation}
From Eqs.~(\ref{eq:31_Metric}) and (\ref{eq:UnitNormalD}) it follows that the spatial part of $\gamma_{\mu\nu}$ equals the three-metric $\gamma_{ij}$, motivating use of the same base symbol.
While contraction of an arbitrary vector with $n^\mu$ yields the portion orthogonal to a spacelike slice, contraction with $\gamma_{\mu\nu} = g_{\mu\nu} + n_\mu n_\nu$ yields the portion tangent to the spacelike slice.
Indeed a trivial calculation confirms that $\gamma_{\mu\nu} n^\nu = 0$. 

The unit normal and orthogonal projector can be used to decompose a stress-energy tensor $T^{\mu\nu}$.
The `Eulerian projections'
\begin{eqnarray}
E &=& n_\mu n_\nu T^{\mu\nu}, \label{eq:EnergyDensity}\\
F^\mu &=& - n_\nu {\gamma^\mu}_\rho T^{\nu\rho}, \label{eq:MomentumDensity}\\
S^{\mu\nu} &=& {\gamma^\mu}_\rho {\gamma^\nu}_\sigma T^{\rho\sigma}  \label{eq:Stress}
\end{eqnarray}
are respectively the energy density, momentum density (or energy flux), and stress measured by an Eulerian observer. 
The momentum density and stress are spacelike, i.e. tangent to the spacelike slice:
\begin{eqnarray}
n_\mu F^\mu  &=& 0, \label{eq:MomentumDensityTangent} \\
n_\mu S^{\mu\nu}  =  n_\nu S^{\mu\nu}&=& 0. \label{eq:StressTangent}
\end{eqnarray}
In terms of Eqs.~(\ref{eq:EnergyDensity})-(\ref{eq:Stress}), a stress-energy tensor can be decomposed as
\begin{equation}
T^{\mu\nu} = E \,n^\mu n^\nu + F^\mu n^\nu + F^\nu n^\mu + S^{\mu\nu}.
\label{eq:StressEnergy_0}
\end{equation}
We call this the `Eulerian decomposition' of $T^{\mu\nu}$.
Emphasizing the spacelike character of $F^\mu$ and $S^{\mu\nu}$ (see Eqs.~(\ref{eq:MomentumDensityTangent}) and (\ref{eq:StressTangent})), it can be re-expressed 
\begin{equation}
T^{\mu\nu} = E \,n^\mu n^\nu + F^i {\gamma^{\mu}}_i n^\nu + F^i {\gamma^{\nu}}_i n^\mu + S^{ij} {\gamma^{\mu}}_i {\gamma^{\nu}}_j
\label{eq:StressEnergy}
\end{equation}
in the lab frame coordinate basis.

Turning from a generic stress-energy tensor to the neutrino radiation in particular,
we similarly define the Eulerian projections and Eulerian decomposition of the monochromatic neutrino stress-energy 
\begin{equation}
\mathcal{T}^{\mu\nu} = {L^\mu}_{\hat{\mu}} {L^\nu}_{\hat{\nu}} \, \mathcal{T}^{{\hat{\mu}}{\hat{\nu}}},
\label{eq:StressEnergyTransformation}
\end{equation} 
whose comoving frame components $\mathcal{T}^{{\hat{\mu}}{\hat{\nu}}}$
were given in Eqs.~(\ref{eq:MonochromaticStressEnergyDefinition}) and (\ref{eq:MonochromaticStressEnergy}). 
The coordinate transformation ${L^\mu}_{\hat{\mu}}$ from the orthonormal comoving frame to the lab frame coordinate basis was discussed in Sec.~\ref{sec:Boltzmann}. 
The Eulerian projections
\begin{eqnarray}
\mathcal{E} &=&  n_\mu n_\nu\mathcal{T}^{\mu\nu}, \label{eq:T_Projection_nn_Eulerian} \\
\mathcal{F}^\mu &=& -n_\nu {\gamma^\mu}_\rho\mathcal{T}^{\nu\rho}, 
\label{eq:T_Projection_n1_Eulerian} \\
\mathcal{S}^{\mu\nu} &=& {\gamma^\mu}_\rho {\gamma^\nu}_\sigma\mathcal{T}^{\rho\sigma} \label{eq:T_Projection_11_Eulerian}
\end{eqnarray}
are respectively the monochromatic neutrino energy density, momentum density, and stress as measured by an Eulerian observer.
The momentum density and stress are spacelike,
\begin{eqnarray}
n_\mu \mathcal{F}^\mu  &=& 0, \label{eq:MomentumDensityTangentMonochromatic} \\
n_\mu \mathcal{S}^{\mu\nu}  =  n_\nu \mathcal{S}^{\mu\nu}&=& 0, \label{eq:StressTangentMonochromatic}
\end{eqnarray}
and 
\begin{equation}
\mathcal{T}^{\mu\nu} = \mathcal{E} \,n^\mu n^\nu + \mathcal{F}^\mu n^\nu + \mathcal{F}^\nu n^\mu + \mathcal{S}^{\mu\nu}
\label{eq:StressEnergyDecompositionEulerian}
\end{equation}
is the Eulerian decomposition of the monochromatic stress-energy.

The results in Appendix~\ref{app:SpacetimeDivergence} for the four components of the spacetime divergence $\nabla_\mu T^{\mu\nu}$ of a stress-energy tensor can be adapted immediately to the spacetime divergence $\left({\mathsf{S}_T}\right)^\nu$ of the monochromatic particle stress-energy in Eq.~(\ref{eq:SpacetimeDivergenceMoments}).
In particular we make the formal replacements
\begin{eqnarray}
T^{\mu\nu} &\rightarrow& \mathcal{T}^{\mu\nu}, \\
E &\rightarrow& \mathcal{E}, \\
F^\mu &\rightarrow& \mathcal{F}^\mu, \\
S^{\mu\nu} &\rightarrow& \mathcal{S}^{\mu\nu}.
\end{eqnarray} 
From Eq.~(\ref{eq:DivergenceOrthogonalFinal}), the projection of the spacetime divergence orthogonal to the spacelike slice---which contributes to the neutrino energy equation---is
\begin{equation}
-n_\nu \left({\mathsf{S}_T}\right)^\nu = \frac{1}{\alpha\sqrt{\gamma}}\left[\frac{\partial \left(\mathsf{D}_{T,n}\right)}{\partial t} + \frac{\partial \left(\mathsf{F}_{T,n}\right)^i }{\partial x^i} - \mathsf{G}_{T,n}\right],
\label{eq:SpacetimeDivergence_31_n}
\end{equation}
where 
\begin{eqnarray}
\mathsf{D}_{T,n} &=& \sqrt{\gamma} \mathcal{E}, 
\label{eq:SpacetimeDivergence_31_n_D}\\
\left(\mathsf{F}_{T,n}\right)^i &=&  \sqrt{\gamma} \left( \alpha \mathcal{F}^i - \beta^i \mathcal{E} \right), 
\label{eq:SpacetimeDivergence_31_n_F}\\
\mathsf{G}_{T,n} &=& -\alpha\sqrt{\gamma} \left( \frac{\mathcal{F}^i}{\alpha} \frac{\partial \alpha}{\partial x^i} - \mathcal{S}^{ij} K_{ij}\right)
\label{eq:SpacetimeDivergence_31_n_G}
\end{eqnarray}
are respectively the `conserved' energy density, energy flux, and gravitational power associated with the neutrinos.
From Eq.~(\ref{eq:DivergenceTangentFinal}), the projection of the spacetime divergence orthogonal to the spacelike slice---which contributes to the neutrino momentum equation---is
\begin{equation}
\gamma_{j \nu} \left({\mathsf{S}_T}\right)^\nu = \frac{1}{\alpha\sqrt{\gamma}}\left[\frac{\partial \left(\mathsf{D}_{T,\gamma}\right)_j}{\partial t} + \frac{\partial {\left(\mathsf{F}_{T,\gamma}\right)^i}_j }{\partial x^i} - \left(\mathsf{G}_{T,\gamma}\right)_j\right],
\label{eq:SpacetimeDivergence_31_gamma}
\end{equation}
where
\begin{eqnarray}
\left(\mathsf{D}_{T,\gamma}\right)_j &=& \sqrt{\gamma} \mathcal{F}_j, 
\label{eq:SpacetimeDivergence_31_gamma_D} \\
{\left(\mathsf{F}_{T,\gamma}\right)^i}_j &=& \sqrt{\gamma} \left( \alpha {\mathcal{S}^i}_j - \beta^i \mathcal{F}_j \right), 
\label{eq:SpacetimeDivergence_31_gamma_F}\\
\left(\mathsf{G}_{T,\gamma}\right)_j &=& -\alpha\sqrt{\gamma}\left( \frac{\mathcal{E}}{\alpha}\frac{\partial\alpha}{\partial x^j} -\frac{\mathcal{F}_i}{\alpha} \,\frac{\partial\beta^i}{\partial x^j} -\frac{\mathcal{S}^{i k}}{2} \frac{\partial \gamma_{ik}}{\partial x^j} \right)\nonumber \\
& &
\label{eq:SpacetimeDivergence_31_gamma_G}
\end{eqnarray}
are respectively the `conserved' momentum density, momentum flux, and gravitational force associated with the neutrinos.
Note that $\mathcal{E}\left(x^\rho,\epsilon \right)$, $\mathcal{F}^\mu\left(x^\rho,\epsilon \right)$, and $\mathcal{S}^{\mu\nu}\left(x^\rho,\epsilon \right)$ are functions of the lab frame spacetime coordinates $x^\rho$ and the neutrino energy $\epsilon$ reckoned in a comoving frame.
%The `conserved' variables are $\sqrt{\gamma}\mathcal{E}$ and $\sqrt{\gamma}\mathcal{F}_i$, whose connection to the primitive variables $\mathcal{J}\left(x^\rho,\epsilon \right)$ and $\mathcal{H}^{\hat\imath}\left(x^\rho,\epsilon \right)$ follow from
Their relations to the primitive variables $\mathcal{J}\left(x^\rho,\epsilon \right)$ and $\mathcal{H}^{\hat\imath}\left(x^\rho,\epsilon \right)$ follow from
Eqs.~(\ref{eq:T_Projection_nn_Eulerian})-(\ref{eq:T_Projection_11_Eulerian}), (\ref{eq:StressEnergyTransformation}), (\ref{eq:CompositeTransformation}), and (\ref{eq:MonochromaticStressEnergy}); but see also the following two subsections and Appendix~\ref{app:EulerianLagrangianProjections}. 
The projections of the spacetime divergence presented here are in accord with the corresponding terms in Eqs.~(3.37) and (3.38) of Shibata et al. \cite{Shibata2011Truncated-Momen}.

\subsection{Four-velocity of Lagrangian observers and the momentum space divergence}
\label{sec:LagrangianObservers}

Before turning to the momentum space divergence, it will be helpful to begin rewriting in covariant form some of the expressions we have given involving comoving frame components, by writing them in terms of the four-velocity $u^\mu$ of Lagrangian observers (i.e. those at rest with respect to the fluid).
In an orthonormal comoving frame we have
\begin{eqnarray}
\left(u^{\hat{\mu}} \right)&=& \left( 1, 0, 0, 0 \right)^T, 
\label{eq:FluidVelocityU}\\
\left( u_{\hat{\mu}} \right) &=& \left( -1, 0, 0, 0 \right). 
  \label{eq:FluidVelocityD}
\end{eqnarray}
Thus for example 
\begin{equation}
\epsilon = -u_{\hat{0}} p^{\hat{0}} = -u_{\hat{\mu}} p^{\hat{\mu}} = -u_{\mu} p^{\mu}
\label{eq:ParticleEnergy}
\end{equation}
is the neutrino energy measured by a Lagrangian observer, expressed in terms of lab frame coordinate basis components $u_{\mu} = u_{\hat{\mu}} {L^{\hat\mu}}_\mu$ and $p^{\mu} = {L^{\mu}}_{\hat\mu} p^{\hat{\mu}}$ 
(the coordinate transformation ${L^\mu}_{\hat{\mu}}$ from the orthonormal comoving frame to the lab frame coordinate basis, and its inverse ${L^{\hat\mu}}_\mu$, were discussed in Sec.~\ref{sec:Boltzmann}).

Turning to the comoving frame angular moments $\mathcal{J}$, $\mathcal{H}^{\hat{\imath}}$, and $\mathcal{K}^{\hat{\imath}\hat{\jmath}}$ defined in Eqs.~(\ref{eq:AngularMoment_0})-(\ref{eq:AngularMoment_3}), we can define covariant versions by extending $\ell^{\hat{\imath}}$ to a unit four-vector $\ell^\mu$ satisfying $\ell_\mu \ell^\mu = 1$.
This vector is spacelike in the comoving frame, with comoving frame components $\left( \ell^{\hat{\mu}}\right) = (0, \ell^{\hat{1}}, \ell^{\hat{2}}, \ell^{\hat{3}})^T$, and is thus orthogonal to $u^\mu$, i.e. 
\begin{equation}
u_{\mu} \ell^\mu = 0
\end{equation} 
in any basis.
Thus $\epsilon\, \ell^\mu$ is a covariant representation of the three-momentum measured by a Lagrangian observer.
Covariant representations of the comoving frame angular moments are  
\begin{eqnarray}
\mathcal{J} &=& \epsilon \int f \, d\Omega, \label{eq:AngularMoment_0_Covariant}\\
\mathcal{H}^{\mu} &=& \epsilon \int \ell^{\mu} \, f \, d\Omega, \label{eq:AngularMoment_1_Covariant}\\
\mathcal{K}^{\mu\nu} &=& \epsilon \int \ell^{\mu} \ell^{\nu} \, f \, d\Omega, \label{eq:AngularMoment_2_Covariant} \\
\mathcal{L}^{\mu\nu\rho} &=& \epsilon \int \ell^{\mu} \ell^{\nu} \ell^{\rho} \, f \, d\Omega, \label{eq:AngularMoment_3_Covariant}
\end{eqnarray}
where the integration is still performed with respect to the comoving frame solid angle.

%While in Eqs.~(\ref{eq:Closure_2})-(\ref{eq:EddingtonTensor_3}) we nominally defined closure of the system in terms of the comoving frame components $\mathcal{K}^{\hat{\imath}\hat{\jmath}}$ and $\mathcal{L}^{\hat{\imath}\hat{\jmath}\hat{k}}$ of the second and third angular moments, in practice we may be able to directly use the lab frame coordinate basis components $\mathcal{K}^{\mu\nu}$ and $\mathcal{L}^{\mu\nu\rho}$.
%That is, it may not be necessary to perform the inverse composite transformations $\mathcal{K}^{\hat{\imath}\hat{\jmath}} = {L^{\hat{\imath}}}_{\mu} {L^{\hat{\jmath}}}_{\nu} \, \mathcal{K}^{{\mu}{\nu}}$ and $\mathcal{L}^{\hat{\imath}\hat{\jmath}\hat{k}} = {L^{\hat{\imath}}}_{\mu} {L^{\hat{\jmath}}}_{\nu}  {L^{\hat{k}}}_{\rho}\, \mathcal{L}^{{\mu}{\nu}{\rho}}$ to apply the closure.
While in Eqs.~(\ref{eq:Closure_2})-(\ref{eq:EddingtonTensor_3}) we nominally defined closure of the system in terms of the comoving frame components $\mathcal{K}^{\hat{\imath}\hat{\jmath}}$ and $\mathcal{L}^{\hat{\imath}\hat{\jmath}\hat{k}}$ of the second and third angular moments, in practice we may be able to obtain the lab frame coordinate basis components $\mathcal{K}^{\mu\nu}$ and $\mathcal{L}^{\mu\nu\rho}$ without ever explicitly transforming any of the moments to and/or from the comoving frame using the transformations ${L^{\hat{\mu}}}_\mu$ and ${L^\mu}_{\hat{\mu}}$ discussed in Sec.~\ref{sec:Boltzmann}.
(Applying the closures without any reference to comoving frame components is desirable because it would be a hassle to have to work explicitly with the transformations ${L^{\hat{\mu}}}_\mu$ and ${L^\mu}_{\hat{\mu}}$. 
These will be quite complicated in the general multidimensional case, because neither the metric $g_{\mu\nu}$, nor therefore the tetrad ${e^\mu}_{\bar{\mu}}$, nor either the boost ${\Lambda^{\bar{\mu}}}_{\hat{\mu}}$, will be diagonal. 
Moreover, it would be necessary to apply the tetrad to the coordinate basis velocity variables obtained with the hydrodynamics solver in order to get the velocity parameters appearing in the boost.
Working only with lab frame coordinate basis components of the comoving frame angular moments avoids these complications.)
Following Ref.~\cite{Anile1992Covariant-Radia}, we note that if $\mathcal{K}^{\mu\nu}$ is regarded as a function of the zeroth and first moments $\mathcal{J}$ and $\mathcal{H}^\mu$, the most general symmetric expression that (a) is spacelike relative to $u^\mu$, and (b) satisfies the trace condition ${\mathcal{K}^\mu}_\mu = \mathcal{J}$ (obvious from Eqs.~(\ref{eq:AngularMoment_0_Covariant}) and (\ref{eq:AngularMoment_2_Covariant})), is of the form
\begin{equation}
\mathcal{K}^{\mu\nu} = \frac{1}{3} \mathcal{J} h^{\mu\nu} + a(\mathcal{J}, \mathcal{H}) \left( \mathcal{H}^\mu \mathcal{H}^\nu -\frac{1}{3} \mathcal{H}^2 h^{\mu\nu}\right),
\label{eq:CovariantClosure_K}
\end{equation}
where $\mathcal{H} = \sqrt{\mathcal{H}_\mu \mathcal{H}^\mu}$, and 
\begin{equation}
h_{\mu\nu} = g_{\mu\nu} + u_\mu u_\nu
\label{eq:OrthogonalProjectorComoving}
\end{equation}
is the orthogonal projector relative to $u^\mu$.
We extend this thinking to the third moment, finding the unique expression
\begin{eqnarray}
\mathcal{L}^{\mu\nu\rho} &=& \frac{1}{5} \left( \mathcal{H}^\mu h^{\nu\rho} + \mathcal{H}^\rho h^{\mu\nu} + \mathcal{H}^\nu h^{\rho\mu} \right) \nonumber\\
& & + b(\mathcal{J}, \mathcal{H}) \left[\vphantom{\frac{1}{5}} \mathcal{H}^\mu \mathcal{H}^\nu \mathcal{H}^\rho \right. \nonumber \\ 
& & \left. - \frac{1}{5} \mathcal{H}^2 \left( \mathcal{H}^\mu h^{\nu\rho} + \mathcal{H}^\rho h^{\mu\nu} + \mathcal{H}^\nu h^{\rho\mu}\right) \right]
\label{eq:CovariantClosure_L}
\end{eqnarray}
satisfying the trace conditions (obvious from Eqs.~(\ref{eq:AngularMoment_1_Covariant}) and (\ref{eq:AngularMoment_3_Covariant}))
\begin{equation}
{{\mathcal{L}^\mu}_\mu}^\rho = \mathcal{H}^\rho,\ \ \ {{\mathcal{L}^{\mu\nu}}_\mu} = \mathcal{H}^\nu,\ \ \  {{\mathcal{L}^{\mu\nu}}_\nu} = \mathcal{H}^\mu.
\end{equation}
The point is that when $\mathcal{K}^{\mu\nu}$ and $\mathcal{L}^{\mu\nu\rho}$ are obtained only from knowledge of $\mathcal{J}$ and $\mathcal{H}^\mu$, in principle the only freedom in the closures are the scalar functions $a(\mathcal{J}, \mathcal{H})$ and $b(\mathcal{J}, \mathcal{H})$, in which the magnitude $\mathcal{H} = \sqrt{\mathcal{H}_\mu \mathcal{H}^\mu}$ can be evaluated in terms of lab frame coordinate basis components.

Consider next the monochromatic stress energy $\mathcal{T}^{\mu\nu}$.
As can be confirmed in the comoving frame, the neutrino momentum $p^\mu$ can be covariantly decomposed as 
\begin{equation}
p^\mu = \epsilon \left( u^\mu + \ell^\mu \right),
\label{eq:MomentumDecomposition}
\end{equation}
i.e. into portions tangent and orthogonal to the Lagrangian observer four-velocity $u^\mu$.
Using this in the monochromatic stress energy 
\begin{equation}
\mathcal{T}^{\mu\nu}
= \frac{1}{\epsilon} \int p^{\mu} p^{\nu}\,f\, d\Omega
\label{eq:MonochromaticStressEnergy_2}
\end{equation}
(see Eqs.~(\ref{eq:MonochromaticStressEnergyDefinition}) and (\ref{eq:StressEnergyTransformation})), one can see that the `Lagrangian projections'
\begin{eqnarray}
\mathcal{J} &=&  u_\mu u_\nu\mathcal{T}^{\mu\nu}, 
\label{eq:T_Projection_uu_Lagrangian} \\
\mathcal{H}^\mu &=& - u_\nu {h^\mu}_\rho\mathcal{T}^{\nu\rho},
\label{eq:T_Projection_u1_Lagrangian} \\
\mathcal{K}^{\mu\nu} &=&  {h^\mu}_\rho {h^\nu}_\sigma\mathcal{T}^{\rho\sigma},
\label{eq:T_Projection_11_Lagrangian} 
\end{eqnarray}
do in fact yield Eqs.~(\ref{eq:AngularMoment_0_Covariant})-(\ref{eq:AngularMoment_2_Covariant}).
(Recall that $h_{\mu\nu}$, given by Eq.~(\ref{eq:OrthogonalProjectorComoving}), is the orthogonal projector relative to $u^\mu$.)
Thus the moments $\mathcal{J}$, $\mathcal{H}^{\mu}$, and $\mathcal{K}^{\mu\nu}$ are respectively the monochromatic neutrino energy density, momentum density, and stress measured by a Lagrangian observer.
Note that $\mathcal{H}^\mu$ and $\mathcal{K}^{\mu\nu}$ are spacelike in the comoving frame:
\begin{eqnarray}
u_\mu \mathcal{H}^\mu  &=& 0, \label{eq:MomentumDensityTangentComoving} \\
u_\mu \mathcal{K}^{\mu\nu}  =  u_\nu \mathcal{K}^{\mu\nu}&=& 0. \label{eq:StressTangentComoving}
\end{eqnarray}
Therefore $\mathcal{T}^{\mu\nu}$ can be written as the `Lagrangian decomposition'
\begin{equation}
\mathcal{T}^{\mu\nu} = \mathcal{J} \,u^\mu u^\nu + \mathcal{H}^\mu u^\nu + \mathcal{H}^\nu u^\mu + \mathcal{K}^{\mu\nu},
\label{eq:StressEnergyDecompositionComoving}
\end{equation}
which provides an alternative to the Eulerian decomposition of Eq.~(\ref{eq:StressEnergyDecompositionEulerian})
in terms of the energy density, momentum density, and stress measured by an Eulerian observer.

Eqs.~(\ref{eq:StressEnergyDecompositionEulerian}) and (\ref{eq:StressEnergyDecompositionComoving}) can be used to write the monochromatic energy density, momentum density, and stress measured by an Eulerian observer ($\mathcal{E}$, $\mathcal{F}^{\mu}$, and $\mathcal{S}^{\mu\nu}$) in terms of their counterparts measured by a Lagrangian observer ($\mathcal{J}$, $\mathcal{H}^{\mu}$, and $\mathcal{K}^{\mu\nu}$), and vice-versa.
In one direction, use Eqs.~(\ref{eq:T_Projection_nn_Eulerian})-(\ref{eq:T_Projection_11_Eulerian})
and substitute Eq.~(\ref{eq:StressEnergyDecompositionComoving}) on the right-hand side.
In the other direction, use Eqs.~(\ref{eq:T_Projection_uu_Lagrangian})-(\ref{eq:T_Projection_11_Lagrangian})
and substitute Eq.~(\ref{eq:StressEnergyDecompositionEulerian}) on the right-hand side.
We will say more about this in the next subsection and in Appendix~\ref{app:EulerianLagrangianProjections}.
%The explicit results are not particularly illuminating and we do not display them.
%These relations will of course be needed in numerical work,
%but it may be best to perform the contractions numerically rather than code tedious analytic expressions.

Next we turn to the third momentum moment $\mathcal{U}^{\rho\mu\nu}$, given by
\begin{equation}
\mathcal{U}^{\mu\nu\rho}
= \frac{1}{\epsilon} \int p^{\mu} p^{\nu} p^{\rho} \,f\, d\Omega
\label{eq:ThirdMoment}
\end{equation}
(see Eq.~(\ref{eq:U_Definition})).
Note that
\begin{equation}
-u_{\rho} \,\mathcal{U}^{\rho\mu\nu} = -u_{\rho}\, \mathcal{U}^{\mu\rho\nu} = -u_{\rho} \,\mathcal{U}^{\mu\nu\rho} = \epsilon \,\mathcal{T}^{\mu\nu}
\label{eq:U_ProjectionComoving}
\end{equation}
is a covariant version of Eq.~(\ref{eq:U_ComponentsTime}).
Using Eq.~(\ref{eq:MomentumDecomposition}) in Eq.~(\ref{eq:ThirdMoment}) and comparing with Eqs.~(\ref{eq:AngularMoment_0_Covariant})-(\ref{eq:AngularMoment_3_Covariant}), we find 
\begin{eqnarray}
\epsilon \,\mathcal{J} &=& -u_\mu u_\nu u_\rho\, \mathcal{U}^{\mu\nu\rho}, 
\label{eq:U_Projection_uuu}\\
\epsilon\, \mathcal{H}^\mu &=&  u_\nu u_\rho {h^\mu}_\sigma \, \mathcal{U}^{\nu\rho\sigma}, 
\label{eq:U_Projection_uu1}\\
\epsilon\, \mathcal{K}^{\mu\nu} &=& - u_\rho {h^\mu}_\sigma {h^\nu}_{\kappa}  \,\mathcal{U}^{\rho\sigma\kappa}, 
\label{eq:U_Projection_u11}\\
\epsilon \,\mathcal{L}^{\mu\nu\rho} &=& {h^\mu}_\sigma {h^\nu}_{\kappa} {h^\rho}_\lambda \mathcal{U}^{\sigma\kappa\lambda}
\label{eq:U_Projection_111_Comoving}
\end{eqnarray}
for the Lagrangian projections of $\mathcal{U}^{\mu\nu\rho}$. 
The associated Lagrangian decomposition is
\begin{eqnarray}
\mathcal{U}^{\mu\nu\rho} &=& \epsilon \left( \mathcal{J} u^\mu u^\nu u^\rho + \mathcal{H}^{\mu} u^\nu u^\rho + \mathcal{H}^{\nu} u^\mu u^\rho + \mathcal{H}^{\rho} u^\mu u^\nu \right. \nonumber \\
& & \left. + \mathcal{K}^{\mu\nu} u^\rho + \mathcal{K}^{\mu\rho} u^\nu + \mathcal{K}^{\rho\nu} u^\mu + \mathcal{L}^{\mu\nu\rho} \right).
\label{eq:U_DecompositionComoving}
\end{eqnarray}
Eqs.~(\ref{eq:U_ProjectionComoving}) and (\ref{eq:U_DecompositionComoving}) are special to contraction and decomposition respectively with respect to the Lagrangian observer four-velocity $u^\mu$. 
That is, the identity (up to a factor $\epsilon$) of the Lagrangian projections $\mathcal{J}$, $\mathcal{H}^\mu$, and $\mathcal{K}^{\mu\nu}$ of $\mathcal{T}^{\mu\nu}$ with the Lagrangian projections of $\mathcal{U}^{\mu\nu\rho}$ (except of course for the irreducible $\mathcal{L}^{\mu\nu\rho}$) holds due to our choice to measure energies and define angular moments in the comoving frame.
We can define (and in fact will use) the Eulerian projections of $\mathcal{U}^{\mu\nu\rho}$,
\begin{eqnarray}
\mathcal{Z} &=& -n_\mu n_\nu n_\rho\, \mathcal{U}^{\mu\nu\rho}, 
\label{eq:U_Projection_nnn}\\
\mathcal{Y}^\mu &=&  n_\nu n_\rho {\gamma^\mu}_\sigma \, \mathcal{U}^{\nu\rho\sigma}, 
\label{eq:U_Projection_nn1}\\
\mathcal{X}^{\mu\nu} &=& - n_\rho {\gamma^\mu}_\sigma {\gamma^\nu}_{\kappa}  \,\mathcal{U}^{\rho\sigma\kappa}, 
\label{eq:U_Projection_n11} \\
\mathcal{W}^{\mu\nu\rho} &=& {\gamma^\mu}_\sigma {\gamma^\nu}_{\kappa} {\gamma^\rho}_\lambda \mathcal{U}^{\sigma\kappa\lambda},
\label{eq:U_Projection_111}
\end{eqnarray}
and construct the associated Eulerian decomposition of $\mathcal{U}^{\mu\nu\rho}$:
\begin{eqnarray}
\mathcal{U}^{\mu\nu\rho} &=& \mathcal{Z} n^\mu n^\nu n^\rho + \mathcal{Y}^{\mu} n^\nu n^\rho + \mathcal{Y}^{\nu} n^\mu n^\rho + \mathcal{Y}^{\rho} n^\mu n^\nu  \nonumber \\
& &  + \mathcal{X}^{\mu\nu} n^\rho + \mathcal{X}^{\mu\rho} n^\nu + \mathcal{X}^{\rho\nu} n^\mu + \mathcal{W}^{\mu\nu\rho}.
\label{eq:U_DecompositionEulerian}
\end{eqnarray}
The relationships between the Eulerian projections $\mathcal{Z}$, $\mathcal{Y}^{\mu}$, $\mathcal{X}^{\mu\nu}$, and $\mathcal{W}^{\mu\nu\rho}$ of $\mathcal{U}^{\mu\nu\rho}$ and the Eulerian projections $\mathcal{E}$, $\mathcal{F}^{\mu}$, and $\mathcal{S}^{\mu\nu}$ of $\mathcal{T}^{\mu\nu}$ are not as simple as the relationships between the coefficients of Eqs.~(\ref{eq:StressEnergyDecompositionComoving}) and (\ref{eq:U_DecompositionComoving}). 
Nevertheless, useful relationships between the Eulerian projections of $\mathcal{U}^{\mu\nu\rho}$ and $\mathcal{T}^{\mu\nu}$ do exist and will be presented in the next subsection.

In a different vein, another comoving frame expression that can be written in covariant form thanks to Eq.~(\ref{eq:FluidVelocityD}) is
\begin{eqnarray}
\Gamma^{\hat{0}}_{\hat{\nu}\hat{\mu}} &=& -\Gamma^{\hat{0}}_{\hat{\nu}\hat{\mu}} u_{\hat{0}} = -\Gamma^{\hat{\rho}}_{\hat{\nu}\hat{\mu}} u_{\hat{\rho}} = \nabla_{\hat{\mu}} u_{\hat{\nu}} - \partial _{\hat{\mu}} u_{\hat{\nu}} \nonumber\\
&=&  \nabla_{\hat{\mu}} u_{\hat{\nu}},
\label{eq:Gamma_0}
\end{eqnarray}
a covariant expression for the connection coefficients appearing in Eq.~(\ref{eq:MomentumDivergenceMoments}).

Having obtained these expressions rewritten covariantly in terms of the Lagrangian observer four-velocity $u^\mu$, we are ready to consider the momentum space divergence.
Using Eq.~(\ref{eq:Gamma_0}),
the angle-integrated momentum space divergence of Eq.~(\ref{eq:MomentumDivergenceMoments}) can be written 
\begin{equation}
\left({\mathsf{M}_T}\right)^\rho = \frac{1}{\epsilon^2} \frac{\partial}{\partial \epsilon}
\left( - \epsilon^2 \, {L^\rho}_{\hat{\rho}} \, \mathcal{U}^{{\hat{\rho}}{\hat{\nu}}{\hat{\mu}}} \nabla_{\hat{\mu}} u_{\hat{\nu}} \right),
\end{equation}
or, taking advantage of the covariant nature of this expression, 
\begin{equation}
\left({\mathsf{M}_T}\right)^\rho  = \frac{1}{\epsilon^2} \frac{\partial}{\partial \epsilon}
\left( - \epsilon^2 \, \mathcal{U}^{\rho\nu\mu} \nabla_{\mu} u_{\nu}  \right),
\label{eq:MomentumDivergence_31}
\end{equation}
where $\mathcal{U}^{\rho\nu\mu}$ is given by Eqs.~(\ref{eq:U_DecompositionComoving}) or (\ref{eq:U_DecompositionEulerian}).
Projecting orthogonal and tangent to the spacelike slice, we have
\begin{eqnarray}
-n_\rho \left({\mathsf{M}_T}\right)^\rho  &=& \frac{1}{\epsilon^2} \frac{\partial}{\partial \epsilon}
\left( \epsilon^2 \, n_\rho \, \mathcal{U}^{\rho\nu\mu} \nabla_{\mu} u_{\nu} \right), 
  \label{eq:MomentumDivergence_31_n} \\
\gamma_{j\rho} \left({\mathsf{M}_T}\right)^\rho  &=& \frac{1}{\epsilon^2} \frac{\partial}{\partial \epsilon}
\left( - \epsilon^2 \, \gamma_{j\rho} \, \mathcal{U}^{\rho\nu\mu} \nabla_{\mu} u_{\nu} \right). \label{eq:MomentumDivergence_31_gamma}
\end{eqnarray}
Up to multiplicative factors of $\epsilon$ in defining the moments, 
the projections of the momentum space divergence (i.e. energy derivative) presented here are in accord with the corresponding terms in Eqs.~(3.37) and (3.38) of Shibata et al. \cite{Shibata2011Truncated-Momen}.

\subsection{Three-velocity of Lagrangian observers and the relation between the lab and comoving frames}
\label{sec:ThreeVelocity}

In Section \ref{sec:Relativistic_VET} we obtained expressions in terms of lab frame coordinate basis spacetime position components $x^\mu$ and comoving frame orthonormal basis momentum components $p^{\hat\imath}$ through application of coordinate transformations ${L^{\mu}}_{\hat{\mu}}$, but in the context of the 3+1 formulation there is a more fruitful way of expressing the relationship between the lab and comoving frames.
In particular the Lagrangian observer four-velocity $u^\mu$ (i.e. the four-velocity of the fluid) can be `Eulerian decomposed' into parts orthogonal and tangent to the spacelike slice, that is to say, parts tangent and orthogonal to the Eulerian observer four-velocity $n^\mu$:
\begin{equation}
u^\mu = \Lambda \left(n^\mu + v^\mu\right).
\label{eq:ThreeVelocity}
\end{equation}
The orthogonality requirement on $v^\mu$,
\begin{equation}
n_\mu v^\mu = 0,
\label{eq:ThreeVelocityOrthogonality}
\end{equation}
implies (see Eq.~(\ref{eq:UnitNormalD})) that $v^\mu$ is spacelike and has components
\begin{equation}
\left(v^\mu \right)= \left(0, v^i \right)^T
\end{equation}
in the lab frame coordinate basis.
The interpretation of $v^\mu$ as the three-velocity of a Lagrangian observer as measured by an Eulerian observer is confirmed by squaring Eq.~(\ref{eq:ThreeVelocity}) to find the expected Lorentz factor
\begin{equation}
\Lambda = \left( 1 - v^\mu v_\mu \right)^{-1/2} = \left( 1 - v^i v_i \right)^{-1/2}
\label{eq:LorentzFactor}
\end{equation}
for a boost between frames with relative three-velocity $v^\mu$. 
(Recall that $u_\mu u^\mu = n_\mu n^\mu = -1$. 
%We note also the algebraic relation
%\begin{equation}
%v^\mu v_\mu = \frac{\Lambda^2 - 1}{\Lambda^2}.
%\end{equation}
The scalar Lorentz factor $\Lambda$ lacks indices and will not be confused with the Lorentz boost ${\Lambda^{\bar\mu}}_{\hat{\mu}}$.)

In the previous subsection we mentioned finding the monochromatic energy density, momentum density, and stress measured by an Eulerian observer ($\mathcal{E}$, $\mathcal{F}^{\mu}$, and $\mathcal{S}^{\mu\nu}$) in terms of their counterparts measured by a Lagrangian observer ($\mathcal{J}$, $\mathcal{H}^{\mu}$, and $\mathcal{K}^{\mu\nu}$), and vice-versa.
In substituting the Lagrangian decomposition of Eq.~(\ref{eq:StressEnergyDecompositionComoving}) in 
Eqs.~(\ref{eq:T_Projection_nn_Eulerian})-(\ref{eq:T_Projection_11_Eulerian}), the factors of $n_\mu$ and ${\gamma^\mu}_\nu$ can be expressed in terms of $u_\mu$ and $v_\mu$ via Eq.~(\ref{eq:ThreeVelocity}). 
Using also Eqs.~(\ref{eq:MomentumDensityTangentComoving}) and (\ref{eq:StressTangentComoving}), this provides an alternate route to the relations obtained more tediously from the transformation $\mathcal{T}^{\mu\nu} = {L^\mu}_{\hat{\mu}} {L^\nu}_{\hat{\nu}} \, \mathcal{T}^{{\hat{\mu}}{\hat{\nu}}}$ and Eq.~(\ref{eq:MonochromaticStressEnergy}). 
The explicit results are analogous to those obtained \cite{Munier1986Radiation-trans} via Lorentz transformations in special relativity. 
They are not particularly illuminating, but for completeness we exhibit them in Appendix~\ref{app:EulerianLagrangianProjections}.
The inverse relations for the Lagrangian projections in terms of the Eulerian projections---obtained by substituting Eq.~(\ref{eq:StressEnergyDecompositionEulerian}) in 
Eqs.~(\ref{eq:T_Projection_uu_Lagrangian})-(\ref{eq:T_Projection_11_Lagrangian})---are even less illuminating, and we do not even bother to display them in an appendix.
Relations between $\mathcal{E}$, $\mathcal{F}^\mu$, $\mathcal{S}^{\mu\nu}$ and $\mathcal{J}$, $\mathcal{H}^\mu$, $\mathcal{K}^{\mu\nu}$ will of course be needed in numerical work,
but it may be best to perform the contractions in Eqs.~(\ref{eq:T_Projection_nn_Eulerian})-(\ref{eq:T_Projection_11_Eulerian}) or (\ref{eq:T_Projection_uu_Lagrangian})-(\ref{eq:T_Projection_11_Lagrangian}) numerically rather than code tedious analytic expressions.
%The explicit results are not particularly illuminating and we do not display them.
%These relations will of course be needed in numerical work,
%but it may be best to perform the contractions numerically rather than code tedious analytic expressions.

We also use Eqs.~(\ref{eq:ThreeVelocity}) and (\ref{eq:ThreeVelocityOrthogonality}) in fulfilling our promise, 
made in the previous subsection, to relate the Eulerian projections of $\mathcal{U}^{\mu\nu\rho}$ in Eq.~(\ref{eq:U_DecompositionEulerian}) to those of $\mathcal{T}^{\mu\nu}$ in Eq.~(\ref{eq:StressEnergyDecompositionEulerian}).
These relations are obtained by plugging Eq.~(\ref{eq:U_DecompositionEulerian}) into Eq.~(\ref{eq:U_ProjectionComoving}) and comparing the results with Eq.~(\ref{eq:StressEnergyDecompositionEulerian}) for the coefficients of outer products of two, one, and zero copies of $n^\mu$. 
The results are
\begin{eqnarray}
\Lambda \left( \mathcal{Z} - v_\mu \mathcal{Y}^\mu \right) &=& \epsilon \,\mathcal{E}, \label{eq:U_T_Projections_11}\\
\Lambda \left( \mathcal{Y}^\mu - v_\nu \mathcal{X}^{\mu\nu} \right) &=& \epsilon \,\mathcal{F^\mu}, \label{eq:U_T_Projections_1n}\\
\Lambda \left( \mathcal{X}^{\mu\nu} - v_\rho \mathcal{W}^{\mu\nu\rho} \right) &=& \epsilon \,\mathcal{S}^{\mu\nu}. \label{eq:U_T_Projections_nn}
\end{eqnarray}
These can be `unraveled' in reverse order to give
\begin{eqnarray}
\Lambda \, \mathcal{X}^{\mu\nu} &=& \epsilon \,\mathcal{S}^{\mu\nu} + \Lambda v_\rho \mathcal{W}^{\mu\nu\rho}, 
\label{eq:F_Unravel} \\
\Lambda \, \mathcal{Y}^{\mu} &=& \epsilon\, \mathcal{F}^\mu + v_\nu \left(\epsilon\,\mathcal{S}^{\mu\nu} + \Lambda v_\rho \mathcal{W}^{\mu\nu\rho} \right), \label{eq:E_Unravel}\\
\Lambda \, \mathcal{Z} &=& \epsilon \,\mathcal{E} + v_\mu \left[
\epsilon\, \mathcal{F}^\mu +  v_\nu \left(\epsilon\,\mathcal{S}^{\mu\nu} + \Lambda v_\rho \mathcal{W}^{\mu\nu\rho} \right)\right]. \nonumber\\
& &
\label{eq:G_Unravel}
\end{eqnarray}
We discussed finding the Eulerian projections $\mathcal{E}$, $\mathcal{F}^\mu$, and $\mathcal{S}^{\mu\nu}$ in terms of the Lagrangian $\mathcal{J}$, $\mathcal{H}^\mu$, and $\mathcal{K}^{\mu\nu}$ (i.e. the comoving frame angular moments) in the previous paragraph; see also Appendix~\ref{app:EulerianLagrangianProjections}.
There remains the third moment $\mathcal{W}^{\mu\nu\rho}$, which can be found in terms of $\mathcal{J}$, $\mathcal{H}^\mu$, $\mathcal{K}^{\mu\nu}$, and $\mathcal{L}^{\mu\nu\rho}$ through Eq.~(\ref{eq:U_Projection_111}), 
using the Lagrangian decomposition of Eq.~(\ref{eq:U_DecompositionComoving}) on the right-hand side.
Similarly, Eqs.~(\ref{eq:U_Projection_nnn})-(\ref{eq:U_Projection_n11}) can be used in lieu of Eqs.~(\ref{eq:F_Unravel})-(\ref{eq:G_Unravel}) to directly obtain $\mathcal{Z}$, $\mathcal{Y}^\mu$, and $\mathcal{X}^{\mu\nu}$ in terms of $\mathcal{J}$, $\mathcal{H}^\mu$, $\mathcal{K}^{\mu\nu}$, and $\mathcal{L}^{\mu\nu\rho}$ as well.
Again we reserve explicit expressions for Appendix~\ref{app:EulerianLagrangianProjections}; and again we also emphasize that it may be best to numerically perform the contractions in Eqs.~(\ref{eq:U_Projection_nnn})-(\ref{eq:U_Projection_111})---or, in the inverse case, Eq.~(\ref{eq:U_Projection_111_Comoving})---rather than code tedious analytic expressions.

These Eulerian projections come into play in making  the momentum space divergence more explicit, 
for we use the Eulerian decomposition of $\mathcal{U}^{\rho\nu\mu}$ given by Eq.~(\ref{eq:U_DecompositionEulerian}) and the Eulerian decomposition of $u_\nu$ given by Eq.~(\ref{eq:ThreeVelocity}) in expanding the expression $\mathcal{U}^{\rho\nu\mu} \nabla_\mu u_\nu$ appearing in Eqs.~(\ref{eq:MomentumDivergence_31_n}) and (\ref{eq:MomentumDivergence_31_gamma}).
At first glance the use of these Eulerian decompositions may seem to complicate things.
With our choice to measure neutrino momentum components and define angular moments in the comoving frame,
the Lagrangian decomposition of $\mathcal{U}^{\rho\nu\mu}$ in Eq.~(\ref{eq:U_DecompositionComoving}) is simpler in the sense described in Section \ref{sec:LagrangianObservers}, and indeed $u_\nu$ is the four-velocity of Lagrangian observers; so why not stay with these Lagrangian expressions?
The problem is that we cannot stay in `Lagrangian world'---in the comoving frame---altogether, much as we might like to, 
because the covariant derivative $\nabla_\mu u_\nu$ is with respect to the lab frame coordinate basis.
We have swept the connection coefficients in Eq.~(\ref{eq:MomentumDivergenceMoments}) temporarily under the rug via Eq.~(\ref{eq:Gamma_0}), but lab frame coordinate basis connection coefficients still lurk in the covariant derivative $\nabla_\mu u_\nu$.

Given the unavoidable necessity of facing the relation between the lab and comoving frames in one way or another, there are significant advantages to consistent use of Eulerian decompositions, in which the relation between frames is focused more in the three-velocity $v^\mu$ of a Lagrangian observer as measured by an Eulerian observer, than in the coordinate transformations ${L^\mu}_{\hat{\mu}}$ of Eq.~(\ref{eq:CompositeTransformation}).

If we stay with Eq.~(\ref{eq:MomentumDivergenceMoments}), we face the unpleasant prospect of evaluating Eq.~(\ref{eq:ConnectionComoving}) for the transformed connection coefficients.
Even rewritten as Eq.~(\ref{eq:MomentumDivergence_31}), we face lab frame coordinate basis connection coefficients when using the Lagrangian decomposition of $\mathcal{U}^{\rho\nu\mu}$ and leaving the Lagrangian observer four-velocity $u_\nu$ as is. 
%But the fact that Eulerian decompositions are most natural in the 3+1 approach allows us to almost completely avoid explicit encounters with connection coefficients,
%the only exception being a simplified case with no inverse metric and only spatial derivatives of three-metric components $\gamma_{ij}$.
But as derivations in Appendices~\ref{app:SpacetimeDivergence} and \ref{app:MomentumSpaceDivergence} show, the fact that Eulerian decompositions are most natural in the 3+1 approach allows us to almost completely avoid explicit encounters with connection coefficients.

Moreover, consistent use of the Eulerian perspective---both in projecting out the portions of the phase space divergence orthogonal and tangent to the spacelike slice, and in Eulerian decompositions of $\mathcal{U}^{\rho\nu\mu}$ and $u_\nu$---also turns out to preclude any appearance of time derivatives of metric functions, even in intermediate steps. Time derivatives of the lapse function $\alpha$ and shift vector $\beta^i$ would be particularly inconvenient in numerical work, as these do not normally have evolution equations associated with them. 
(Unfortunately, we shall see that time derivatives of the Lorentz factor $\Lambda$ and three-velocity $v^i$ remain; these are something of a nuisance, but at least in principle they could be written in terms of spatial derivatives via hydrodynamics evolution equations.)

Finally, the Eulerian decomposition of $\mathcal{U}^{\rho\nu\mu}$ is more readily tied to the Eulerian decomposition of $\mathcal{T}^{\nu\mu}$, i.e. to the energy density $\mathcal{E}$, momentum density $\mathcal{F}^\mu$, and stress $\mathcal{S}^{\mu\nu}$ measured by an Eulerian observer. 
This is advantageous in relating four-momentum conservation to lepton number conservation, for cancellations must occur between the spacetime and momentum space divergences, and it is $\mathcal{E}$, $\mathcal{F}^\mu$, and $\mathcal{S}^{\mu\nu}$ that appear in the spacetime divergence in Eqs.~(\ref{eq:SpacetimeDivergence_31_n}) and (\ref{eq:SpacetimeDivergence_31_gamma}).

Details of the calculation of $\mathcal{U}^{\rho\nu\mu} \nabla_\mu u_\nu$, using the Eulerian decompositions of $\mathcal{U}^{\rho\nu\mu}$ and $\nabla_\mu u_\nu$, are given in Appendix~\ref{app:MomentumSpaceDivergence}.
Using those results for the projection orthogonal to the spacelike slice, we have
\begin{equation}
-n_\nu \left({\mathsf{M}_T}\right)^\nu  = \frac{1}{\alpha\sqrt{\gamma}}\frac{1}{\epsilon^2} \frac{\partial}{\partial \epsilon}
\left[ \epsilon^2 \left( \mathsf{R}_{T,n} + \mathsf{O}_{T,n} \right) \right], 
  \label{eq:MomentumDivergence_31_n_2}
\end{equation}
where
\begin{eqnarray}
\mathsf{R}_{T,n} &=&  \alpha\sqrt{\gamma}\, \Lambda
\left[ \frac{ \left(\mathcal{Z} v^i - \mathcal{Y}^i \right) }{\alpha} \frac{\partial\alpha}{\partial x^i}
-  \frac{\mathcal{Y}_k v^i}{\alpha} \frac{\partial \beta^k}{\partial x^i}\right. \nonumber \\
& & \left. - \frac{\mathcal{X}^{k i}  v^m}{2}  \frac{\partial \gamma_{ki}}{\partial x^m}
+  \mathcal{X}^{k i} K_{k i}
\right], 
\label{eq:MomentumDivergence_31_n_2_R} \\
\mathsf{O}_{T,n} &=&  \alpha\sqrt{\gamma} \left[\mathcal{Z} n^\mu \frac{\partial\Lambda}{\partial x^\mu}
+ \mathcal{Y}^i \frac{\partial\Lambda}{\partial x^i} \right. \nonumber \\
& & \left. - \mathcal{Y}_k n^\mu \frac{\partial \left( \Lambda v^k\right) }{\partial x^\mu} 
- {\mathcal{X}_k}^i  \frac{\partial\left( \Lambda v^k\right)}{\partial x^i}
\right],
\label{eq:MomentumDivergence_31_n_2_O} 
\end{eqnarray}
arise from changes in the neutrino energy as measured in the comoving frame due to gravitational redshift $\left(\mathsf{R}_{T,n}\right)$ and the acceleration of the observer riding along with the fluid $\left(\mathsf{O}_{T,n}\right)$.
Similarly,
\begin{equation}
\gamma_{j\nu} \left({\mathsf{M}_T}\right)^\nu = \frac{1}{\alpha\sqrt{\gamma}}\frac{1}{\epsilon^2} \frac{\partial}{\partial \epsilon}
\left\{ \epsilon^2 \left[ \left(\mathsf{R}_{T,\gamma}\right)_j + \left(\mathsf{O}_{T,\gamma} \right)_j\right] \right\}, 
  \label{eq:MomentumDivergence_31_gamma_2}
\end{equation}
with
\begin{eqnarray}
\left(\mathsf{R}_{T,\gamma}\right)_j &=&  \alpha\sqrt{\gamma}\, \Lambda
\left[ \frac{ \left(\mathcal{Y}_j v^i - {\mathcal{X}_j}^i \right) }{\alpha} \frac{\partial\alpha}{\partial x^i}
-  \frac{\mathcal{X}_{jk} v^i}{\alpha} \frac{\partial \beta^k}{\partial x^i}\right. \nonumber \\
& & \left. - \frac{{\mathcal{W}_j}^{k i}  v^m}{2}  \frac{\partial \gamma_{ki}}{\partial x^m}
+  {\mathcal{W}_j}^{k i} K_{k i}
\right], 
\label{eq:MomentumDivergence_31_gamma_2_R} \\
\left(\mathsf{O}_{T,\gamma}\right)_j &=&  \alpha\sqrt{\gamma} \left[\mathcal{Y}_j n^\mu \frac{\partial\Lambda}{\partial x^\mu}
+ {\mathcal{X}_j}^i \frac{\partial\Lambda}{\partial x^i} \right. \nonumber \\
& & \left. - \mathcal{X}_{jk} n^\mu \frac{\partial \left( \Lambda v^k\right) }{\partial x^\mu} 
- {\mathcal{W}_{jk}}^i  \frac{\partial\left( \Lambda v^k\right)}{\partial x^i}
\right]
\label{eq:MomentumDivergence_31_gamma_2_O} 
\end{eqnarray}
for the projection of the momentum space divergence tangent to the spacelike slice.
That the structures of $\mathsf{R}_{T,n}$ and $\mathsf{O}_{T,n}$ parallel those of $\left(\mathsf{R}_{T,\gamma}\right)_j$ and $\left(\mathsf{O}_{T,\gamma}\right)_j$
is simply a reflection of the parallel structure of Eqs.~(\ref{eq:U_Projection_n}) and (\ref{eq:U_Projection_gamma}).
As given here they are expressed in terms of the Eulerian projections $\mathcal{Z}$, $\mathcal{Y}^\mu$, $\mathcal{X}^{\mu\nu}$, and $\mathcal{W}^{\mu\nu\rho}$ of $\mathcal{U}^{\rho\nu\mu}$ (see Eq.~(\ref{eq:U_DecompositionEulerian})).
In numerical work these could be further expressed in terms of the Eulerian projections of $\mathcal{T}^{\nu\mu}$ (see Eq.~(\ref{eq:StressEnergyDecompositionEulerian})), i.e. the energy density $\mathcal{E}$, momentum density $\mathcal{F}^\mu$, and stress $\mathcal{S}^{\mu\nu}$ measured by an Eulerian observer, via Eqs.~(\ref{eq:F_Unravel})-(\ref{eq:G_Unravel}).
They could also be expressed directly in terms of the Lagrangian projections, as shown in Appendix~\ref{app:EulerianLagrangianProjections}.

\subsection{Four-momentum and lepton number exchange}
\label{sec:MomentumNumberExchange_31}

The relationship of lepton number and energy exchange at a high level is readily seen.
Revisiting Section \ref{sec:MomentumNumberExchange} in terms of covariant expressions involving the Lagrangian observer four-velocity $u^\mu$ (see Section \ref{sec:LagrangianObservers}), it is easy to see that Eq.~(\ref{eq:MonochromaticNumberFlux}) becomes
\begin{equation}
\mathcal{N}^{\mu} = -\frac{1}{\epsilon} u_\nu \mathcal{T}^{\mu\nu},
\label{eq:MonochromaticNumberFluxCovariant}
\end{equation}
that Eq.~(\ref{eq:MomentumDivergenceMomentsNumber}) becomes
\begin{equation}
\mathsf{M}_N = \frac{1}{\epsilon^2} \frac{\partial}{\partial \epsilon}
\left( - \epsilon^2 \, \mathcal{T}^{\nu\mu} \nabla_\mu u_\nu \right),
\label{eq:MomentumDivergenceNumberCovariant}
\end{equation}
and that Eq.~(\ref{eq:Identity}) becomes
\begin{equation}
\mathsf{S}_N + \mathsf{M}_N = -\frac{1}{\epsilon} u_{\nu} \left[ \left({\mathsf{S}_T}\right)^\nu + \left({\mathsf{M}_T}\right)^\nu\right]. 
\label{eq:IdentityCovariant}
\end{equation}
The first term on the right-hand side is
\begin{eqnarray}
-\frac{1}{\epsilon} u_{\nu} \left({\mathsf{S}_T}\right)^\nu &=& -\frac{1}{\epsilon} u_{\nu} \nabla_\mu \mathcal{T}^{\mu\nu} \\
&=&  \nabla_\mu \left(-\frac{1}{\epsilon} u_{\nu} \mathcal{T}^{\mu\nu} \right) 
+ \frac{\mathcal{T}^{\mu\nu}}{\epsilon} \nabla_\mu u_{\nu} \\
&=& \mathsf{S}_N + \frac{\mathcal{T}^{\mu\nu}}{\epsilon} \nabla_\mu u_{\nu},
\label{eq:ST_Contraction}
\end{eqnarray}
thanks to Eq.~(\ref{eq:MonochromaticNumberFluxCovariant}).
The second term on the right-hand side of Eq.~(\ref{eq:IdentityCovariant}) is
\begin{eqnarray}
-\frac{1}{\epsilon} u_{\rho} \left({\mathsf{M}_T}\right)^\rho &=& -\frac{1}{\epsilon} u_{\rho} \left[\frac{1}{\epsilon^2} \frac{\partial}{\partial \epsilon}
\left( - \epsilon^2 \, \mathcal{U}^{\rho\nu\mu} \nabla_\mu u_\nu \right)\right] \\
&=& \frac{1}{\epsilon^2} \frac{\partial}{\partial \epsilon}
\left( - \epsilon^2 \, \mathcal{T}^{\nu\mu} \nabla_\mu u_\nu \right) \nonumber \\
& &- \epsilon \,\mathcal{T}^{\nu\mu} \nabla_\mu u_\nu \frac{\partial}{\partial \epsilon} \left( -\frac{1}{\epsilon}\right) \\
&=& \mathsf{M}_N - \frac{\mathcal{T}^{\nu\mu}}{\epsilon} \nabla_\mu u_{\nu},
\label{eq:MT_Contraction}
\end{eqnarray}
by virtue of Eqs.~(\ref{eq:U_ProjectionComoving}) and (\ref{eq:MomentumDivergenceNumberCovariant}).
The sum of Eqs.~(\ref{eq:ST_Contraction}) and (\ref{eq:MT_Contraction}) gives Eq.~(\ref{eq:IdentityCovariant}) as required.

This consistency between lepton number and energy exchange applies not only at this high level, but also to the detailed form of the moment equations as we have most expressly written them, which has implications for their discretization. 
In terms of the decomposition of $u^\mu$ in Eq.~(\ref{eq:ThreeVelocity}), we have
\begin{equation}
\mathsf{S}_N + \mathsf{M}_N = -\frac{\Lambda}{\epsilon} \left(n_\nu + v_\nu \right) \left[ \left({\mathsf{S}_T}\right)^\nu + \left({\mathsf{M}_T}\right)^\nu\right]
\label{eq:IdentityDecomposed}
\end{equation}
for Eq.~(\ref{eq:IdentityCovariant}).
This is, naturally, closely related to our projections of the spacetime and momentum space divergences orthogonal and tangent to the spacelike slice.

We consider first the spacetime divergence on the right-hand side of Eq.~(\ref{eq:IdentityDecomposed}).
Using Eqs.~(\ref{eq:SpacetimeDivergence_31_n})-(\ref{eq:SpacetimeDivergence_31_n_F}) and (\ref{eq:UnitNormalU}), we have
\begin{eqnarray}
-\frac{\Lambda}{\epsilon} n_\nu \left({\mathsf{S}_T}\right)^\nu &=& \left\{
\frac{\partial }{\partial t}\left[\frac{\Lambda \left(\mathsf{D}_{T,n}\right)}{\epsilon}\right]  + \frac{\partial  }{\partial x^i} \left[ \frac{\Lambda\left(\mathsf{F}_{T,n}\right)^i}{\epsilon}\right] \right. \nonumber \\
& &\left.  - \frac{1 }{\epsilon}\left[\Lambda\left(\mathsf{G}_{T,n}\right) + \mathsf{E}_{T,n} \right]   \right\} \frac{1}{\alpha\sqrt{\gamma}},
\label{eq:IdentitySpacetimeDivergence_n}
\end{eqnarray}
where 
\begin{equation}
\mathsf{E}_{T,n} = \alpha\sqrt{\gamma} \left(\mathcal{E}n^\mu   \frac{\partial\Lambda}{\partial x^\mu}  
+ \mathcal{X}^i  \frac{\partial\Lambda}{\partial x^i}\right)
\label{eq:Extra_n}
\end{equation}
are the `extra' terms that arise from pulling the factor $\Lambda $ inside the time and space derivatives. 
Thus the discretized form of $\mathsf{E}_{T,n}$ will be dictated by the discretization chosen for the first two terms of Eq.~(\ref{eq:SpacetimeDivergence_31_n}).
Similarly, using Eqs.~(\ref{eq:SpacetimeDivergence_31_gamma})-(\ref{eq:SpacetimeDivergence_31_gamma_F}) and (\ref{eq:UnitNormalU}), we have
\begin{eqnarray}
-\frac{\Lambda v^j}{\epsilon} \gamma_{j \nu} \left({\mathsf{S}_T}\right)^\nu &=& \left\{\!
\frac{\partial }{\partial t}\!\left[\!\frac{\Lambda v^j \left(\mathsf{D}_{T,\gamma}\right)_j}{\epsilon}\!\right]\! + \!\frac{\partial  }{\partial x^i}\! \left[\! \frac{\Lambda v^j {\left(\mathsf{F}_{T,\gamma}\right)^i}_j }{\epsilon}\!\right] \right. \nonumber \\
& &\left.  - \frac{1}{\epsilon}\left[\Lambda v^j \left(\mathsf{G}_{T,\gamma}\right)_j  + \mathsf{E}_{T,\gamma}\right] \right\} \! \left(\! -\frac{1}{\alpha\sqrt{\gamma}} \! \right)\!,
\nonumber \\
&&\label{eq:IdentitySpacetimeDivergence_gamma}
\end{eqnarray}
where
\begin{equation}
\mathsf{E}_{T,\gamma} 
= \alpha\sqrt{\gamma} \left[ \mathcal{F}_j n^\mu   \frac{\partial \left( \Lambda v^j \right)}{\partial x^\mu}  
+ {\mathcal{S}^i}_j \frac{\partial \left( \Lambda v^j \right)}{\partial x^i} \right]
\label{eq:Extra_gamma}
\end{equation}
are the `extra' terms that arise from pulling the factor $\Lambda v^j$ inside the time and space derivatives. 
Thus the discretized form of $\mathsf{E}_{T,\gamma}$ will be dictated by the discretization chosen for the first two terms of Eq.~(\ref{eq:SpacetimeDivergence_31_gamma}).
Adding Eqs.~({\ref{eq:IdentitySpacetimeDivergence_n}) and ({\ref{eq:IdentitySpacetimeDivergence_gamma}) gives
\begin{eqnarray}
-\frac{\Lambda}{\epsilon} \left( \right. &n_\nu& \left. + v^j \gamma_{j \nu} \right) \left({\mathsf{S}_T}\right)^\nu \nonumber \\
&=& \mathsf{S}_N  
- \left\{ \frac{\Lambda }{\epsilon} \left[ \left(\mathsf{G}_{T,n}\right)
- v^j \left(\mathsf{G}_{T,\gamma}\right)_j \right]\right.  \nonumber \\
&& \left.
\ \ \ \ \ \ \ \ + \frac{1}{\epsilon}\left(\mathsf{E}_{T,n}- \mathsf{E}_{T,\gamma} \right)\right\}\frac{1}{\alpha\sqrt{\gamma}}
\label{eq:u_S}
\end{eqnarray}
for the contribution of the spacetime divergence to the right-hand side of Eq.~(\ref{eq:IdentityDecomposed}).

Turning to the momentum space divergence, from Eqs.~(\ref{eq:MomentumDivergence_31_n_2}) and (\ref{eq:MomentumDivergence_31_gamma_2}) we have
\begin{eqnarray}
-\frac{\Lambda}{\epsilon} \left(\right.&n_\nu &\left. + \ v^j \gamma_{j \nu} \right) \left({\mathsf{M}_T}\right)^\nu \nonumber \\
&=& {\mathsf{M}_N} 
- \frac{1 }{\epsilon^2} \frac{\partial}{\partial \epsilon}\left( \frac{1}{\epsilon} \right)
\left\{ \Lambda \epsilon^2 \left[ \left(\mathsf{R}_{T,n}\right)
- v^j \left(\mathsf{R}_{T,\gamma}\right)_j \right] \right.  \nonumber \\
&& \left.
\ \ \ \ \ \ \ \ + \Lambda \epsilon^2\left[\left(\mathsf{O}_{T,n}\right) - v^j \left(\mathsf{O}_{T,\gamma}\right)_j \right] 
\right\}\frac{1}{\alpha\sqrt{\gamma}}
\label{eq:u_M}
\end{eqnarray}
for the contribution of the momentum space divergence to the right-hand side of Eq.~(\ref{eq:IdentityDecomposed}).
Note that the long term following $\mathsf{M}_N$ results from pulling $1/\epsilon$ through the energy derivative; therefore its discretized form is dictated by the discretization chosen for Eqs.~(\ref{eq:MomentumDivergence_31_n_2}) and (\ref{eq:MomentumDivergence_31_gamma_2}).

The sum of Eqs.~(\ref{eq:u_S}) and (\ref{eq:u_M}) equals Eq.~(\ref{eq:IdentityDecomposed}), as required, because (a) the gravitational redshift terms from the momentum space divergence cancel the gravitational force and power terms from the spacetime divergence, 
and (b) the observer corrections from the momentum space divergence cancel the `extra' terms from pulling $\Lambda$ and $v^k$ through the time and space derivatives.
That is, (a)
\begin{equation}
0 = \frac{\Lambda }{\epsilon} \left[ \left(\mathsf{G}_{T,n}\right)
- v^j \left(\mathsf{G}_{T,\gamma}\right)_j \right]
- \frac{\Lambda }{\epsilon^2} \left[ \left(\mathsf{R}_{T,n}\right)
- v^j \left(\mathsf{R}_{T,\gamma}\right)_j \right],
\label{eq:CancellationGravity}
\end{equation}
and (b)
\begin{equation}
0 = \frac{1}{\epsilon}\left(\mathsf{E}_{T,n}- \mathsf{E}_{T,\gamma} \right)
- \frac{\Lambda}{\epsilon^2} \left[\left(\mathsf{O}_{T,n}\right) - v^j \left(\mathsf{O}_{T,\gamma}\right)_j \right]. 
\label{eq:CancellationObserver}
\end{equation}
These cancellations emerge in a surprisingly tractable way: the  gravitational redshift and observer correction terms from the normal and tangent projections of the momentum space divergence combine in just the right way to make use of Eqs.~(\ref{eq:U_T_Projections_11})-(\ref{eq:U_T_Projections_nn}), which relate the Eulerian projections $\mathcal{Z}$, $\mathcal{Y}^\mu$, $\mathcal{X}^{\mu\nu}$, and $\mathcal{W}^{\mu\nu\rho}$ of $\mathcal{U}^{\rho\nu\mu}$ (see Eq.~(\ref{eq:U_DecompositionEulerian})) to the Eulerian projections of $\mathcal{T}^{\nu\mu}$ (see Eq.~(\ref{eq:StressEnergyDecompositionEulerian})), i.e. the energy density $\mathcal{E}$, momentum density $\mathcal{F}^\mu$, and stress $\mathcal{S}^{\mu\nu}$ measured by an Eulerian observer.
In particular, combining Eqs.~(\ref{eq:MomentumDivergence_31_n_2_R}) and (\ref{eq:MomentumDivergence_31_gamma_2_R}) using Eqs.~(\ref{eq:U_T_Projections_11})-(\ref{eq:U_T_Projections_nn}) yields
\begin{eqnarray}
\left(\mathsf{R}_{T,n}\right)  -  v^j \!\left(\mathsf{R}_{T,\gamma}\right)_j 
&=& \epsilon\, \alpha\sqrt{\gamma} \! \left[ \! \frac{\left(\mathcal{E} v^i \!-\! \mathcal{F}^i \right) }{\alpha} \frac{\partial\alpha}{\partial x^i} 
-  \frac{\mathcal{F}_k v^i}{\alpha} \frac{\partial \beta^k}{\partial x^i} \right. \nonumber \\
& &\left. - \frac{\mathcal{S}^{k i} v^m}{2}  \frac{\partial \gamma_{ki}}{\partial x^m}
+  \mathcal{S}^{k i} K_{k i}
\right].
\end{eqnarray}
Plugging this and Eqs.~(\ref{eq:SpacetimeDivergence_31_n_G}) and (\ref{eq:SpacetimeDivergence_31_gamma_G}) into Eq.~(\ref{eq:CancellationGravity}), the term-by-term cancellations are apparent.
Similarly, combining Eqs.~(\ref{eq:MomentumDivergence_31_n_2_O}) and (\ref{eq:MomentumDivergence_31_gamma_2_O}) using Eqs.~(\ref{eq:U_T_Projections_11})-(\ref{eq:U_T_Projections_nn}) yields
\begin{eqnarray}
\left(\mathsf{O}_{T,n}\right) - v^j \left(\mathsf{O}_{T,\gamma}\right)_j
&=& \frac{\epsilon}{\Lambda} \alpha\sqrt{\gamma} 
\left[\mathcal{E} n^\mu \frac{\partial\Lambda}{\partial x^\mu}
+ \mathcal{F}^i \frac{\partial\Lambda}{\partial x^i} \right. \nonumber \\
& & \left. - \mathcal{F}_k n^\mu \frac{\partial \left( \Lambda v^k\right) }{\partial x^\mu} 
- {\mathcal{S}_k}^i  \frac{\partial\left( \Lambda v^k\right)}{\partial x^i} \right]. \nonumber \\
& &
\end{eqnarray}
Plugging this and Eqs.~(\ref{eq:Extra_n}) and (\ref{eq:Extra_gamma}) into Eq.~(\ref{eq:CancellationObserver}), once again the term-by-term cancellations are apparent.

This analytic demonstration of the consistency of our conservative four-momentum moment equations with a conservative number moment equation ideally should be repeated in the discretized case in order to discover discretizations that are faithful to this consistency.
We do not present a full discretization of the moment equations in this paper, but make some additional comments in Appendix~\ref{app:TowardsDiscretization}.

\section{Conclusion}
\label{sec:Conclusion}

\begin{table}[b]
\caption{\label{tab:Spacetime}
%Some important variables in the conservative 3+1 general relativistic Variable Eddington Tensor radiation moments formalism.}
Some spacetime and fluid variables.}
\begin{ruledtabular}
\begin{tabular}{lll}
$x^\mu$ & Spacetime position & Sec.~\ref{sec:Geometry} \\
${L^\mu}_{\hat{\mu}}$ & Transformation between comoving & Sec.~\ref{sec:Boltzmann}; \\
 & and lab frames & Eq.~(\ref{eq:CompositeTransformation}) \\
$\alpha$ & Lapse function & Eq.~(\ref{eq:31_Metric}) \\
$\beta^i$ & Shift vector & Eq.~(\ref{eq:31_Metric}) \\
$\gamma_{ij} $ & Three-metric & Eq.~(\ref{eq:31_Metric}) \\
$K_{ij} $ & Extrinsic curvature & Sec.~\ref{sec:Geometry}; \\
& & Eq.~(\ref{eq:ExtrinsicCurvature}) \\
$n^\mu$ & Unit normal to spacelike slice; &   Eq.~(\ref{eq:UnitNormalU}) \\
     & four-velocity of Eulerian observers & \\
$u^\mu$ & Fluid four-velocity; & Eq.~(\ref{eq:FluidVelocityU}) \\
	& four-velocity of Lagrangian observers & \\ 
$v^\mu$ & Three-velocity of Lagrangian observers & Eq.~(\ref{eq:ThreeVelocity}) \\
$\Lambda $ & Lorentz factor of Lagrangian observers & Eq.~(\ref{eq:LorentzFactor}) \\
$\gamma_{\mu\nu}$ & Projector orthogonal to $n^\mu$ & Eq.~(\ref{eq:OrthogonalProjector}) \\
$h_{\mu\nu}$ & Projector orthogonal to $u^\mu$ & Eq.~(\ref{eq:OrthogonalProjectorComoving}) \\
\end{tabular}
\end{ruledtabular}
\end{table}

\begin{table}[b]
\caption{\label{tab:Moments}
Particle momentum, and particle distribution and its monochromatic (energy-dependent) moments. 
The angular moments are taken with respect to the three-momentum direction unit vector $\ell^\mu$ reckoned by Lagrangian (comoving) observers.
The momentum moments are taken with respect to the particle four-momentum $p^\mu$.
All moments are functions of the particle energy $\epsilon$ measured by a Lagrangian observer.}
\begin{ruledtabular}
\begin{tabular}{lll}
$p^\mu$ & Particle four-momentum & Sec.~\ref{sec:Relativistic_VET} \\
$\epsilon$ & Particle energy measured by & Eq.~(\ref{eq:ComovingMomentum}); \\ 
   & Lagrangian observers  & Eq.~(\ref{eq:ParticleEnergy})  \\
$\Omega$ & Particle momentum direction & Sec.~\ref{sec:Relativistic_VET} \\
& measured by Lagrangian observers & \\
$\ell^\mu$ & Three-momentum direction unit & Eq.~(\ref{eq:ComovingMomentum}); \\
 & vector measured by Lagrangian & Eq.~(\ref{eq:MomentumDecomposition}) \\
  & observers & \\
$f(x^\mu, \epsilon, \Omega)$ & Particle distribution function & Sec.~\ref{sec:Relativistic_VET} \\
$\mathcal{J}(x^\mu, \epsilon)$ & Zeroth angular moment & Eq.~(\ref{eq:AngularMoment_0}); \\
 & & Eq.~(\ref{eq:AngularMoment_0_Covariant}) \\
$\mathcal{H}^\nu(x^\mu, \epsilon)$ & First angular moment & Eq.~(\ref{eq:AngularMoment_1}); \\
 & & Eq.~(\ref{eq:AngularMoment_1_Covariant}) \\
$\mathcal{K}^{\nu\rho}(x^\mu, \epsilon)$ & Second angular moment & Eq.~(\ref{eq:AngularMoment_2}); \\
 & & Eq.~(\ref{eq:AngularMoment_2_Covariant}); \\
 & & Eq.~(\ref{eq:CovariantClosure_K}) \\
$\mathcal{L}^{\nu\rho\sigma}(x^\mu, \epsilon)$ & Third angular moment & Eq.~(\ref{eq:AngularMoment_3}); \\
 & & Eq.~(\ref{eq:AngularMoment_3_Covariant}); \\
 & & Eq.~(\ref{eq:CovariantClosure_L}) \\
 $ \epsilon \mathcal{N}^\nu(x^\mu, \epsilon)$ & First momentum moment; & Eq.~(\ref{eq:MonochromaticNumberFlux_0}); \\
  & comoving energy times number flux & Eq.~(\ref{eq:MonochromaticNumberFluxCovariant}) \\
 $ \mathcal{T}^{\nu\rho}(x^\mu, \epsilon) $ & Second momentum moment; & Eq.~(\ref{eq:MonochromaticStressEnergyDefinition}); \\
  & stress-energy tensor & Eq.~(\ref{eq:MonochromaticStressEnergy_2}) \\ 
$\mathcal{U}^{\nu\rho\sigma}(x^\mu, \epsilon)$ & Third momentum moment & Eq.~(\ref{eq:U_Definition}); \\
 & & Eq.~(\ref{eq:ThirdMoment}) \\
\end{tabular}
\end{ruledtabular}
\end{table}

\begin{table}[b]
\caption{\label{tab:Projections_T}
Lagrangian and Eulerian decompositions and projections of the second monochromatic (energy-dependent) momentum moment $\mathcal{T}^{\mu\nu}$ (stress-energy).
Both the Lagrangian and Eulerian projections are functions of the particle energy $\epsilon$ measured by a Lagrangian observer. 
Note that the Lagrangian projections are the angular moments (see Table~\ref{tab:Moments}), as the latter are defined with respect to the comoving frame.}
\begin{ruledtabular}
\begin{tabular}{lll}
$ \mathcal{T}^{\nu\rho}(x^\mu, \epsilon)$ & Lagrangian decomposition &  Eq.~(\ref{eq:StressEnergyDecompositionComoving}) \\
$ \mathcal{J}(x^\mu, \epsilon)$ & Energy density measured by & Eq.~(\ref{eq:T_Projection_uu_Lagrangian}) \\
 & Lagrangian observers \\
$ \mathcal{H}^\nu(x^\mu, \epsilon)$ & Energy flux measured by & Eq.~(\ref{eq:T_Projection_u1_Lagrangian}) \\
 & Lagrangian observers \\
$ \mathcal{K}^{\nu\rho}(x^\mu, \epsilon)$ & Stress measured by & Eq.~(\ref{eq:T_Projection_11_Lagrangian}) \\
 & Lagrangian observers \\
$ \mathcal{T}^{\nu\rho}(x^\mu, \epsilon)$ & Eulerian decomposition &  Eq.~(\ref{eq:StressEnergyDecompositionEulerian}) \\
$ \mathcal{E}(x^\mu, \epsilon)$ & Energy density measured by & Eq.~(\ref{eq:T_Projection_nn_Eulerian}) \\
 & Eulerian observers \\
$ \mathcal{F}^\nu(x^\mu, \epsilon)$ & Energy flux measured by & Eq.~(\ref{eq:T_Projection_n1_Eulerian}) \\
 & Eulerian observers \\
$ \mathcal{S}^{\nu\rho}(x^\mu, \epsilon)$ & Stress measured by & Eq.~(\ref{eq:T_Projection_11_Eulerian}) \\
 & Eulerian observers \\
\end{tabular}
\end{ruledtabular}
\end{table}

\begin{table}[b]
\caption{\label{tab:Projections_U}
Lagrangian and Eulerian decompositions and projections of the third monochromatic (energy-dependent) momentum moment $\mathcal{U}^{\mu\nu\rho}$.
Both the Lagrangian and Eulerian projections are functions of the particle energy $\epsilon$ measured by a Lagrangian observer. 
Note that the Lagrangian projections are the angular moments (see Table~\ref{tab:Moments}), and also (except for the third) are the same as the projections of $ \mathcal{T}^{\nu\rho}$, up to a factor of $\epsilon$ (see Table~\ref{tab:Projections_T}), thanks to Eq.~(\ref{eq:U_ProjectionComoving}).}
\begin{ruledtabular}
\begin{tabular}{lll}
$ \mathcal{U}^{\nu\rho\sigma}(x^\mu, \epsilon)$ & Lagrangian decomposition &  Eq.~(\ref{eq:U_DecompositionComoving}) \\
$ \epsilon \mathcal{J}(x^\mu, \epsilon)$ & Zeroth projection measured by & Eq.~(\ref{eq:U_Projection_uuu}) \\
 & Lagrangian observers \\
$ \epsilon \mathcal{H}^\nu(x^\mu, \epsilon)$ & First projection measured by & Eq.~(\ref{eq:U_Projection_uu1}) \\
 & Lagrangian observers \\
$ \epsilon \mathcal{K}^{\nu\rho}(x^\mu, \epsilon)$ & Second projection measured by & Eq.~(\ref{eq:U_Projection_u11}) \\
 & Lagrangian observers \\
$ \epsilon \mathcal{L}^{\nu\rho\sigma}(x^\mu, \epsilon)$ & Third projection measured by & Eq.~(\ref{eq:U_Projection_111_Comoving}) \\
 & Lagrangian observers \\
$ \mathcal{U}^{\nu\rho\sigma}(x^\mu, \epsilon)$ & Eulerian decomposition &  Eq.~(\ref{eq:U_DecompositionEulerian}) \\
$ \mathcal{Z}(x^\mu, \epsilon)$ & Zeroth projection measured by & Eq.~(\ref{eq:U_Projection_nnn}) \\
 & Eulerian observers \\
$ \mathcal{Y}^\nu(x^\mu, \epsilon)$ & First projection measured by & Eq.~(\ref{eq:U_Projection_nn1}) \\
 & Eulerian observers \\
$ \mathcal{X}^{\nu\rho}(x^\mu, \epsilon)$ & Second projection measured by & Eq.~(\ref{eq:U_Projection_n11}) \\
 & Eulerian observers \\
$ \mathcal{W}^{\nu\rho\sigma}(x^\mu, \epsilon)$ & Third projection measured by & Eq.~(\ref{eq:U_Projection_111}) \\
 & Eulerian observers \\
\end{tabular}
\end{ruledtabular}
\end{table}

\begin{table}[b]
\caption{\label{tab:Equations}
Major entities appearing in Eqs.~(\ref{eq:EnergyConservation}) and (\ref{eq:MomentumConservation}), the energy-dependent conservative 3+1 general relativistic Variable Eddington Tensor radiation moments equations.}
\begin{ruledtabular}
\begin{tabular}{lll}
$ \mathsf{D}_{T,n}$ & Conserved energy density &  Eq.~(\ref{eq:SpacetimeDivergence_31_n_D}) \\
$\left(\mathsf{D}_{T,\gamma}\right)_j$ & Conserved momentum density & Eq.~(\ref{eq:SpacetimeDivergence_31_gamma_D}) \\
$\left(\mathsf{F}_{T,n}\right)^i$ & Flux of conserved energy & Eq.~(\ref{eq:SpacetimeDivergence_31_n_F}) \\
${\left(\mathsf{F}_{T,\gamma}\right)^i}_j$ & Flux of conserved momentum & Eq.~(\ref{eq:SpacetimeDivergence_31_gamma_F}) \\
$ \mathsf{R}_{T,n} $ & Gravitational shifts, energy eqn. & Eq.~(\ref{eq:MomentumDivergence_31_n_2_R}) \\
$\left(\mathsf{R}_{T,\gamma}\right)_j $ & Gravitational shifts, momentum eqn. & Eq.~(\ref{eq:MomentumDivergence_31_gamma_2_R}) \\
$\mathsf{O}_{T,n}$ & Observer corrections, energy eqn. & Eq.~(\ref{eq:MomentumDivergence_31_n_2_O}) \\
$\left(\mathsf{O}_{T,\gamma} \right)_j$ & Observer corrections, momentum eqn. & Eq.~(\ref{eq:MomentumDivergence_31_gamma_2_O}) \\
$\mathsf{G}_{T,n}$ & Gravitational energy source & Eq.~(\ref{eq:SpacetimeDivergence_31_n_G}) \\
$\left(\mathsf{G}_{T,\gamma}\right)_j$ & Gravitational momentum source & Eq.~(\ref{eq:SpacetimeDivergence_31_gamma_G}) \\
$\mathsf{C}_{T,n}$ & Collision energy source & Eq.~(\ref{eq:Collision_n}) \\
$\left(\mathsf{C}_{T,\gamma}\right)_j$ & Collision momentum source & Eq.~(\ref{eq:Collision_gamma}) \\
\end{tabular}
\end{ruledtabular}
\end{table}

%We have obtained a full elaboration of conservative 3+1 general relativistic variable Eddington tensor radiation transport equations (though leaving aside specific closure prescriptions and contributions to the collision integral, which are stories unto themselves).
We now assemble the expressions obtained in Secs.~\ref{sec:EulerianObservers}-\ref{sec:ThreeVelocity} into conservative 3+1 general relativistic variable Eddington tensor radiation transport equations. 
These four equations are conservation laws for the energy and momentum carried by the neutrino radiation:
\begin{eqnarray}
\frac{\partial \left(\mathsf{D}_{T,n}\right)}{\partial t} + \frac{\partial \left(\mathsf{F}_{T,n}\right)^i }{\partial x^i} 
&+& \frac{1}{\epsilon^2} \frac{\partial}{\partial \epsilon}
\left[ \epsilon^2 \left( \mathsf{R}_{T,n} + \mathsf{O}_{T,n} \right) \right] \nonumber \\
&=& \mathsf{G}_{T,n} + \mathsf{C}_{T,n}, 
\label{eq:EnergyConservation}\\
\frac{\partial \left(\mathsf{D}_{T,\gamma}\right)_j}{\partial t} + \frac{\partial {\left(\mathsf{F}_{T,\gamma}\right)^i}_j }{\partial x^i} 
&+& \frac{1}{\epsilon^2} \frac{\partial}{\partial \epsilon}
\left\{\epsilon^2 \left[ \left(\mathsf{R}_{T,\gamma}\right)_j + \left(\mathsf{O}_{T,\gamma} \right)_j\right] \right\} \nonumber \\
&=& \left(\mathsf{G}_{T,\gamma}\right)_j + \left(\mathsf{C}_{T,\gamma}\right)_j.
\label{eq:MomentumConservation}
\end{eqnarray}
Tables~\ref{tab:Spacetime}-\ref{tab:Equations} present overviews of the many variables that have been assembled into the major entities appearing in these equations.
The global `lab frame' spacetime coordinates $t$ and $x^i$ are those associated with the 3+1 formulation of general relativity, in which the line element and metric components $g_{\mu\nu}$ are given by Eqs.~(\ref{eq:SpacetimeInterval})-(\ref{eq:ProperLength}).
Equations~(\ref{eq:EnergyConservation}) and (\ref{eq:MomentumConservation}), expressing energy and momentum conservation respectively, come from the projections of the phase space divergence of the monochromatic neutrino stress energy orthogonal and tangent to the spacelike slice.
The projection orthogonal to the spacelike slice is via contraction with $n^\mu$, the unit normal to the spacelike slice, which is also the four-velocity of Eulerian observers (see Eqs.~(\ref{eq:UnitNormalU}) and (\ref{eq:UnitNormalD})).
The projection tangent to the spacelike slice is via contraction with the orthogonal projector $\gamma_{\mu\nu} = g_{\mu\nu} + n_\mu n_\nu$.
The momentum space coordinate in these angle-integrated moment equations is the energy $\epsilon = - u_\mu p^\mu$ measured by a Lagrangian observer, whose four-velocity $u^\mu$ is that of the fluid. 
These coordinate choices allow particle/fluid interactions to be evaluated in the comoving frame in the context of Eulerian grid-based approaches to multidimensional spatial dependence.

A rough analogy with conservative formulations of hydrodynamics is evident.
The `conserved' monochromatic energy and momentum densities $\mathsf{D}_{T,n}$ and $\left(\mathsf{D}_{T,\gamma}\right)_j$ are given by Eqs.~(\ref{eq:SpacetimeDivergence_31_n_D}) and (\ref{eq:SpacetimeDivergence_31_gamma_D}).
The energy and momentum fluxes $\left(\mathsf{F}_{T,n}\right)^i$ and ${\left(\mathsf{F}_{T,\gamma}\right)^i}_j$ are given by Eqs.~(\ref{eq:SpacetimeDivergence_31_n_F}) and (\ref{eq:SpacetimeDivergence_31_gamma_F}).
These are expressed in terms of the `Eulerian projections' (see Eq.~(\ref{eq:StressEnergyDecompositionEulerian})) of the monochromatic neutrino stress energy $\mathcal{T}^{\mu\nu}$---the second momentum angular moment of the distribution function $f$ (see Eq.~(\ref{eq:MonochromaticStressEnergy_2})).
These Eulerian projections are the monochromatic energy density $\mathcal{E}$, the momentum density $\mathcal{F}^\mu$, and the stress $\mathcal{S}^{\mu\nu}$ measured by an Eulerian observer (i.e. in the lab frame).
These may be expressed (see Appendix~\ref{app:EulerianLagrangianProjections}) in terms of the `Lagrangian projections' of $\mathcal{T}^{\mu\nu}$ (see Eq.~(\ref{eq:StressEnergyDecompositionComoving})), which are the energy density $\mathcal{J}$, the momentum density $\mathcal{H}^\mu$, and the stress $\mathcal{K}^{\mu\nu}$ measured by a Lagrangian observer (i.e. in the comoving frame).
(Once again, we emphasize that the energy dependence of not only the Lagrangian projections, but also the Eulerian projections, is on the energy $\epsilon$ measured by a Lagrangian observer in the comoving frame.)
The closure relations---that is, the Eddington tensors---are defined in terms of these Lagrangian projections.
Roughly speaking, the Eulerian and Lagrangian projections are respectively like the `conserved' and `primitive' variables in hydrodynamics, while the Eddington tensor closure relation between Lagrangian projections is analogous to the equation of state relating primitive (comoving frame) hydrodynamics variables.
Rather than code tedious analytic expressions like those in Appendix~\ref{app:EulerianLagrangianProjections}, it may be best to obtain the Eulerian projections in terms of the Lagrangian projections by numerically performing the contractions in Eqs.~(\ref{eq:T_Projection_nn_Eulerian})-(\ref{eq:T_Projection_11_Eulerian}), using Eq.~(\ref{eq:StressEnergyDecompositionComoving}) on the right-hand side;
and in Eqs.~(\ref{eq:U_Projection_nnn})-(\ref{eq:U_Projection_111}), using Eq.~(\ref{eq:U_DecompositionComoving}) on the right-hand side.
Similarly, the Lagrangian projections can be obtained numerically in terms of the Eulerian projections using the contractions in Eqs.~(\ref{eq:T_Projection_uu_Lagrangian})-(\ref{eq:T_Projection_11_Lagrangian}) and (\ref{eq:U_Projection_111_Comoving}), using Eqs.~(\ref{eq:StressEnergyDecompositionEulerian}) and (\ref{eq:U_DecompositionEulerian}) on the right-hand sides.

As with self-gravitating hydrodynamics, the presence of gravitational source terms makes Eqs.~(\ref{eq:EnergyConservation}) and (\ref{eq:MomentumConservation}) more properly `balance equations' rather than strict conservation laws.
The terms $\mathsf{G}_{T,n}$ and $\left(\mathsf{G}_{T,\gamma}\right)_j$, given by 
Eqs.~(\ref{eq:SpacetimeDivergence_31_n_G}) and (\ref{eq:SpacetimeDivergence_31_gamma_G}), represent the energy and momentum exchange between the neutrinos and the gravitational field as embodied in the spacetime geometry.
(If spherical or cylindrical spatial coordinates were used, these source terms also would include the fictitious forces associated with these. 
For the purpose of discretizing the spatial flux, it would be desirable in this case to factor out the portion of the spatial metric determinant arising from these coordinate choices, such that it would appear (a) in the denominator outside the spatial derivative, and (b) in the numerator inside the derivative, separate from the flux.)

As monochromatic (i.e. energy-dependent) radiation transport equations, Eqs.~(\ref{eq:EnergyConservation}) and (\ref{eq:MomentumConservation}) contain terms beyond those present in self-gravitating hydrodynamics.
First, the source terms
\begin{eqnarray}
\mathsf{C}_{T,n} &=& - \alpha \sqrt{\gamma} \, n_\mu \,q^\mu, 
\label{eq:Collision_n} \\
\left(\mathsf{C}_{T,\gamma}\right)_j &=& \alpha \sqrt{\gamma} \, \gamma_{j\mu} \,q^\mu
\label{eq:Collision_gamma}
\end{eqnarray}
represent the energy and momentum exchange with the fluid, with the relationship of the source $q^\mu$ to the collision integral given in Eq.~(\ref{eq:MonochromaticMomentumSource}).
Second, the energy divergences on the left-hand sides of Eqs.~(\ref{eq:EnergyConservation}) and (\ref{eq:MomentumConservation}) arise from changes in the neutrino energy as measured in the comoving frame due to gravitational redshift---$\mathsf{R}_{T,n}$ and $\left(\mathsf{R}_{T,\gamma}\right)_j$, see Eqs.~(\ref{eq:MomentumDivergence_31_n_2_R}) and (\ref{eq:MomentumDivergence_31_gamma_2_R})---and the acceleration of the observer riding along with the fluid, $\mathsf{O}_{T,n}$ and $\left(\mathsf{O}_{T,\gamma}\right)_j$; see Eqs.~(\ref{eq:MomentumDivergence_31_n_2_O}) and (\ref{eq:MomentumDivergence_31_gamma_2_O}).
These are given in terms of the Eulerian projections (see Eq.~(\ref{eq:U_DecompositionEulerian})) of the monochromatic third momentum angular moment $\mathcal{U}^{\mu\nu\rho}$ (see Eq.~(\ref{eq:ThirdMoment})).
The Eulerian projections of $\mathcal{U}^{\mu\nu\rho}$ can be expressed in terms of the Eulerian projections of $\mathcal{T}^{\mu\nu}$ (Section~\ref{sec:ThreeVelocity}), or more directly in terms of the Lagrangian projections (Appendix~\ref{app:EulerianLagrangianProjections}).
Finally, we note that the energy divergence is conservative with respect to integration over the differential energy volume $\epsilon^2\, d\epsilon$.

While the conservative variable Eddington tensor moment equations express four-momentum exchange between the neutrinos and the fluid, we also have examined in detail the relationship of these to the conservative number exchange equation (see Sections~\ref{sec:MomentumNumberExchange} and \ref{sec:MomentumNumberExchange_31}).
This can be done in a tractable way on a term-by-term basis thanks to our greater elaboration of the momentum space divergence than in previous work. 
Important conceptual features of our approach include 
(a) consistent use of what we call `Eulerian decompositions' and `Eulerian projections,' which are natural to the 3+1 formulation; and relatedly, (b) a shift from conceptualizing the relationship between the lab and comoving frames from coordinate transformations ${L^\mu}_{\hat{\mu}}$ to the (covariant) relative three-velocity $v^\mu$ connecting the four-velocities $n^\mu$ and $u^\mu$ of Eulerian and Lagrangian observers.
Our approach is geometric, in conception if not notation, to an extent that allows us to obtain explicit results while almost completely avoiding encounters with connection coefficients.
This understanding of the relationship between conservative four-momentum exchange and conservative number exchange can guide the determination of discretizations of the variable Eddington tensor moment equations that facilitate simultaneous energy and lepton number conservation in numerical simulations, yielding greater confidence in simulation outcomes.
We make a few comments on discretization of the moment equations in Appendix~\ref{app:TowardsDiscretization}, and look forward to implementation in simulations, perhaps initially in limits that only partially include relativistic effects \cite{Endeve2012Conservative-Mu}.

\appendix

\section{Spacetime divergence}
\label{app:SpacetimeDivergence}

%\section{Appendix: Spacetime divergence}

In this appendix we compute the spacetime divergence of a stress-energy tensor with the 3+1 metric.
The Eulerian decomposition of a stress-energy tensor is given by Eq.~(\ref{eq:StressEnergy_0}) or (\ref{eq:StressEnergy}).
In particular, we derive contributions to `conservative' evolution equations for the energy density $E$ and momentum density $F_i$ measured by an Eulerian observer (whose four-velocity is the unit normal $n^u$ to a spacelike slice).

Some relations involving derivatives of the unit normal $n^\mu$ and the orthogonal projector $\gamma_{\mu\nu}$ will prove useful.
The gradient of the unit normal is related to the extrinsic curvature and lapse function by \cite{Gourgoulhon200731-Formalism-an}
\begin{equation}
\nabla_\mu n_\nu = -K_{\nu\mu} - \frac{n_\mu}{\alpha} \frac{\partial \alpha}{\partial x^\nu}.
\label{eq:ExtrinsicCurvature}
\end{equation}
Because $n^\nu \nabla_\mu n_\nu = \nabla_\mu ( n^\nu n_\nu ) / 2 = 0$; and because $K_{\mu\nu}$ is tangent to the spacelike slice, i.e. spacelike in the lab frame coordinate basis ($n^\mu K_{\mu\nu} = n^\nu K_{\mu\nu} = 0$), the nonvanishing projections of this equation are
\begin{eqnarray}
n^\mu \nabla_\mu n_\nu &=& \frac{1}{\alpha} \frac{\partial \alpha}{\partial x^\nu},
\label{eq:nGradientTime} \\
{\gamma^\mu}_i {\gamma^\nu}_j \nabla_\mu n_\nu &=& -K_{ij}
\label{eq:nGradientSpace}
\end{eqnarray}
in the lab frame coordinate basis.
Eq.~(\ref{eq:nGradientTime}) relates the four-acceleration of an Eulerian observer to the gradient of the lapse function.
Eq.~(\ref{eq:nGradientSpace}) relates the spatial part of the gradient of the unit normal to the extrinsic curvature, expressing the fact that the direction of the normal varies with the warp of the slice as embedded in spacetime.
Another relation valid in the lab frame coordinate basis for vectors $z^\mu$ tangent to the spacelike slice ($z^0 = 0$) is
\begin{equation}
z_\mu \frac{\partial n^\mu}{\partial x^\nu} = -\frac{z_i}{\alpha} \frac{\partial \beta^i}{\partial x^\nu}. \ \ \  (z^\mu \ \mathrm{spacelike})
\label{eq:nGradientU}
\end{equation}
This follows from writing 
\begin{eqnarray}
z_\mu \frac{\partial n^\mu }{ \partial x^\nu } &=& z_0 \frac{ \partial n^0 }{ \partial x^\nu } + z_i \frac{ \partial n^i }{ \partial x^\nu } \\
&=&  g_{0i} z^i \frac{\partial n^0 }{ \partial x^\nu } + z_i \frac{ \partial n^i }{ \partial x^\nu }
\end{eqnarray}
and using $g_{0i} = \beta_i$ and Eq.~(\ref{eq:UnitNormalU}).
Finally, and more straightforwardly, the gradient of the orthogonal projector is
\begin{equation}
\nabla_\mu \gamma_{\nu\rho} = n_\rho \nabla_\mu n_\nu + n_\nu \nabla_\mu n_\rho,
\label{eq:gammaGradient}
\end{equation}
thanks to the vanishing covariant derivative of the four-metric $g_{\mu\nu}$.
 
The projection of the spacetime divergence of a stress-energy tensor orthogonal to the spacelike slice contributes to an energy equation.
Contracting the divergence with $n_\nu$ and taking it inside the derivative, we have
\begin{equation}
-n_\nu \nabla_\mu T^{\mu\nu} = -\nabla_\mu \left( n_\nu T^{\mu\nu} \right) + T^{\mu\nu} \nabla_\mu n_\nu.
\label{eq:DivergenceOrthogonal}
\end{equation}
The first term on the right-hand side is
\begin{eqnarray}
-\nabla_\mu \left( n_\nu T^{\mu\nu} \right) &=& \nabla_\mu \left( E n^\mu + F^i {\gamma^{\mu}}_i\right) \\
&=&\! \frac{1}{\sqrt{-g}}\frac{\partial}{\partial x^\mu} \!\!\left[\! \sqrt{-g} \!\left(\! E n^\mu \!+\! F^i {\gamma^{\mu}}_i\right)\!\right] \\
&=& \frac{1}{\alpha\sqrt{\gamma}}\frac{\partial}{\partial t} \left( \sqrt{\gamma} E \right) \nonumber \\
&& \!+\! \frac{1}{\alpha\sqrt{\gamma}}\frac{\partial}{\partial x^i}\!\! \left[\! \sqrt{\gamma} \!\left(\! \alpha F^i \!-\! \beta^i E \right)\!\right],
\label{eq:DivergenceOrthogonal_1}
\end{eqnarray} 
where we have used Eqs.~(\ref{eq:MetricDeterminant}), (\ref{eq:UnitNormalU}), and (\ref{eq:StressEnergy}).
In substituting Eq.~(\ref{eq:StressEnergy}) into the second term on the right-hand side of Eq.~(\ref{eq:DivergenceOrthogonal}), the first two terms vanish:
\begin{equation}
\left( E \,n^\mu +  F^i {\gamma^{\mu}}_i \right) n^\nu \nabla_\mu n_\nu = 0,
\end{equation}
because $n^\nu \nabla_\mu n_\nu = \nabla_\mu ( n^\nu n_\nu ) / 2 = 0$.
The remaining terms give
\begin{equation}
T^{\mu\nu} \nabla_\mu n_\nu = \frac{F^i}{\alpha} \frac{\partial \alpha}{\partial x^i} - S^{ij} K_{ij},
\label{eq:DivergenceOrthogonal_2}
\end{equation}
thanks to Eqs.~(\ref{eq:nGradientTime}) and (\ref{eq:nGradientSpace}).
Putting Eqs.~(\ref{eq:DivergenceOrthogonal_1}) and (\ref{eq:DivergenceOrthogonal_2}) together,
\begin{eqnarray}
-n_\nu \nabla_\mu T^{\mu\nu}&=& 
\frac{1}{\alpha\sqrt{\gamma}}\frac{\partial}{\partial t} \left( \sqrt{\gamma} E \right) \nonumber \\
& &+ \frac{1}{\alpha\sqrt{\gamma}}\frac{\partial}{\partial x^i} \left[ \sqrt{\gamma} \left( \alpha F^i - \beta^i E \right)\right] \nonumber \\
& &+\frac{F^i}{\alpha} \frac{\partial \alpha}{\partial x^i} - S^{ij} K_{ij}
\label{eq:DivergenceOrthogonalFinal}
\end{eqnarray}
is the portion of the spacetime divergence orthogonal to the spacelike slice.

The projection of the spacetime divergence tangent to the spacelike slice, which contributes to a momentum equation, is a bit more involved.
Contracting with the orthogonal projector and taking it inside the derivative,
\begin{equation}
\gamma_{j \nu} \nabla_\mu T^{\mu\nu} = \nabla_\mu \left( \gamma_{j \nu} T^{\mu\nu} \right) - T^{\mu\nu} \nabla_\mu \gamma_{j \nu}.
\label{eq:DivergenceTangent}
\end{equation}
The first term on the right-hand side is
\begin{equation}
\nabla_\mu \left( \gamma_{j \nu} T^{\mu\nu} \right) 
= \frac{1}{\sqrt{-g}}\frac{\partial}{\partial x^\mu} \left(\sqrt{-g} \,\gamma_{j \nu} T^{\mu\nu}\right) 
-\Gamma^\rho_{j\mu} \gamma_{\rho \nu} T^{\mu\nu}.
\label{eq:DivergenceTangent_1}
\end{equation}
The first term on the right-hand side is
\begin{eqnarray}
\frac{1}{\sqrt{-g}} \frac{\partial}{\partial x^\mu}\!\!&\!\! \left( \right. \!\!&\!\! \left.\sqrt{-g} \,\gamma_{j \nu} T^{\mu\nu}\right) \nonumber \\
&=& \frac{1}{\sqrt{-g}}\frac{\partial}{\partial x^\mu} \!\!\left[\! \sqrt{-g} \left( F_j n^\mu \!+\! {S^i}_j {\gamma^{\mu}}_i\right)\right] \\
&=& \frac{1}{\alpha\sqrt{\gamma}}\frac{\partial}{\partial t} \left( \sqrt{\gamma} F_j \right) \nonumber\\
& &+ \frac{1}{\alpha\sqrt{\gamma}}\frac{\partial}{\partial x^i} \left[ \sqrt{\gamma} \left( \alpha {S^i}_j - \beta^i F_j \right)\right].
\label{eq:DivergenceTangent_1a}
\end{eqnarray}
The second term on the right-hand side of Eq.~(\ref{eq:DivergenceTangent_1}) is
\begin{equation}
-\Gamma^\rho_{j\mu} \gamma_{\rho \nu} T^{\mu\nu} 
= -\Gamma^\rho_{j\mu} \left( F_\rho n^\mu + {S^\mu}_\rho \right).
\label{eq:DivergenceTangent_1b}
\end{equation}
The first term on the right-hand side is
\begin{eqnarray}
-\Gamma^\rho_{j\mu} F_\rho n^\mu  
&=& F_\rho \left( \partial_j n^\rho - \nabla_j n^\rho \right) \\
&=& -\frac{F_i}{\alpha} \,\frac{\partial\beta^i}{\partial x^j} + F^i K_{ji},
\label{eq:DivergenceTangent_1ba}
\end{eqnarray}
where the first and second terms follow from Eqs.~(\ref{eq:nGradientU}) and (\ref{eq:nGradientSpace}) respectively.
The second term on the right-hand side of Eq.~(\ref{eq:DivergenceTangent_1b}) is
\begin{eqnarray}
-\Gamma^\rho_{j\mu} {S^\mu}_\rho &=& -\Gamma^\rho_{j i}\, g_{\rho k}\, S^{i k} \\
&=& -\frac{S^{i k}}{2} \left( \frac{\partial \gamma_{ki}}{\partial x^j} + \frac{\partial \gamma_{jk}}{\partial x^i} - \frac{\partial \gamma_{ji}}{\partial x^k}\right).
\end{eqnarray}
The last two terms in parentheses are antisymmetric in $i,k$ and vanish upon contraction with the symmetric $S^{i k}$, leaving
\begin{equation}
-\Gamma^\rho_{j\mu} {S^\mu}_\rho = -\frac{S^{i k}}{2} \frac{\partial \gamma_{ik}}{\partial x^j}.
\label{eq:DivergenceTangent_1bb}
\end{equation} 
With Eqs.~(\ref{eq:DivergenceTangent_1ba}) and (\ref{eq:DivergenceTangent_1bb}), Eq.~(\ref{eq:DivergenceTangent_1b}) becomes
\begin{equation}
-\Gamma^\rho_{j\mu} \gamma_{\rho \nu} T^{\mu\nu} = -\frac{F_i}{\alpha} \,\frac{\partial\beta^i}{\partial x^j} + F^i K_{ji} -\frac{S^{i k}}{2} \frac{\partial \gamma_{ik}}{\partial x^j}.
\end{equation}
This, together with Eq.~(\ref{eq:DivergenceTangent_1a}), yields
\begin{eqnarray}
\nabla_\mu \left( \gamma_{j \nu} T^{\mu\nu} \right) 
&=& \frac{1}{\alpha\sqrt{\gamma}}\frac{\partial}{\partial t} \left( \sqrt{\gamma} F_j \right) \nonumber \\
& &+ \frac{1}{\alpha\sqrt{\gamma}}\frac{\partial}{\partial x^i} \left[ \sqrt{\gamma} \left( \alpha {S^i}_j - \beta^i F_j \right)\right] \nonumber \\
& & -\frac{F_i}{\alpha} \,\frac{\partial\beta^i}{\partial x^j} + F^i K_{ji} -\frac{S^{i k}}{2} \frac{\partial \gamma_{ik}}{\partial x^j}
\label{eq:DivergenceTangent_1_Final}
\end{eqnarray}
for Eq.~(\ref{eq:DivergenceTangent_1}), the first term of Eq.~(\ref{eq:DivergenceTangent}).
The second term of Eq.~(\ref{eq:DivergenceTangent}) is less complicated.
Using Eq.~(\ref{eq:gammaGradient}),
\begin{eqnarray}
- T^{\mu\nu} \nabla_\mu \gamma_{j \nu} &=& - T^{\mu\nu} \left(n_\nu \nabla_\mu n_j + n_j \nabla_\mu n_\nu \right) \\
&=& - T^{\mu\nu} n_\nu \nabla_\mu n_j, 
\end{eqnarray}
in which the second term has vanished due to Eq.~(\ref{eq:UnitNormalD}).
This can be rewritten
\begin{eqnarray}
- T^{\mu\nu} \nabla_\mu \gamma_{j \nu} &=& \left( E n^\mu + F^i {\gamma^{\mu}}_i\right) \nabla_\mu n_j \\
&=& \frac{E}{\alpha}\frac{\partial\alpha}{\partial x^j} - F^i K_{ij},
\label{eq:DivergenceTangent_2_Final}
\end{eqnarray}
where we successively have used Eq.~(\ref{eq:StressEnergy}) and Eqs.~(\ref{eq:nGradientTime}) and (\ref{eq:nGradientSpace}).
Adding Eqs.~(\ref{eq:DivergenceTangent_1_Final}) and (\ref{eq:DivergenceTangent_2_Final}) and allowing for the symmetry of $K_{ij}$, we have
\begin{eqnarray}
\gamma_{j \nu} \nabla_\mu T^{\mu\nu}
&=& \frac{1}{\alpha\sqrt{\gamma}}\frac{\partial}{\partial t} \left( \sqrt{\gamma} F_j \right) \nonumber \\
& &+ \frac{1}{\alpha\sqrt{\gamma}}\frac{\partial}{\partial x^i} \left[ \sqrt{\gamma} \left( \alpha {S^i}_j - \beta^i F_j \right)\right] \nonumber \\
& & +\frac{E}{\alpha}\frac{\partial\alpha}{\partial x^j} -\frac{F_i}{\alpha} \,\frac{\partial\beta^i}{\partial x^j} -\frac{S^{i k}}{2} \frac{\partial \gamma_{ik}}{\partial x^j}
\label{eq:DivergenceTangentFinal}
\end{eqnarray}
for the projection of the spacetime divergence tangent to the spacelike slice.

\section{Eulerian and Lagrangian projections}
\label{app:EulerianLagrangianProjections}

In this appendix we express the Eulerian projections $\mathcal{E}$, $\mathcal{F}^\mu$, and $\mathcal{S}^{\mu\nu}$ of $\mathcal{T}^{\mu\nu}$ (see Eqs.~(\ref{eq:MonochromaticStressEnergy_2}) and  (\ref{eq:StressEnergyDecompositionEulerian})), and $\mathcal{Z}$, $\mathcal{Y}^\mu$, $\mathcal{X}^{\mu\nu}$, and $\mathcal{W}^{\mu\nu\rho}$ of $\mathcal{U}^{\mu\nu\rho}$ (see Eqs.~(\ref{eq:ThirdMoment}) and  (\ref{eq:U_DecompositionEulerian})), in terms the angular moments $\mathcal{J}$, $\mathcal{H}^\mu$, $\mathcal{K}^{\mu\nu}$, and $\mathcal{L}^{\mu\nu\rho}$ (see Eqs.~(\ref{eq:AngularMoment_0_Covariant})-(\ref{eq:AngularMoment_3_Covariant})), which are also the Lagrangian projections of $\mathcal{T}^{\mu\nu}$ and---up to a factor $\epsilon$---of $\mathcal{U}^{\mu\nu\rho}$ (see Eqs.~(\ref{eq:StressEnergyDecompositionComoving}) and (\ref{eq:U_DecompositionComoving})). 
Because the Lagrangian projections are spacelike in the comoving frame, we eliminate $u^\mu$ in favor of $v^\mu$ and (in free indices) $n^\mu$.
In the case of the neutrino energy density, flux, and stress, we also show how the relations are consistent with published results in special relativity obtained with Lorentz transformations \cite{Munier1986Radiation-trans}, rather than with projections as done here.

We use four basic types of contractions, which follow from Eq.~(\ref{eq:ThreeVelocity}): 
\begin{eqnarray}
n_\mu u^\mu &=& -\Lambda, \label{eq:Contraction_1A} \\
n_\mu Z^\mu &=& - v_\mu Z^\mu, \ \ \ (u_\mu Z^\mu = 0) \\
\gamma_{\rho\mu} u^\mu &=& \Lambda v_\rho, \\
\gamma_{\rho\mu} Z^\mu &=& \left(g_{\rho\mu} - n_\rho v_\mu \right) Z^\mu, \ \ \ (u_\mu Z^\mu = 0) 
\label{eq:Contraction_4A}
\end{eqnarray}
where $Z^\mu$ is a stand-in for any spacelike index relative to the Lagrangian observer four-velocity $u^\mu$ (i.e. $u_\mu Z^\mu = 0$).

The Eulerian projections of $\mathcal{T}^{\mu\nu}$ are
\begin{eqnarray}
\mathcal{E} &=& n_\mu n_\nu \mathcal{T}^{\mu\nu}, \\
\mathcal{F}_\rho &=& -\gamma_{\rho\mu} n_\nu \mathcal{T}^{\mu\nu}, \\
\mathcal{S}_{\rho\sigma} &=& \gamma_{\rho\mu} \gamma_{\sigma\nu} \mathcal{T}^{\mu\nu}.
\end{eqnarray}
Using Eqs.~(\ref{eq:StressEnergyDecompositionComoving}) and (\ref{eq:Contraction_1A})-(\ref{eq:Contraction_4A}) on the right-hand sides,
we find
\begin{eqnarray}
\mathcal{E} &=& \Lambda^2 \mathcal{J} + 2\Lambda v_\mu \mathcal{H}^\mu + v_\mu v_\nu \mathcal{K}^{\mu\nu}, 
\label{eq:EulerianLagrangian_E}\\
\mathcal{F}_\rho &=& \Lambda^2 v_\rho \mathcal{J} + \Lambda \left( g_{\rho\mu} - n_\rho v_\mu \right) \mathcal{H}^\mu \nonumber \\
& & +  \Lambda v_\rho v_\mu \mathcal{H}^\mu + \left( g_{\rho\mu} - n_\rho v_\mu \right) v_\nu \mathcal{K}^{\mu\nu}, 
\label{eq:EulerianLagrangian_F}\\
\mathcal{S}_{\rho\sigma} &=& \Lambda^2 v_\rho v_\sigma \mathcal{J} 
+ \Lambda \left( g_{\rho\mu} - n_\rho v_\mu \right) v_\sigma \mathcal{H}^\mu \nonumber \\
& & + \Lambda \left( g_{\sigma\mu} - n_\sigma v_\mu \right) v_\rho \mathcal{H}^\mu \nonumber \\
& &+ \left(g_{\rho\mu} - n_\rho v_\mu \right) \left(g_{\sigma\nu} - n_\sigma v_\nu \right) \mathcal{K}^{\mu\nu}.
\label{eq:EulerianLagrangian_S}
\end{eqnarray}

The Eulerian projections of $\mathcal{U}^{\mu\nu\rho}$ are
\begin{eqnarray}
\mathcal{Z} &=& -n_\mu n_\nu n_\rho\, \mathcal{U}^{\mu\nu\rho}, \\
\mathcal{Y}_\sigma &=& \gamma_{\sigma\mu} n_\nu n_\rho\, \mathcal{U}^{\mu\nu\rho}, \\
\mathcal{X}_{\sigma\kappa} &=& -\gamma_{\sigma\mu} \gamma_{\kappa\nu} n_\rho \,\mathcal{U}^{\mu\nu\rho}, \\
\mathcal{W}_{\sigma\kappa\lambda} &=& \gamma_{\sigma\mu} \gamma_{\kappa\nu} \gamma_{\lambda\rho} \mathcal{U}^{\mu\nu\rho}.
\end{eqnarray}
Using Eqs.~(\ref{eq:U_DecompositionComoving}) and (\ref{eq:Contraction_1A})-(\ref{eq:Contraction_4A}) on the right-hand sides,
we find
\begin{eqnarray}
\mathcal{Z} &=& \epsilon\left(\Lambda^3 \mathcal{J} + 3\Lambda^2 v_\mu \mathcal{H}^{\mu} + 3\Lambda v_\mu v_\nu \mathcal{K}^{\mu\nu} \right. \nonumber \\
& &\left. + v_\mu v_\nu v_\rho\mathcal{L}^{\mu\nu\rho}, \right) \\
\mathcal{Y}_\sigma &=& \epsilon \left[ \Lambda^3 v_\sigma \mathcal{J} 
+ \Lambda^2 \left(g_{\sigma\mu} - n_\sigma v_\mu \right) \mathcal{H}^\mu + 2 \Lambda^2 v_\sigma v_\mu \mathcal{H}^\mu \right. \nonumber \\
& & \left. + 2 \Lambda \left( g_{\sigma\mu} - n_\sigma v_\mu \right) v_\nu \mathcal{K}^{\mu\nu} + \Lambda v_\sigma v_\mu v_\nu \mathcal{K}^{\mu\nu} \right. \nonumber \\
& & \left. + \left( g_{\sigma\mu} - n_\sigma v_\mu \right) v_\nu v_\rho \mathcal{L}^{\mu\nu\rho}
 \right], \\
\mathcal{X}_{\sigma\kappa} &=& \epsilon \left[ \Lambda^3 v_\sigma v_\kappa \mathcal{J} + \Lambda^2 \left(g_{\sigma\mu} - n_\sigma v_\mu \right) v_\kappa \mathcal{H}^\mu \right. \nonumber \\
& &\left. + \Lambda^2 \left( g_{\kappa\mu} - n_\kappa v_\mu \right) v_\sigma \mathcal{H}^\mu + \Lambda^2 v_\sigma v_\kappa v_\mu \mathcal{H}^\mu \right. \nonumber \\
& & \left. + \Lambda \left(g_{\sigma\mu} - n_\sigma v_\mu \right) \left(g_{\kappa\nu} - n_\kappa v_\nu \right) \mathcal{K}^{\mu\nu} \right. \nonumber \\
& & \left. + \Lambda \left(g_{\sigma\mu} - n_\sigma v_\mu \right) v_\nu \mathcal{K}^{\mu\nu} + \Lambda \left(g_{\kappa\mu} - n_\kappa v_\mu \right) v_\nu \mathcal{K}^{\mu\nu} \right. \nonumber \\
& & \left. + \left(g_{\sigma\mu} - n_\sigma v_\mu \right) \left(g_{\kappa\nu} - n_\kappa v_\nu \right) v_\rho \mathcal{L}^{\mu\nu\rho}
\right], \\
\mathcal{W}_{\sigma\kappa\lambda} &=& \epsilon \left[ \Lambda^3 v_\sigma v_\kappa v_\lambda \mathcal{J}
+ \Lambda^2 \left( g_{\sigma\mu} - n_\sigma v_\mu \right) v_\kappa v_\lambda \mathcal{H}^\mu \right. \nonumber \\
& & \left. + \Lambda^2 \left( g_{\kappa\mu} - n_\kappa v_\mu \right) v_\lambda v_\sigma \mathcal{H}^\mu \right. \nonumber \\
& & \left. + \Lambda^2 \left( g_{\lambda\mu} - n_\lambda v_\mu \right) v_\sigma v_\kappa \mathcal{H}^\mu \right. \nonumber \\
& & \left. + \Lambda \left( g_{\sigma\mu} - n_\sigma v_\mu \right) \left( g_{\kappa\nu} - n_\kappa v_\nu \right) v_\lambda \mathcal{K}^{\mu\nu} \right. \nonumber \\
& & \left. + \Lambda \left( g_{\kappa\mu} - n_\kappa v_\mu \right) \left( g_{\lambda\nu} - n_\lambda v_\nu \right) v_\sigma \mathcal{K}^{\mu\nu} \right. \nonumber \\
& & \left. + \Lambda \left( g_{\lambda\mu} - n_\lambda v_\mu \right) \left( g_{\sigma\nu} - n_\sigma v_\nu \right) v_\kappa \mathcal{K}^{\mu\nu} \right. \nonumber \\
& & \left. + \left( g_{\sigma\mu} - n_\sigma v_\mu \right) \left( g_{\kappa\nu} - n_\kappa v_\nu \right) \right. \nonumber \\
& & \left. \ \ \ \times \left( g_{\lambda\rho} - n_\lambda v_\rho \right) \mathcal{L}^{\mu\nu\rho} 
\right].
\end{eqnarray}

We now prepare to compare the (raised index) spatial components of Eqs.~(\ref{eq:EulerianLagrangian_E})-(\ref{eq:EulerianLagrangian_S}) with the special relativistic results in Eqs.~(182)-(184) of Ref.~\cite{Munier1986Radiation-trans}.
In flat spacetime the unit normal of Eq.~(\ref{eq:UnitNormalU}) becomes
\begin{equation}
\left(n^\mu\right) = \left( 1, 0, 0, 0 \right)^T.
\label{eq:UnitNormalU_Flat}
\end{equation} 
The tetrad ${e^\mu}_{\bar{\mu}}$ is a Kronecker delta in flat spacetime Cartesian coordinates, so that the composite transformation ${L^\mu}_{\hat{\mu}}$ of Eq.~(\ref{eq:CompositeTransformation}) becomes
\begin{equation}
{L^\mu}_{\hat{\mu}} = {\delta^\mu}_{\bar{\mu}} {\Lambda^{\bar{\mu}}}_{\hat{\mu}}.
\label{eq:CompositeFlat}
\end{equation}
An explicit expression for the Lorentz boost between the orthonormal lab frame and the comoving frame is
\begin{equation}
\begin{pmatrix}
{\Lambda^{\bar{0}}}_{\hat{0}} & {\Lambda^{\bar{0}}}_{\hat{\imath}} \\
{\Lambda^{\bar{\imath}}}_{\hat{0}} & {\Lambda^{\bar{\imath}}}_{\hat{\imath}}
\end{pmatrix}
=
\begin{pmatrix}
\Lambda & \Lambda V_{\hat{\imath}} \\
\Lambda V^{\bar{\imath}} & {\delta^{\bar{\imath}}}_{\hat{\imath}} + \frac{(\Lambda - 1 )}{V^2} V^{\bar{\imath}} V_{\hat{\imath}} 
\end{pmatrix},
\label{eq:BoostExplicit}
\end{equation}
where
\begin{equation}
V^2 = V^{\bar{1}} V_{\hat{1}} + V^{\bar{2}} V_{\hat{2}} + V^{\bar{3}} V_{\hat{3}}.
\end{equation}
It can be shown that
\begin{equation}
\Lambda = \left( 1 - V^2 \right)^{-1/2} = \left( 1 - v^\mu v_\mu \right)^{-1/2} = \left( 1 - v^i v_i \right)^{-1/2}
\end{equation} 
is equal to the Lorentz boost we have been using.
The quantities 
\begin{equation}
V^{\bar{1}} = V_{\hat{1}}, \ V^{\bar{2}} = V_{\hat{2}}, \ V^{\bar{3}} = V_{\hat{3}} \ 
\end{equation} 
are not to be regarded as components of a four-vector, but simply as the three-velocity parameters appearing in the Lorentz boost, expressed in a manner consistent with our index conventions.
In flat spacetime the spatial components of the (covariant) three-velocity four-vector   $v^\mu$ are related to these boost velocity parameters by
\begin{equation}
v^i = {\delta^{i}}_{\bar{\imath}} V^{\bar{\imath}},
\label{eq:ThreeVelocity_U_L}
\end{equation}
a perhaps expected result that follows from Eqs.~(\ref{eq:ThreeVelocity}), (\ref{eq:UnitNormalU_Flat}), (\ref{eq:CompositeFlat}), (\ref{eq:BoostExplicit}), and (\ref{eq:FluidVelocityU}).
However, we caution that the perhaps less-expected result for the lowered-index comoving frame spatial components of $v^\mu$ is
\begin{equation}
v_{\hat{\imath}} = \Lambda V_{\hat{\imath}}.
\end{equation}
Below we use this to evaluate contractions of the form
\begin{eqnarray}
v_\mu Z^\mu &=& v_{\hat{\imath}} Z^{\hat{\imath}} \\
&=& \Lambda V_{\hat{\imath}} Z^{\hat{\imath}}, \ \ \  (u_\mu Z^\mu = 0)
\label{eq:Contraction_v_Z}
\end{eqnarray}
i.e. contraction with an index that is spacelike relative to the Lagrangian observer four-velocity $u^\mu$.

We are now ready to compare the (raised index) spatial components of Eqs.~(\ref{eq:EulerianLagrangian_E})-(\ref{eq:EulerianLagrangian_S}) with the special relativistic results in Eqs.~(182)-(184) of Ref.~\cite{Munier1986Radiation-trans}, using Eqs.~(\ref{eq:UnitNormalU_Flat}), (\ref{eq:CompositeFlat}), (\ref{eq:BoostExplicit}), and (\ref{eq:Contraction_v_Z}). 
Equation~(\ref{eq:EulerianLagrangian_E}) can be expressed as
\begin{equation} 
\mathcal{E} = \Lambda^2 \mathcal{J} + 2\Lambda v_{\hat{\imath}} \mathcal{H}^{\hat{\imath}} + v_{\hat{\imath}} v_{\hat{\jmath}} \mathcal{K}^{{\hat{\imath}}{\hat{\jmath}}},
\end{equation}
or
\begin{equation} 
\mathcal{E} = \Lambda^2 \mathcal{J} + 2\Lambda^2 V_{\hat{\imath}} \mathcal{H}^{\hat{\imath}} + \Lambda^2 V_{\hat{\imath}} V_{\hat{\jmath}} \mathcal{K}^{{\hat{\imath}}{\hat{\jmath}}},
\end{equation}
which agrees with Eq.~(182) of Ref.~\cite{Munier1986Radiation-trans}.
The spatial components of Eq.~(\ref{eq:EulerianLagrangian_F}) are 
\begin{equation}
\mathcal{F}^i = \Lambda^2 v^i \mathcal{J} + \Lambda {L^i}_{\hat{\imath}}  \mathcal{H}^{\hat{\imath}} +  \Lambda v^i v_{\hat{\imath}} \mathcal{H}^{\hat{\imath}} + {L^i}_{\hat{\imath}}  v_{\hat{\jmath}} \mathcal{K}^{{\hat{\imath}}{\hat{\jmath}}}, 
\end{equation}
or
\begin{eqnarray}
\mathcal{F}^i &=&  {\delta^i}_{\bar{\imath}} \left\{\Lambda^2 V^{\bar{\imath}} \mathcal{J} + \Lambda \left[{\delta^{\bar{\imath}}}_{\hat{\imath}} + \frac{(\Lambda - 1 )}{V^2} V^{\bar{\imath}} V_{\hat{\imath}}  \right] \mathcal{H}^{\hat{\imath}} \right. \nonumber \\
& & \left. +  \Lambda^2 V^{\bar{\imath}} V_{\hat{\imath}} \mathcal{H}^{\hat{\imath}} + \Lambda \left[ {\delta^{\bar{\imath}}}_{\hat{\imath}} + \frac{(\Lambda - 1 )}{V^2} V^{\bar{\imath}} V_{\hat{\imath}}  \right]  V_{\hat{\jmath}} \mathcal{K}^{{\hat{\imath}}{\hat{\jmath}}} \right\}, \nonumber \\
& &
\end{eqnarray}
which agrees with Eq.~(183) of Ref.~\cite{Munier1986Radiation-trans}.
The spatial components of Eq.~(\ref{eq:EulerianLagrangian_F}) are 
\begin{equation}
\mathcal{S}^{ij} = \Lambda^2 v^i v^j \mathcal{J} 
+ \Lambda {L^i}_{\hat{\imath}} v^j \mathcal{H}^{\hat{\imath}} 
 + \Lambda {L^j}_{\hat{\jmath}} v^i \mathcal{H}^{\hat{\jmath}} 
 + {L^i}_{\hat{\imath}}  {L^j}_{\hat{\jmath}} \mathcal{K}^{\hat{\imath}\hat{\jmath}},
\end{equation}
or
\begin{eqnarray}
\mathcal{S}^{ij} &=& {\delta^i}_{\bar{\imath}}{\delta^j}_{\bar{\jmath}} \left\{ \vphantom{\frac{(\Lambda - 1 )}{V^2} } \Lambda^2 V^{\bar{\imath}} V^{\bar{\jmath}} \mathcal{J} \right. 
+ \Lambda \left({\delta^{\bar{\imath}}}_{\hat{\imath}} V^{\bar{\jmath}} \mathcal{H}^{\hat{\imath}} + {\delta^{\bar{\jmath}}}_{\hat{\jmath}} V^{\bar{\imath}}  \mathcal{H}^{\hat{\jmath}} \right) \nonumber \\
& & \left.
+ 2\Lambda \left[ \frac{(\Lambda - 1 )}{V^2} V^{\bar{\imath}} V_{\hat{\imath}}  \right] V^{\bar{\jmath}} \mathcal{H}^{\hat{\imath}}  \right. \nonumber \\
 && \left.
 + \, \left[{\delta^{\bar{\imath}}}_{\hat{\imath}} + \frac{(\Lambda - 1 )}{V^2} V^{\bar{\imath}} V_{\hat{\imath}}  \right] 
 \!\! \left[{\delta^{\bar{\jmath}}}_{\hat{\jmath}} + \frac{(\Lambda - 1 )}{V^2} V^{\bar{\jmath}} V_{\hat{\jmath}}  \right] \! \mathcal{K}^{\hat{\imath}\hat{\jmath}} \right\}, \nonumber \\
 & &
\end{eqnarray}
which agrees with Eq.~(184) of Ref.~\cite{Munier1986Radiation-trans}.

%\section{Appendix: Momentum space divergence}
\section{Momentum space divergence}
\label{app:MomentumSpaceDivergence}

In this appendix we compute, with the 3+1 metric, expressions entering the momentum space divergence term of the phase space divergence of a monochromatic stress-energy tensor.
In particular, we derive contributions to `conservative' evolution equations for the monochromatic energy density $\mathcal{E}$ and momentum density $\mathcal{F}_i$ measured by an Eulerian observer (whose four-velocity is the unit normal $n^\mu$ to a spacelike slice).

We project the momentum space divergence into portions orthogonal and tangent to the spacelike slice, and find it helpful to also use Eulerian decompositions of the tensors that appear.
Our starting points are Eqs.~(\ref{eq:MomentumDivergence_31_n}) and (\ref{eq:MomentumDivergence_31_gamma}) for the portions of the momentum space divergence orthogonal and tangent to the spacelike slice:
\begin{eqnarray}
-n_\rho \left({\mathsf{M}_T}\right)^\rho &=& \frac{1}{\epsilon^2} \frac{\partial}{\partial \epsilon}
\left( \epsilon^2 \, n_\rho \, \mathcal{U}^{\rho\nu\mu} \nabla_{\mu} u_{\nu} \right), 
  \label{eq:MomentumDivergence_31_n_A} \\
\gamma_{j\rho} \left({\mathsf{M}_T}\right)^\rho &=& \frac{1}{\epsilon^2} \frac{\partial}{\partial \epsilon}
\left( - \epsilon^2 \, \gamma_{j\rho} \, \mathcal{U}^{\rho\nu\mu} \nabla_{\mu} u_{\nu} \right). \label{eq:MomentumDivergence_31_gamma_A}
\end{eqnarray}
The Eulerian decomposition of $\mathcal{U}^{\rho\nu\mu}$ is given by Eq.~(\ref{eq:U_DecompositionEulerian}). 
The projections orthogonal and tangent to the spacelike slice, which appear in the above two equations, are
\begin{eqnarray}
- n_\rho \, \mathcal{U}^{\rho\nu\mu} &=& \mathcal{Z}\, n^\nu n^\mu + \mathcal{Y}^\nu n^\mu + \mathcal{Y}^\mu n^\nu + \mathcal{X}^{\nu\mu}, \label{eq:U_Projection_n}\\
\gamma_{j\rho} \, \mathcal{U}^{\rho\nu\mu} &=& \mathcal{Y}_j n^\nu n^\mu + {\mathcal{X}_j}^\nu n^\mu + {\mathcal{X}_j}^\mu n^\nu + {\mathcal{W}_j}^{\nu\mu}.
\label{eq:U_Projection_gamma}
\end{eqnarray}
We also use Eq.~(\ref{eq:ThreeVelocity}),
\begin{equation}
u^\mu = \Lambda \left(n^\mu + v^\mu\right),
\end{equation}
i.e. the Eulerian decomposition of $u_\nu$, in expressing $\nabla_\mu u_\nu$.

There are four types of contractions appearing when either Eq.~(\ref{eq:U_Projection_n}) or (\ref{eq:U_Projection_gamma}) is contracted with $\nabla_\mu u_\nu$; we consider them in turn, beginning with
\begin{equation}
n^\nu n^\mu \nabla_\mu u_\nu 
= n^\nu n^\mu \left[ \left(n_\nu + v_\nu \right) \partial_\mu\Lambda + \Lambda \,\nabla_\mu n_\nu  + \Lambda\, \nabla_\mu v_\nu \right]. 
\end{equation}
The first term gives
\begin{equation}
n^\nu n^\mu \left(n_\nu + v_\nu \right) \partial_\mu\Lambda = -n^\mu \partial_\mu\Lambda
\end{equation}
because of the orthogonality relation $n^\mu v_\mu = 0$.
The second term vanishes because $n^\mu n_\mu = -1 = \mathrm{constant}$ implies
\begin{equation}
0 = \nabla_\mu \left( n^\nu n_\nu \right) = 2 n^\nu \nabla_\mu n_\nu.
\label{eq:d_nSquared}
\end{equation}
In the third term we use
\begin{equation}
n^\nu \nabla_\mu v_\nu = - v^\nu \nabla_\mu n_\nu, 
\label{eq:d_n_v}
\end{equation}
which follows from $0 = \nabla_\mu \left( n^\nu v_\nu \right)$.
The third term becomes
\begin{equation}
n^\nu n^\mu \left( \Lambda\, \nabla_\mu v_\nu \right) = -\frac{\Lambda\, v^i}{\alpha} \frac{\partial\alpha}{\partial x^i}, 
\end{equation}
where we have used Eq.~(\ref{eq:nGradientTime}).
All together,
\begin{equation}
n^\nu n^\mu \nabla_\mu u_\nu = -\frac{\Lambda\, v^i}{\alpha} \frac{\partial\alpha}{\partial x^i} -n^\mu \frac{\partial\Lambda}{\partial x^\mu}  \label{eq:Contraction_1}
\end{equation}
for this type of contraction.

Consider next a contraction of the form
\begin{equation}
A^\nu n^\mu \nabla_\mu u_\nu 
= A^\nu n^\mu \left[ n_\nu \, \partial_\mu\Lambda + \Lambda \,\nabla_\mu n_\nu  + \nabla_\mu \left( \Lambda v_\nu\right) \right],
\end{equation}
where $A^\nu$ is spacelike (i.e. $n_\nu A^\nu=0$, such that $A^0 = 0$ in the coordinate basis), and we break up the right-hand side differently than in the previous paragraph.
The first term vanishes because $A^\nu$ is spacelike.
The second term gives 
\begin{equation}
A^\nu n^\mu \left( \Lambda \,\nabla_\mu n_\nu \right) = \frac{\Lambda\, A^i}{\alpha} \frac{\partial\alpha}{\partial x^i}
\end{equation}
thanks again to Eq.~(\ref{eq:nGradientTime}).
In the third term it will turn out best to raise the index on $v_\nu$ before turning the gradient back, in a sense, on $n^\mu$:
\begin{eqnarray}
A^\nu n^\mu \left[ \nabla_\mu \left( \Lambda v_\nu\right) \right] &=& A_\nu n^\mu \left[ \nabla_\mu \left( \Lambda v^\nu\right) \right] \\
&=& A_\nu n^\mu \!\left[ \partial_\mu \left( \Lambda v^\nu\right)\! +\! \Gamma^\nu_{\rho\mu} \Lambda  v^\rho \right] \\
&=& A_k n^\mu \partial_\mu \left( \Lambda v^k\right)  \nonumber \\
& &+ \Lambda  A_\nu v^m \left( \Gamma^\nu_{\mu m} n^\mu \right) \\
&=& A_k n^\mu \partial_\mu \left( \Lambda v^k\right)  \nonumber \\
& &+ \Lambda  A_\nu v^m \left( \nabla_m n^\nu - \partial_m n^\nu \right) \\
&=& A_k n^\mu \partial_\mu \left( \Lambda v^k\right)  - \Lambda  A^k v^m  K_{mk} \nonumber \\
& &+ \Lambda  A_k \alpha^{-1} v^m\partial_m \beta^k,
\end{eqnarray}
where we have used Eqs.~(\ref{eq:nGradientSpace}) and (\ref{eq:nGradientU}) in the last step.
All together, 
\begin{eqnarray}
A^\nu n^\mu \nabla_\mu u_\nu &=& \frac{\Lambda A^i}{\alpha} \frac{\partial\alpha}{\partial x^i} 
+ \frac{\Lambda  A_k v^i}{\alpha} \frac{\partial \beta^k}{\partial x^i}
- \Lambda  A^k v^i  K_{ik} \nonumber\\
& &+ A_k n^\mu \frac{\partial \left( \Lambda v^k\right) }{\partial x^\mu}  
\label{eq:Contraction_2}
\end{eqnarray}
for this type of contraction.

For the next contraction we return to the first split we used on the right-hand side:
\begin{equation}
A^\mu n^\nu \nabla_\mu u_\nu =  \left[ \left(n_\nu + v_\nu \right) \partial_\mu\Lambda + \Lambda \,\nabla_\mu n_\nu  + \Lambda\, \nabla_\mu v_\nu \right],
\end{equation}
where again $A^\mu$ is spacelike.
The first term gives simply
\begin{equation}
A^\mu n^\nu \left[ \left(n_\nu + v_\nu \right) \partial_\mu\Lambda \right] 
= - A^i \frac{\partial\Lambda}{\partial x^i}.
\end{equation}
From Eq.~(\ref{eq:d_nSquared}), the second term vanishes.
In the third term we use Eqs.~(\ref{eq:d_n_v}) and (\ref{eq:nGradientSpace}), whence
\begin{eqnarray}
A^\mu n^\nu  \left( \Lambda\, \nabla_\mu v_\nu \right) &=& -\Lambda A^\mu v^\nu  \nabla_\mu n_\nu \\
&=& \Lambda A^i v^k K_{ki}.
\end{eqnarray}
Altogether, we have
\begin{equation}
A^\mu n^\nu \nabla_\mu u_\nu = \Lambda A^i v^k K_{ki} - A^i \frac{\partial\Lambda}{\partial x^i}
\label{eq:Contraction_3}
\end{equation}
for this type of contraction.

For the final contraction, one with a spacelike tensor $B^{\nu\mu}$, we return to the alternative
\begin{equation}
B^{\nu\mu} \nabla_\mu u_\nu 
= B^{\nu \mu} \left[ n_\nu \, \partial_\mu\Lambda + \Lambda \,\nabla_\mu n_\nu  + \nabla_\mu \left( \Lambda v_\nu\right) \right]
\end{equation}
on the right-hand side.
The first term vanishes because $B^{\nu\mu}$ is spacelike.
The second term immediately yields, via Eq.~(\ref{eq:nGradientSpace}),
\begin{equation}
B^{\nu\mu} \left( \Lambda \,\nabla_\mu n_\nu \right)
= \Lambda B^{k i} \nabla_i n_k = - \Lambda B^{k i} K_{k i}.
\end{equation} 
In the third term it once again will turn out best to raise the index on $v_\nu$.
We find
\begin{eqnarray}
B^{\nu \mu} \left[ \nabla_\mu \left( \Lambda v_\nu\right) \right]
&=& {B_\nu}^\mu \nabla_\mu \left( \Lambda v^\nu\right) \\
&=&  {B_\nu}^\mu \left[ \partial_\mu \left( \Lambda v^\nu\right) + \Gamma^\nu_{\sigma\mu} \left( \Lambda v^\sigma \right)
\right] \\
&=&   {B_k}^i \partial_i \left( \Lambda v^k\right) \nonumber \\
& &+ \Lambda B^{k i}  v^m g_{k \nu} \Gamma^\nu_{m i} \\
&=&   {B_k}^i \partial_i \left( \Lambda v^k\right) \nonumber\\
& &+ \frac{\Lambda B^{k i}  v^m}{2} \left( \partial_i \gamma_{km} \right. \nonumber\\ 
& & \left.\ \ \ \ \ \ \ \ \ \ \ \  + \partial_m \gamma_{ki} - \partial_k \gamma_{mi} \right) \\
&=&   {B_k}^i \partial_i \left( \Lambda v^k\right) \nonumber \\
& &+ \frac{\Lambda B^{k i}  v^m}{2}  \partial_m \gamma_{ki},
\end{eqnarray}
where in the last step, the first and third terms---antisymmetric in $i$ and $k$---vanish upon contraction with the symmetric $B^{k i}$.
Altogether, we have
\begin{equation}
B^{\nu\mu} \nabla_\mu u_\nu =  \frac{\Lambda B^{k i}  v^m}{2}  \frac{\partial \gamma_{ki}}{\partial x^m} - \Lambda B^{k i} K_{k i} + {B_k}^i  \frac{\partial\left( \Lambda v^k\right)}{\partial x^i}
\label{eq:Contraction_4}
\end{equation}
for this type of contraction.

In summary, Eqs.~(\ref{eq:Contraction_1}), (\ref{eq:Contraction_2}), (\ref{eq:Contraction_3}), and (\ref{eq:Contraction_4}) exhibit the four types of contractions appearing when either Eq.~(\ref{eq:U_Projection_n}) or (\ref{eq:U_Projection_gamma}) is contracted with $\nabla_\mu u_\nu$.

\section{Towards Discretization}
\label{app:TowardsDiscretization}

Elaboration of a full discretization of the conservative four-momentum moment Eqs.~(\ref{eq:EnergyConservation}) and (\ref{eq:MomentumConservation}) is beyond the scope of this paper, but we here comment briefly on the possibility of discretizations that are faithful to the analytic connection with the conservative number equation outlined in Sec.~\ref{sec:MomentumNumberExchange_31}.
Such consistent differencing has been addressed in some detail in spherical symmetry by Liebend\"{o}rfer et al. \cite{Liebendorfer2004A-Finite-Differ}; see also Mezzacappa et al. \cite{Mezzacappa2006Neutrino-Transp}.
We expect the consistency demonstrated by Liebend\"{o}rfer et al. \cite{Liebendorfer2004A-Finite-Differ} to be possible also for our equations beyond spherical symmetry.

A key analytic step in Sec.~\ref{sec:MomentumNumberExchange_31} is the use of the product rule for derivatives to pull factors inside the spacetime and momentum space divergences, leaving `extra' terms whose cancellation is required; see Eqs.~(\ref{eq:IdentitySpacetimeDivergence_n}), (\ref{eq:IdentitySpacetimeDivergence_gamma}), and (\ref{eq:u_M}).  
This operation---referred to as ``integration by parts'' by Liebend\"{o}rfer et al. \cite{Liebendorfer2004A-Finite-Differ}---is also central to the cancellations discussed in Secs.~3.1 and 3.3 of that work.
An example of a finite difference version of the product rule in the single spatial variable in spherical symmetry is shown in the unnumbered equation following Eq.~(58) of Liebend\"{o}rfer et al. \cite{Liebendorfer2004A-Finite-Differ}.
The structure of that representation of the product rule is not directly applicable to our multidimensional case, because they use the cell face value of the variable they `pull through' the derivative, and in multiple spatial dimensions there are a corresponding number of different cell face values.

However, we here demonstrate a finite difference representation of the product rule that works with a multidimensional spatial divergence.
In particular we give a representation of
\begin{equation}
g \frac{\partial F^i}{\partial x^i} = \frac{\partial \left( g F^i \right)}{\partial x^i} 
- F^i \frac{\partial g}{\partial x^i}
\label{eq:ProductRuleContinuum}
\end{equation}
for arbitrary $g$ and $F^i$.
Begin for example with a finite-volume-inspired discretization
\begin{equation}
\left( g \frac{\partial F^i}{\partial x^i} \right)_\lrar = \frac{g_\lrar}{V}
\sum_q \left[\left( A_q F^q \right)_\qrar - \left( A_q F^q \right)_\qlar\right]
\label{eq:ProductRule1}
\end{equation}
of the left-hand side.
Different from Liebend\"{o}rfer et al. \cite{Liebendorfer2004A-Finite-Differ}, we use the cell center value of the function $(g_\lrar)$ that initially is outside the divergence; thus it stands on an equal footing with respect to all dimensions. 
The summation convention on repeated indices $i$ on the left-hand side has given way to an explicit sum over dimensions $q$ on the right-hand side.
Cell-centered values are denoted by a double-headed arrow subscript $(\lrar)$.
The single-headed arrow subscripts $(\qrar)$ and $(\qlar)$ denote values on outer and inner cell faces in dimension $q$ respectively.
The cell volume and face areas are $V$ and $A_q$. 
We split each of the two terms in Eq.~(\ref{eq:ProductRule1}) in half, add and subtract terms involving center values of $g$ from the next $(\qplus)$ and previous $(\qminus)$ cells in direction $q$, and rearrange to find
\begin{equation}
\left( g \frac{\partial F^i}{\partial x^i} \right)_\lrar = \left[ \frac{\partial \left( g F^i \right)}{\partial x^i} \right]_\lrar
- \left( F^i \frac{\partial g}{\partial x^i} \right)_\lrar.
\label{eq:ProductRule2}
\end{equation}
Here
\begin{equation}
\left[ \frac{\partial \left( g F^i \right)}{\partial x^i} \right]_\lrar 
= \frac{1}{V}
\sum_q \left[\left( A_q\, g F^q \right)_\qrar - \left( A_q\, g F^q \right)_\qlar\right],
\end{equation}
in which we are representing the values of $g$ on the cell faces as
\begin{eqnarray}
g_\qrar &=& \frac{g_\lrar + g_\qplus}{2}, \\
g_\qlar &=& \frac{g_\qminus + g_\lrar}{2}.
\label{eq:gFace}
\end{eqnarray}
As for the second term in Eq.~(\ref{eq:ProductRule2}), it can be regarded as the average of values on opposing cell faces:
\begin{equation}
\left( F^i \frac{\partial g}{\partial x^i} \right)_\lrar 
= \frac{1}{2} \sum_q \left[ \left( F^q \frac{\partial g}{\partial x^q} \right)_\qrar + \left( F^q \frac{\partial g}{\partial x^q} \right)_\qlar \right],
\label{eq:gExtra1}
\end{equation}
where
\begin{eqnarray}
\left( F^q \frac{\partial g}{\partial x^q} \right)_\qrar &=& \frac{\left( A_q F^q \right)_\qrar}{V} \left(g_\qplus - g_\lrar \right), \label{eq:gExtra2} \\
\left( F^q \frac{\partial g}{\partial x^q} \right)_\qlar &=& \frac{\left( A_q F^q \right)_\qlar}{V} \left(g_\lrar - g_\qminus \right). \label{eq:gExtra3}
\end{eqnarray}
Equations~(\ref{eq:gExtra1})-(\ref{eq:gExtra3}) are not the most obvious discretization of $\left( F^i \partial g / \partial x^i \right)_\lrar $ one would think to write down, but inspection shows that it is not unreasonable.
This is the sort of thing we have in mind when we say, following Eq.~(\ref{eq:Extra_n}) for example, that ``the discretized form of $\mathsf{E}_{T,n}$ will be dictated by the discretization chosen for the first two terms of Eq.~(\ref{eq:SpacetimeDivergence_31_n}).''
The discretized forms of the derivatives in $\mathsf{E}_{T,n}$ and $\mathsf{E}_{T,\gamma}$ derived from the above sort of procedure can then be used to represent the spatial derivatives that appear in the momentum space divergence, i.e. in $\left(\mathsf{O}_{T,n}\right)$ and $\left(\mathsf{O}_{T,\gamma}\right)_j$, so that the cancellations in Eq.~(\ref{eq:CancellationObserver}) can be effected even at modest resolution.

With this key step in the connection between discretized four-momentum and number conservation generalized to the multidimensional case, we do not see any showstopping impediment to carrying a discretization to completion along the lines of Liebend\"{o}rfer et al. \cite{Liebendorfer2004A-Finite-Differ}.
We have not specified representations of the face values $\left( F^q \right)_\qrar$ and $\left( F^q \right)_\qlar$ in Eq.~(\ref{eq:ProductRule1}), so these can be discretized according the considerations given in Liebend\"{o}rfer et al. \cite{Liebendorfer2004A-Finite-Differ} and references therein, with the results being carried forward into Eqs.~(\ref{eq:gExtra2}) and (\ref{eq:gExtra3}). 
We do not foresee any overconstraints, or other serious issues beyond those faced and addressed by Liebend\"{o}rfer et al. \cite{Liebendorfer2004A-Finite-Differ}; 
but of course this remains to be seen with a complete implementation and numerical testing.

%\begin{eqnarray}
%\left( g \frac{\partial F^i}{\partial x^i} \right)_\lrar &=& \frac{1}{V}
%\sum_q \left[\left( A_q F^q \right)_\qrar \left( \frac{g_\lrar + g_\qplus}{2} \right)
%+ \frac{\left( A_q F^q \right)_\qrar g_\qplus}{2} \right. \nonumber \\
%& & \left. - \frac{\left( A_q F^q \right)_\qlar g_\lrar}{2}\right]
%\end{eqnarray}

\begin{acknowledgments}
We thank Evan O'Connor for useful discussions and corrections.
This research was supported by the Office of Advanced Scientific Computing Research and the Office of Nuclear Physics, U.S. Department of Energy. 
\end{acknowledgments}

\def\apjl{Astrophys. J. Lett.}
\def\apjs{Astrophys. J. Suppl. Ser. }
\def\mnras{Mon. Not. Roy. Ast. Soc. }
\def\aap{Astron. Astrophys. }
%\bibliography{bibliography10}


\end{document}